\documentclass[11pt,letterpaper,usenames,dvipsnames]{article}
\usepackage{jheppub}

\RequirePackage{pdf14}
\usepackage{enumitem}
\usepackage[utf8]{inputenc}
\usepackage{graphicx,tikz}
\usepackage[framemethod=TikZ]{mdframed} 
\usepackage{amsfonts,amsmath,amssymb,amsthm,esint}
\usepackage{array,setspace,mathrsfs,yfonts,dsfont,bbm}
\usepackage{colonequals,amscd,mathtools}
\usepackage{relsize,suffix,cancel,bbm,capt-of,upgreek,verbatim,xcolor}
\usepackage[T1]{fontenc}
\usepackage{enumitem}
\usepackage{hhline}
\usepackage{arydshln}
\usepackage{arydshln}
\usepackage{multirow}
\usepackage{blkarray}
\usepackage{floatrow,subfig}
\usepackage{tcolorbox}
\usepackage{blindtext}
\usetikzlibrary{patterns,arrows,decorations.markings}

\usepackage{tikz}
\usetikzlibrary{arrows,snakes,shapes.arrows,decorations.markings}
     \tikzset{>=triangle 90}
     \tikzstyle{bbc}=[draw,circle,fill=black,scale=.75]
     \tikzstyle{rc}=[circle,fill=red,scale=.6]
     \tikzstyle{wc}=[draw,circle,scale=.75]

\def\red#1{{\color{red}{#1}}}

\def\del{{\partial}}

\def\bar{\overline}
\def\til{\widetilde}

\def\vev#1{{\langle{#1}\rangle}}

\def\^{\wedge}
\def\I{\mathds{1}}

\def\Im{{\rm Im}}

\def\U{{\rm U}}
\def\SU{{\rm SU}}

\def\SL{{\rm SL}}
\def\GL{{\rm GL}} 
\def\Sp{{\rm Sp}}

\def\a{{\alpha}}

\def\g{{\gamma}}

\def\G{{\Gamma}}

\def\D{{\Delta}}
\def\Dsi{\D^{\rm sing}}

\def\th{{\theta}}

\def\l{{\lambda}}

\def\L{{\Lambda}}
\def\Lt{{\til\L}}

\def\r{{\rho}}
\def\s{{\sigma}}
\def\sb{{\bar{\sigma}}}
\def\S{{\Sigma}}

\def\t{{\tau}}

\def\f{{\phi}}
\def\vf{{\varphi}}

\def\w{{\omega}}

\def\ba{{\boldsymbol a}}

\def\bh{{\boldsymbol h}}
\def\bm{{\boldsymbol m}}
\def\bp{{\boldsymbol p}}
\def\bq{{\boldsymbol q}}
\def\bu{{\boldsymbol u}}

\def\af{\mathfrak{a}}
\def\cf{\mathfrak{c}}
\def\df{\mathfrak{d}}
\def\ef{\mathfrak{e}}
\def\ff{\mathfrak{f}}
\def\gf{\mathfrak{g}}

\def\nf{\mathfrak{n}}

\def\slf{\mathfrak{sl}}

\def\spf{\mathfrak{sp}}
\def\suf{\mathfrak{su}}

\def\mT{\mathsf{T}}

\def\cC{{\mathcal C}}
\def\cCb{\overline\cC}
\def\cCrg{\cC_{\rm reg}}

\def\cE{{\mathcal E}}

\def\cH{{\mathcal H}}

\def\cM{{\mathcal M}}

\def\cN{{\mathcal N}}

\def\cS{{\mathcal S}}
\def\cSb{{\bar\cS}}
\def\cSbme{\cSb_{\rm metr}}
\def\cSbcp{\cSb_{\rm cplx}}
\def\cSpq{\cS_{(p,q)}}
\def\cSbpq{\bar{\cS}_{(p,q)}}
\def\cT{{\mathcal T}}

\def\C{\mathbb{C}} 
\def\DD{\mathbb{D}} 
\def\H{\mathbb{H}}
\def\N{\mathbb{N}} 
\def\O{\mathbb{O}}

\def\R{\mathbb{R}} 
 
\def\Z{\mathbb{Z}} 

\def\beq{\begin{equation}}
\def\eeq{\end{equation}}
\def\nn{\nonumber}
\newcommand{\bpmat}{\begin{pmatrix}}
\newcommand{\epmat}{\end{pmatrix}}
\newcommand{\bsmat}{\begin{smallmatrix}}
\newcommand{\esmat}{\end{smallmatrix}}
\newfloatcommand{capbtabbox}{table}[][\FBwidth]
\newtheorem{fact}{Fact}
\newtheorem*{conjecture}{Conjecture}

\newcounter{ex}[section]
\renewcommand{\theex}{\arabic{section}.\arabic{ex}}

\newcounter{example}[section]
\renewcommand{\theexample}{\arabic{section}.\arabic{example}}

\newcounter{dig}[section]
\renewcommand{\thedig}{\arabic{section}.\arabic{dig}}

\newcounter{definition}[section]
\renewcommand{\thedefinition}{\arabic{section}.\arabic{definition}}

\newcounter{solution}[section]
\renewcommand{\thesolution}{\arabic{section}.\arabic{solution}}

\title{Towards a classification of rank $r$ $\mathcal{N}=2$ SCFTs\\ 
\Large{Part II: special Kahler stratification of the Coulomb branch}}
\author[1]{Philip C. Argyres}
\author[2]{and Mario Martone}
\affiliation[1]{University of Cincinnati, Physics Department, PO Box 210011, Cincinnati OH 45221}
\affiliation[2]{University of Texas, Austin, Physics Department, Austin TX 78712}
\emailAdd{philip.argyres@gmail.com}
\emailAdd{mariomartone@utexas.edu}

\abstract{We study the stratification of the singular locus of four dimensional $\cN=2$ Coulomb branches.  We present a set of self-consistency conditions on this stratification which can be used to extend the classification of scale-invariant rank 1 Coulomb branch geometries to two complex dimensions, and beyond.  The calculational simplicity of the arguments presented here stems from the fact that the main ingredients needed --- the rank 1 deformation patterns and the pattern of inclusions of rank 2 strata --- are discrete topological data which satisfy strong self-consistency conditions through their relationship to the central charges of the SCFT.  This relationship of the stratification data to the central charges is used here, but is derived and explained in a companion paper \cite{Martone:2020nsy} by one of the authors.  We illustrate the use of these conditions by re-analyzing many previously-known examples of rank 2 SCFTs, and also by finding examples of new theories.  The power of these conditions stems from the fact that for Coulomb branch stratifications a conjecturally complete list of physically allowed ``elementary slices'' is known.  By contrast, constraining the possible elementary slices of symplectic singularities relevant for Higgs branch stratifications remains an open problem.
}
  
\begin{document}
\maketitle

\section{Introduction}

After many decades of investigation, the study of superconformal field theories (SCFTs) continues to provide new valuable lessons. Recently substantial progress has been achieved in the systematic understanding of SCFTs in both five and six dimensions, see e.g. \cite{Heckman:2018jxk,Bhardwaj:2019hhd,Apruzzi:2019enx,Apruzzi:2019opn,Apruzzi:2019kgb,Bhardwaj:2020gyu} and reference therein.  But, perhaps surprisingly, the four dimensional case, despite tremendous effort still escapes a complete understanding and even a conjectural classification is missing.  The case of minimal supersymmetry in 4d seems at the moment to be out of reach,\footnote{Cf.\ \cite{Razamat:2020pra} for some interesting progress in $\cN=1$ SCFTs with a weak-coupling limit.} so the $\cN\ge2$ arena is likely the best place to test our systematic understanding of SCFTs in four dimensions.  For $\cN=4$ (maximal) global supersymmetry in 4d a picture which could provide a full classification has been presented in \cite{Aharony:2013hda} and \cite{Bourget:2018ond,Argyres:2018wxu}.  $\cN=3$ theories \cite{Ferrara:1998zt,Garcia-Etxebarria:2015wns,Aharony:2015oyb,Aharony:2016kai} seem almost as constrained as $\cN=4$ and some progress towards systematically understanding them has been achieved \cite{Argyres:2019ngz,Argyres:2019yyb}.  But it is difficult to test this understanding at present, and better tests await the development of new techniques in string constructions, the superconformal bootstrap, or in $\cN=2$ RG flows.  The classification of 4d $\cN=2$ SCFTs, being less constrained than their $\cN=3$ brethren, would seem a much more difficult task.  In this paper we will argue the opposite. 

It is both authors' conviction that one of the most promising routes to a classification of $\cN \ge 2$ SCFTs in four dimensions is via a systematic study of the moduli space of these theories and in particular of their Coulomb branches.\footnote{See \cite{Argyres:2020nrr} for a very short summary of the overall philosophy of our approach.} These moduli space geometries and their fate under deformation of the SCFT by $\cN=2$-preserving relevant operators (whose analogs do not exist in $\cN=3$ theories), carry a tremendous amount of information about the underlying field theory.  Furthermore, they carry a variety of mathematical structures which constrain the allowed possibilities and which, leveraged by a good dose of physics intuition, can bring a full understanding of the possible cases.  This was the case for the complete classification of rank 1 Coulomb branch geometries, which stands still strong to date, achieved by the authors as well as some of their closest collaborators \cite{Argyres:2015ffa, Argyres:2015gha, Argyres:2016xua, Argyres:2016xmc}.  The rank 1 case is in many ways non-generic and thus one might think that it is the only case in which a complete classification can be achieved.  But,  by taking advantage of a kind of inductive structure relating SCFTs with moduli spaces of different dimensions, as well as of the existence of $\cN=2$-preserving relevant deformations, we will be able to present calculational evidence for the efficacy of a simple approach to the classification of 4d $\cN=2$ SCFTs beyond rank 1.  In particular, this paper, using the results of a companion paper \cite{Martone:2020nsy}, highlights a tight structure of self-consistency conditions on the topology of rank 2 and higher Coulomb branch geometries. The result is that the rank 1 geometries act as a kind of essential building block of higher-dimensional Coulomb branch geometries, and so the rank 1 classification opens the door to the exploration of a set of relatively simple algebraic constraints on higher-rank geometries.  We do not solve these constraints in any generality here, but do provide many examples to illustrate the tightness of these constraints.\footnote{The examples presented here --- chosen to illustrate various points --- are a subset of all the ones which we have worked out.  We did not include all our examples out of sheer impatience; the interested reader can request from the authors a Mathematica{\texttrademark} notebook with all examples to date.}

The central idea of this paper is a systematic understanding of the singular locus of higher complex dimensional Coulomb branches. Since the seminal work of Seiberg and Witten \cite{Seiberg:1994rs, Seiberg:1994aj}, it has been clear that the richness of the Coulomb branch geometry lies in fact in its singularities rather than in its smooth points. Here we will formalize the fact that the singular locus, aside from strata of specific kind clarified below, provides the Coulomb branch with a \emph{special Kahler stratification}.\footnote{It is not the first time that the existence of a special Kahler stratification of the Coulomb branch is discussed and it was partially present in previous papers by the authors \cite{Argyres:2018zay,Argyres:2018urp} and was fully elaborated in the context of $\cN=3$ theories in \cite{Argyres:2019yyb}.  This concept has also been studied in detail recently in \cite{Beem:2019tfp, Bourget:2019rtl, Bourget:2019aer, Grimminger:2020dmg} for Higgs branches, which are holomorphic symplectic varieties and enjoy a hyperkahler stratification \cite{beauville1999symplectic,kaledin2006symplectic}
} The analysis of this stratification elucidates that the Coulomb branch singularity has more structure than was previously appreciated, and that this structure can be leveraged considerably to classify $\cN=2$ SCFTs.  The stratification of the Coulomb branch singular locus is at same time more constraining and richer than its Higgs branch counterpart.  The stronger constraints come from the simple fact that the complex dimension of the Coulomb branch strata jump by exactly one at every step which implies that, following a terminology introduced in \cite{Bourget:2019rtl}, the \emph{elementary slices} of the stratification of the Coulomb branch are all entries in an appropriately extended Kodaira list (see below). In other words the $r-1$ dimensional singular locus of a rank $r$ $\cN=2$ SCFT can be ``decomposed'' into extremely simple building blocks, namely a nested series of one complex-dimensional spaces each with a single point-like singularity, and which have been thoroughly understood by now. In a restricted set of cases, the strata only inherit a weaker version of special Kahler geometry allowing for a richer behavior. 

In addition to explaining this special Kahler stratification, one goal of this paper is to make this picture more concrete in the context of rank 2 $\cN=2$ SCFTs.  The rank 2 moduli spaces which we present in our examples pass a variety of consistency checks.  Certainly the most stringent consistency condition on the Coulomb branch structure in our examples is a new relation between the stratification that we present in this paper and the conformal and flavor central charges of the underlying microscopic SCFT.   This relation is derived and discussed in a separate paper \cite{Martone:2020nsy} by one of the authors, but here we put it to work. We are able to beautifully reproduce all the properties of rank-2 theories from their rank-1 building blocks. In particular we find perfect agreement with the $\cN=2$ \emph{UV-IR simple flavor condition} \cite{Martone:2020nsy}.

We also discuss preliminary self-consistency conditions on the way the Coulomb, mixed, and Higgs branch structures come together.  This idea has inspired a forthcoming work which revisits the rank 1 classification from the chiral algebra side \cite{CCLMW2020}, and was recently developed in the context of $\cN=4$ three dimensional theories \cite{Grimminger:2020dmg}. A tool that we will utilize to present these results, very much along the lines of \cite{Grimminger:2020dmg},  is the \emph{Hasse diagram}, which is an efficient way to pictorially present a partially ordered set, in this case the poset provided by the stratification.  We will introduce Hasse diagrams for Coulomb branches of four dimensional $\cN=2$ SCFTs below and extend them, in a few specific examples, to the full moduli space Hasse diagrams. 

In our analysis we also discover a series of striking facts about the IR behavior of higher rank $\cN=2$ SCFTs. First, some non-discretely gauged rank 2 geometries naturally flow, for special values of their Coulomb branch parameters, to discretely gauged rank 1 theories \cite{Argyres:2016yzz}, see in particular sec \ref{subsec:Discrete}.  Depending on the reader's perspective this might be more or less surprising.  In either case, the appearance of a discretely gauged theory on a sub-locus of a non-discretely gauged moduli space of vacua implies that discretely gauged theories can be embedded in larger theories. This provides a potentially interesting way to learn more about these theories, particularly given that starting at rank 3 we would be able to probe some of the exotic phenomena which take place in the context of discretely gauged rank 2 theories \cite{Argyres:2018wxu}.

Secondly, in analyzing the elementary transverse slices of the Coulomb branch stratification, we find, unexpectedly, that this list extends beyond the list of positive curvature geometries often associated with the Kodaira classification of elliptic singularities to include the seemingly unphysical negative curvature  \emph{irregular geometries} also associated with the Kodaira list and discussed in \cite{Argyres:2017tmj}. Irregular geometries present the enticing phenomenon of apparent violation of the conformal unitarity bound.  This naive violation is evaded by the occurrence of non-trivial relations among the generators of the Coulomb branch coordinate ring. It is still unclear to the authors the extent to which the appearance of these geometries is relevant for the IR physics of $\cN=2$ SCFTs. It certainly largely confirms the authors' belief that to fully understand higher rank geometries, we need to utilize the full range of mathematically allowed lower rank behaviors.  

Finally, the stratification, as we will thoroughly explain below, is at its core made of strata and their transverse slices. We find that, in almost all cases, they both inherit a special Kahler structure from the ambient space (the full Coulomb branch geometry).  We characterize the special cases in which strata do not inherit a full special Kahler structure, and show that they inherit, instead, a weaker version which we call a \emph{loose special Kahler structure}.  This is simply a special Kahler structure in which $\Sp(2r,\R)$ electric-magnetic duality monodromies (as opposed to $\Sp(2r,\Z)$ ones) are allowed.  While the interpretation of the transverse slice as a special Kahler space is clear from the physics, at least locally, the significance of the (perhaps loose) special Kahler structure on the strata remains somewhat obscure.  We are able to leverage some of the constraints that arise from this structure when we discuss the stratification of the full moduli space. But we believe that our understanding of the strata as special Kahler spaces might still reserve important physics implications.

This paper is organized as follows. The next section constructs mathematically the stratification of Coulomb branch as special Kahler space for arbitrary ranks and clarifies when the constraint is instead weaker. Section \ref{sec:rank2} specifies the discussion to rank 2 where the picture somewhat simplifies. After these two fairly mathematical sections, we illustrate the constraining power of the stratification structure of rank 2 Coulomb branches with plenty of examples in section \ref{sec:examp}. Hasse diagrams for Coulomb branches of $\cN=2$ SCFTs appear prominently.  Section \ref{sec:FullModuli} outlines the idea of the stratification of the full moduli space of vacua (i.e., also including any Higgs and mixed branches) in a few examples. The discussion there is neither complete nor systematic. We close in section \ref{sec:Conclusions} with a brief discussion of some of the novel behaviors encountered in our examples, outlining the next natural steps towards a systematic classification of rank 2 (and higher) Coulomb branch geometries. 

\section{Stratification of Coulomb branch singularities}\label{sec:CBstrat}

The general branch of the moduli space of an $\cN=2$ field theory, see, e.g., \cite{Argyres:1996eh, Argyres:2016xmc}, is one where there are both $n_v$ massless vector multiplets and $n_h$ massless neutral hypermultiplets.  This follows from the fact that the vector multiplet and the hypermultiplet are the only free $\cN=2$ superconformal multiplets which could contain the dilaton associated with the spontaneously broken conformal symmetry.  If both $n_v$ and $n_h$ are non-zero, these are called mixed branches.  A branch whose generic point has only massless vector multiplets ($n_h=0$) is a special Kahler variety called the Coulomb branch ($\cC$), while a branch whose generic point has no vector multiplets ($n_v=0$) is a hyperkahler variety called a Higgs branch.  A mixed branch with $(n_v, n_h) = (n_v^\text{mixed}, n_h^\text{mixed})$ intersects the Coulomb branch along an $n_v^\text{mixed}$-complex-dimensional special Kahler subvariety.  It can likewise intersect a Higgs branch along an $n_h^\text{mixed}$-quaternionic-dimensional hyperkahler subvariety.  (Also, mixed branches can intersect each other in both special Kahler and hyperkahler directions.) 

After reminding the reader about the generalities of the Coulomb branch structure, we will outline how much of the information about Coulomb branch singularities, at any rank, can be reduced to well known information about rank-1 scale invariant geometries and  we will look closely at the details of the stratification of the Coulomb branch singular locus.

\subsection{Coulomb branch generalities}

We now briefly recall the ingredients of the special Kahler geometry of the Coulomb branch and their connections to low energy physics.  Along the way we will introduce a few assumptions.  These are that the charge lattice is principally polarized, that the Coulomb branch has a $\C^*$ complex homothety inherited from a microscopic superconformal invariance, and that the Coulomb branch chiral ring is freely generated.  We will explain these assumptions below as they come up.  They are in the nature of simplifying assumptions meant to make the discussion and exposition easier; our central result on the special Kahler stratification of the Coulomb branch holds without them.

As previously mentioned, the low-energy theory on a generic point of $\cC$ is simply a free $\cN=2$ supersymmetric $\U(1)^r$ gauge theory with no massless charged states. $r$ is called the \emph{rank} of the theory and coincides with the complex dimensionality of $\cC$, ${\rm dim}_\C\cC=r$. $\cC$ is a singular space and its singular locus will be denoted as $\cSb$.  ($\cS$ stands for ``singular" and also ``stratum" as we will see.)  $\cSb$ is a closed subset of $\cC$, since there is no consistent physical interpretation of the IR effective action at boundary points of $\cSb$ which are not in $\cSb$ \cite{Argyres:2018zay}.  The smooth part of the Coulomb branch is $\cCrg := \cC \setminus \cSb$.  Thus $\cCrg$ is an open subset of $\cC$.  Note that without further assumptions, $\cCrg$ need not be connected.

The Coulomb branch is both a complex space and a metric space, and so $\cC$ can have singularities in each of these structures \cite{Argyres:2018wxu}.  Its set of \emph{metric} singularities is denoted $\cSbme$.  From basic physical principles --- that there cannot be a transition among two inequivalent vacua at zero energy cost --- the moduli space is necessarily a metric space.  This means  that it has a well-defined distance function, measuring local energy costs, but this distance function may not be derived from a smooth Riemannian metric.  So $\cSbme$ is the locus of metric non-analyticities in $\cSb$.  We call the set of \emph{complex structure} singularities $\cSbcp$.  The singular locus is the union of the loci of the two types of singularities, $\cSb = \cSbme \cup \cSbcp$.  Typically, $\cSbcp$ is a subset of $\cSbme$, though we will see a physical example where we are led to consider a Coulomb branch geometry with a metrically smooth point which is nevertheless a complex singularity. 

The physics interpretations of metric and complex singularities of the Coulomb branch are remarkably different.  $\cSbme$ is the locus of $\cC$ where extra charged states become massless or, in other words, where the low-energy physics is not captured solely by a bunch of free $\cN=2$ vector multiplets: it may still be free in the IR, or it may correspond to an interacting IR fixed point. $\cSbcp$ is instead the locus of vacua for which the operators generating the corresponding Coulomb branch chiral ring satisfy non-trivial relations.  This means that the chiral ring is not freely generated at points in $\cSbcp$.  

A central fact about Coulomb branch geometry is that there is no globally defined lagrangian description of the low energy $\cN=2$ $\U(1)^r$ gauge theory, and non-trivial monodromies have to be considered to describe the physics on $\cCrg$. These are specific elements of the electric-magnetic duality group and which depend on the physics at the singular loci, and in particular on $\cSbme$, and can therefore be used to characterize it. 

Because of the unbroken low energy $\U(1)^r$ gauge invariance on $\cC$, states in the low-energy theory are labeled by a set of $2r$ integral electric and magnetic charges, $Q\in\Z^{2r} = \L$, the \emph{charge lattice}.  The pairing induced by the Dirac-Zwanziger-Schwinger quantization condition on $\L$ gives the charge lattice an integral symplectic structure.  We denote the Dirac pairing by $\langle Q, Q'\rangle:=Q^T\DD Q'$, where $\DD$ is an integer non-degenerate skew-symmetric $2r\times 2r$ matrix, and we are using a matrix notation where we consider $Q$ as a $2r$-component column vector.  We call $\DD$ the \emph{polarization} of the charge lattice.

Low energy $\U(1)^r$ electric-magnetic duality is reflected in the fact that while the charge lattice remains the same over all points in $\cCrg$, upon dragging a given $Q\in\L$ along a closed path $\g\subset \cCrg$ it need not return to the same value, but may suffer a \emph{monodromy} of the form 
\begin{align}\label{}
Q &\overset{\g}{\rightsquigarrow} Q' = MQ &
&\text{with} &
M &\in \Sp_\DD(2r,\Z).
\end{align}
Here $\Sp_\DD(2r,\Z)$ is the electric-magnetic \emph{duality group}; it is the subgroup of $\GL(2r,\Z)$ matrices satisfying $M\DD M^T=\DD$, since the Dirac pairing of charges is preserved under monodromies.  This is summarized by saying that the set of $\L$'s fibered over $\cCrg$ forms a linear system with structure group $\Sp_\DD(2r,\Z)$.

A full understanding of the physical meaning of the polarization remains largely an open question which we will not address here. In fact we will assume in the rest of this paper that the polarization $\DD$ is one which can be brought by a choice of lattice basis to the canonical symplectic form
\begin{align}\label{prinpol}
\DD=\bpmat 0 & \I_r\\-\I_r & 0 \epmat 
\end{align}
in terms of $r\times r$ blocks. Then $\Sp_\DD(2r,\Z)\cong \Sp(2r,\Z)$ and $\DD$ is called \emph{principal}.  In a canonical basis in which a principal $\DD$ is given by \eqref{prinpol}, the charge vector can be written as
\begin{align}\label{}
Q &= \bpmat \bp \\ \bq \epmat, &
&\text{with} &
\bp &= \bpmat p^1\\ \vdots\\ p^r \epmat, &
&\text{and} &
\bq &= \bpmat q_1\\ \vdots\\ q_r \epmat ,
\end{align}
where (conventionally) $\bp$ are the magnetic and $\bq$ the electric charges.

The complex central charge, $Z_Q$, of the low energy $\cN=2$ supersymmetry algebra of a vacuum in $\cCrg$ acting on the superselection sector of states with charge $Q \in \L$ is a locally holomorphic function on $\cCrg$ and depends linearly on $Q$.  This means that the central charge is given by a holomorphic section, $\s$, of the rank-$2r$ complex $\Sp(2r,\Z)$ vector bundle dual to the charge lattice bundle.  We write this section as a $2r$-component column vector of \emph{special coordinates}, 
\begin{align}\label{specCo}
\s &:= \bpmat \ba^D\\ \ba\ \ \epmat,& 
&\text{with}&
\ba^D & := \bpmat a^D_1\\ \vdots\\ a_r^D \epmat, &
&\text{and}&
\ba &:=\bpmat a^1\\ \vdots\\a^r\epmat .
\end{align}
The Dirac pairing $\DD$ induces a symplectic product on the space of $\s$'s, and the splitting of $\s$ into $\ba^D$ and $\ba$ shown in \eqref{specCo} reflects the canonical symplectic form of $\DD$ chosen in \eqref{prinpol}.  The central charge is then
\beq\label{ccBPS}
Z_Q := Q^T\s ,
\eeq
where the dual pairing between the charges and the special coordinates is given by the matrix transpose.  It follows from the $\cN=2$ supersymmetry algebra that $|Z_Q|$ is a lower bound on the mass of any state with charge $Q$.

The special coordinates on $\cCrg$ satisfy the additional constraint that
\begin{align}\label{SKconds}
\vev{d\s \, \overset{\^}{,} \, d\s} = 0, 
\end{align}
where $d$ is the exterior derivative on $\cCrg$ and $\vev{\cdot,\cdot}$ is the symplectic product induced by the Dirac pairing.  \eqref{SKconds} implies that $\t_{ij} := \frac{\del a^D_i}{\del a^j} = \t_{ji}$, and that $\ba^D$ and $\ba$ are separately good holomorphic coordinates on $\cCrg$.  $\t_{ij}$ is the matrix of low energy $\U(1)^r$ gauge couplings, and 
\begin{align}\label{SKmetric}
ds^2 
= i \vev{d\sb , d\s}
= \Im ( da^D_j d\bar a^j)
\end{align}
is a Kahler metric on $\cCrg$.  Positivity of this metric implies $\Im(\t_{ij})$ is positive definite.

All this together equips $\cCrg$ with a \emph{rigid special Kahler} structure.  This structure can be extended in a natural way to the singular locus, $\cSb$, of the Coulomb branch, as we will shortly elaborate on.  

Our main objects of study in this paper are $\cN=2$ superconformal field theories.  This superconformal symmetry group includes an $\R^+ \times \U(1)_R\times \SU(2)_R$ dilatation plus R-symmetry subgroup which can be spontaneously broken, and so acts nontrivially on the space of vacua.  There is a single vacuum of the entire moduli space which is invariant under dilatations.  We call this superconformal vacuum the origin of the moduli space.  While the $\SU(2)_R$ is unbroken on $\cC$ and therefore acts trivially, the $\U(1)_R$ symmetry is spontaneously broken away from the origin, and its action combines with the $\R^+$ dilation action of the conformal algebra to give a $\C^*$ action on $\cC$. The entire structure of $\cC$ has to be compatible with this $\C^*$ action and in particular $\cSb$ and $\cCrg$ have to be closed under said action. We will often refer to the set of constraints arising from the compatibility with the $\C^*$ action as the constraints coming from scale invariance. In this language we will only consider here scale invariant Coulomb branch geometries. 

Although there are known examples of $\cN=2$ SCFTs with non-empty locus, $\cSbcp$, of complex singularities \cite{Bourget:2018ond,Argyres:2018urp}, for simplicity we will henceforth assume that $\cSbcp=\varnothing$ and that the Coulomb branch chiral ring of the SCFT we are analyzing is freely generated. This implies, in particular, that there is a set of $r$ complex coordinates, $\bu$, with definite scaling dimensions, which are globally defined on $\cCrg$ and which we will call the \emph{scaling coordinates} of the Coulomb branch.  Note that this assumption implies that $\cCrg$ is a connected smooth $r$-dimensional special Kahler manifold, though it is neither compact nor metrically complete.

An interesting phenomenon which we will touch upon below, is that even in this case, some of the strata, $\cSb_i$ (see below), of $\cC$ can nevertheless have complex singular locus $\cSb_{i,\rm cplx} \neq \varnothing$.  Thus even though we start with a Coulomb branch with a regular complex structure, this property need not be preserved by its special Kahler stratification.

\subsection{Complex analytic structure of the singular locus}
\label{ssCplxAnal}

Having established the generalities of the Coulomb branch structure, we can delve into the details of the singular locus, with the aim of extracting interesting constraints on the geometry of $\cSb$.  

$\cSb$ is a complex analytic subspace of $\cC$ whose irreducible components, $\cSb_i$, have codimension 1 and are each associated to a sublattice $\L_i \subset \L$ of electric and magnetic charges.  This is the sublattice spanned by the charges of those states becoming massless along $\cSb_i$.  This is because the $\cSb_i$ are defined by the vanishing of the holomorphic central charge given by \eqref{ccBPS}.  We will review this argument briefly in this subsection; a more detailed discussion of parts of this argument are given in sections 2.2 and 4.2 of \cite{Argyres:2018zay}, and in \cite{Argyres:2018urp}.

We assume that pathological behaviors which allow $\cSb$ to have accumulation points becoming dense in $\cC$ do not occur.\footnote{\cite{Argyres:2018zay} gives a more rigorous formulation of this assumption.}  Then there is a collection of open sets $U \subset \cSb$ covering a dense subset of $\cSb$, and for which each $U$ is described physically as the set of vacua where some particular charged states become massless.  Denote the set of electric and magnetic $\U(1)^r$ charges of these massless states by $\Phi_U \subset \L$.  The locus of $\cC$ where these states could become massless is given by the zeros of their central charges, $\cSb_U := \{\bu\in\cC \,|\, Z_Q(\s(\bu))=0, \ \ \forall\, Q\in\Phi_U \}$.  

We now show that $\cSb_U \subset \cSb$.  This might fail to be true if walls of marginal stability across which the spectrum of BPS states changes discontinuously divide $\cSb_U$ into separate components.  Note first that the set of charges of massless states, $\Phi_U$, need not be a sublattice.  Denote by $\L(\Phi_U)$ the sublattice integrally spanned by $\Phi_U$.  Since $Z_Q(\s) = Q^T\s$ is linear in $Q$, the set $\cSb_U$ only depends on the span $\L(\Phi_U)$ and not on $\Phi_U$ itself. 

As one continuously varies a point in $\cSb_U$, the set of charges of massless states, $\Phi_U$, cannot decrease unless one crosses a wall of marginal stability.  Say the sets of massless charged states on the two sides of a wall are $\Phi$ and $\Phi'$.  Walls of marginal stability can occur at loci where $|Z_Q| = |Z_{Q'_1}|+|Z_{Q'_2}|$ with $Q=Q'_1+Q'_2$, and $Q\in\Phi$, $Q'_1,Q'_2\in\Phi'$.  Thus $\L(\Phi) \subset \L(\Phi')$.  But since the argument also works with $\Phi$ and $\Phi'$ interchanged, $\L(\Phi') \subset \L(\Phi)$, so the two lattices are equal, and therefore their associated locus of vanishing masses are the same, $\cSb_U = \cSb_{U'}$.   We therefore conclude that $\cSb_U \subset \cSb$ irrespective of any intervening walls of marginal stability.

The section $\s$ does not diverge anywhere on $\cC$ for otherwise there would be a subsector of the theory which unphysically decouples at all scales \cite{Argyres:2018zay}.  $\cSb_U$ is defined by a finite number of equations, namely $Z_Q(\s(\bu))=0$ for a set of $Q$ spanning $\L(\Phi_U)$.  The $Z_Q(\s(\bu))$ are holomorphic away from $\cSb_U$, but may have branch points along $\cSb_U$ (corresponding to non-trivial electric-magnetic duality monodromies around $\cSb_U$).  This, together with the finiteness of $\s$, is nevertheless enough to show that $\cSb_U$ is a complex analytic subvariety of $\cC$ \cite{Forstneric:1992}.

The vanishing central charge equations defining $\cSb_U$ must, in fact, be proportional to one another for all $Q\in\Lt(\Phi_U)$, and $\cSb_U$ must therefore be of codimension 1 in $\cC$.  For suppose that $Z_Q=0$ and $Z_{Q'}=0$ were independent for two charges in $\L(\Phi_U)$.  They then define different codimension 1 subspaces of $\cC$ and so cannot both vanish on the open subset $U$ of $\cSb$.

We thus have that $\cSb$ is a complex codimension 1 analytic subvariety of $\cC$.  Each $\cSb_U$ defines an irreducible component of $\cSb$, though many different choices of $U$ will define the same component.  So we denote the irreducible components of the singular locus instead by
\beq\label{stra1}
\cSb^{(r-1)} := \bigcup_{i\in I^{(r-1)}}\cSb^{(r-1)}_i,\qquad 
\cSb^{(r-1)}_i := \Bigl\{\bu\in\cC \,\Big|\, Z_Q\left(\s(\bu)\right)=0, \quad \forall \, Q\in\L_i^{(r-1)} \Bigr\}.
\eeq
The distinct components are indexed by $i$ which runs over some finite index set $I^{(r-1)}$, and the component $\cSb^{(r-1)}_i$ is defined by the vanishing of the central charge for charges in the sublattice $\L_i^{(r-1)}$.  We have added the ``$(r-1)$'' superscripts to remind us of the complex dimension of these components; they will be useful momentarily when we discuss the stratification of $\cC$.

\subsection{Stratification and Hasse diagrams}

An analytic space, such as $\cSb^{(r-1)}$, admits a \emph{stratification}, essentially a decomposition into a set of disjoint lower-dimensional connected complex manifolds, called \emph{strata}.  We will extend this to a stratification of the whole Coulomb branch, $\cC$, by defining the strictly descending sequence of closed analytic spaces
\begin{align}\label{stra2}
\cC := \cSb^{(r)} \supset \cSb^{(r-1)} \supset \cdots 
\supset \cSb^{(d_j)} \supset \cSb^{(d_{j-1})} \supset
\cdots \supset \cSb^{(0)} \supset \cSb^{(-\infty)} \equiv \varnothing,
\end{align}
where the $(d)$ superscript denotes the complex dimension of each space and $d_j > d_{j-1}$.  This is a stratification of $\cC$ if the differences between neighboring spaces in this sequence, 
\begin{align}\label{stra3}
\cS^{(d_j)} := \cSb^{(d_j)} \setminus \cSb^{(d_{j-1})} ,
\end{align}
are smooth complex \emph{manifolds}.  Then $\cS^{(d)}$ has dimension $d$ and $\cSb^{(d)}$ is the closure of $\cS^{(d)}$ in $\cC$, justifying the notation.  Thus $\cS^{(r)} = \cC\setminus \cSb^{(r-1)} = \cCrg$, the regular subset of the Coulomb branch.  (For a more precise definition, discussion, and examples of topological and Whitney stratifications, see, e.g., \cite{Kirwan:2006}.)  

The manifolds $\cS^{(d)}$ may not be connected, and we denote their disjoint connected components by a subscript
\begin{align}\label{stra4}
\cS^{(d)} = \coprod_{i\in I^{(d)}}\cS^{(d)}_i,
\end{align}
where $I^{(d)}$ is some finite index set.  The connected $d$-dimensional complex manifolds $\cS^{(d)}_i$ are the \emph{strata} of the stratification.  By comparison, \eqref{stra1} is the closure of \eqref{stra4} for $d=r-1$; the spaces in the union on the right side of \eqref{stra1} are not disjoint, but intersect at lower-dimensional subvarieties.  Indeed, the set of all strata form a partially ordered set under inclusion of their closures.  Denote this partial ordering by inequalities such as $i < j$ for comparable indices $i,j$ in the combined index set $P := \coprod_d I^{(d)}$ of all strata.  Thus
\begin{align}\label{stra5}
i \le j   \qquad \Leftrightarrow \qquad \cS^{(d')}_i \subset \cSb^{(d)}_j .
\end{align}
Furthermore, the poset $P$ is graded by the dimension of the strata.  This means that there is a dimension function, ${\rm dim}: P \to \Z_{\ge0}$ given by ${\rm dim}(i) = d$ if $i \in I^{(d)}$, which is compatible with the ordering: ${\rm dim}(i) > {\rm dim}(j)$ if $i>j$.

It will be useful to talk about the properties of both the strata and their closures.  So, to make the discussion easier, we introduce some terminology: the closure of a stratum is the \emph{component} associated with the stratum, so
\begin{align}\label{}
\cSb^{(d)}_i \ \text{is the \emph{component} associated to the \emph{stratum}}\ \cS^{(d)}_i.
\nn
\end{align}
Note that a component is the disjoint union of its associated stratum with finitely many strata of lower dimension.

\begin{figure}[ht]
\centering
\begin{tikzpicture}
\begin{scope}[scale=.7,xshift=+0cm]
\fill[color=blue!20] (0,0) circle (3.35);
\draw[thick,draw=black!45,fill=black!10] (0,0) -- (1.5,-2.5) arc (375:165:1.5cm and .5cm) -- cycle;
\draw[thick,dashed,black!25] (1.5,-2.5) arc (15:165:1.5cm and .5cm);
\draw[rotate=150,thick,draw=black!45,fill=black!10] (0,0) -- (1.5,-2.5) arc (375:165:1.5cm and .5cm) -- cycle;
\draw[rotate=150,thick,black!45] (1.5,-2.5) arc (15:165:1.5cm and .5cm);
\draw[rotate=210,thick,draw=black!45,fill=black!10] (0,0) -- (1.5,-2.5) arc (375:165:1.5cm and .5cm) -- cycle;
\draw[rotate=210,thick,black!45] (1.5,-2.5) arc (15:165:1.5cm and .5cm);
\draw[ultra thick,draw=red] (0,0) -- (-.04,3.0);
\draw[rotate=50,ultra thick,draw=red] (0,0) -- (2.15,0);
\draw[rotate=260,ultra thick,draw=red] (0,0) -- (3.15,0);
\node[circle,fill=green!50!black,scale=.5] at (0,0) {};
\node[green!50!black] at (-.73,-0.12) {$\cS^{(0)}_0$};
\node[red] at (-.5,1.55) {\footnotesize $\cS^{(1)}_x$};
\node[red] at (0.7,1.4) {\footnotesize $\cS^{(1)}_y$};
\node[red] at (+.25,-1.6) {\small $\cS^{(1)}_z$};
\node[black!65] at (+.5,-2.6) {$\cS^{(2)}_c$};
\node[black!65] at (1.3,2.3) {$\cS^{(2)}_b$};
\node[black!65] at (-1.3,2.3) {$\cS^{(2)}_a$};
\node[blue!85] at (2.3,-1) {$\cS^{(3)}_1$};
\node at (0,-4.5) {\bf (a)};
\end{scope}
\begin{scope}[scale=.7,xshift=+9cm]
	\node [circle,scale=2,color=blue!40,draw,fill,inner sep=1pt, 
	label={[blue]above:$\cS^{(3)}_1$}] 
	at (0,3) [] (3) {};
	\node [circle,scale=2,color=black!20,draw,fill,inner sep=1pt, 
	label=right:$\cS^{(2)}_c$] 
	at (+2,1) [] (2c) {};
	\node [circle,scale=2,color=black!20,draw,fill,inner sep=1pt, 
	label=right:$\cS^{(2)}_b$] 
	at (0,1) [] (2b) {};
	\node [circle,scale=2,color=black!20,draw,fill,inner sep=1pt, 
	label=left:$\cS^{(2)}_a$] 
	at (-2,1) [] (2a) {};
	\node [circle,scale=2,color=red,draw,fill,inner sep=1pt, 
	label={[red]right:$\cS^{(1)}_z$}] 
	at (+2,-1) [] (1z) {};
	\node [circle,scale=2,color=red,draw,fill,inner sep=1pt, 
	label={[red]right:$\cS^{(1)}_y$}] 
	at (0,-1) [] (1y) {};
	\node [circle,scale=2,color=red,draw,fill,inner sep=1pt, 
	label={[red]left:$\cS^{(1)}_x$}] 
	at (-2,-1) [] (1x) {};
	\node [circle,scale=2,color=green!70!black,draw,fill,inner sep=1pt, 
	label={[green!50!black]below:$\cS^{(0)}_0$}] 
	at (0,-3) [] (0) {};
	\draw (3) edge (2c);
	\draw (3) edge (2b);
	\draw (3) edge (2a);
	\draw (2c) edge (1z);
	\draw (2b) edge (1y);
	\draw (2b) edge (1x);
	\draw (2a) edge (1x);
	\draw (1z) edge (0);
	\draw (1y) edge (0);
	\draw (1x) edge (0);
	\draw[black!70, dashed, rounded corners] (0.2,-3.2) -- (0.2,1.2) -- (-0.2,1.2) -- (-2.2,-.8) -- (-2.2,-1.2) -- (-0.2,-3.2) -- cycle;
\node at (0,-4.5) {\bf (b)};
\end{scope}
\end{tikzpicture}
\caption{Cartoon of a 3-dimensional scale-invariant Coulomb branch, where each real dimension in the figure represents 1 complex dimension.  The cones and lines are meant to extend to infinity; they are truncated here due to lack of space.  Figure (a)  shows a space where $\cS^{(d)}_i$ are the strata of dimension $d$ and $i$ is a unique label.  $\cS^{(3)}_1$ is the manifold of non-singular points of the Coulomb branch, $\cS^{(0)}_0$ is the unique superconformal vacuum, and the other strata are manifolds of metric and/or complex structure singularities.  The partial ordering by inclusion under closure among the strata is given in the Hasse diagram shown in figure (b).  The union of the strata enclosed by the dashed line is the component $\cSb^{(2)}_b$ associated to the stratum $\cS^{(2)}_b$.}
\label{fig:hasse}
\end{figure}
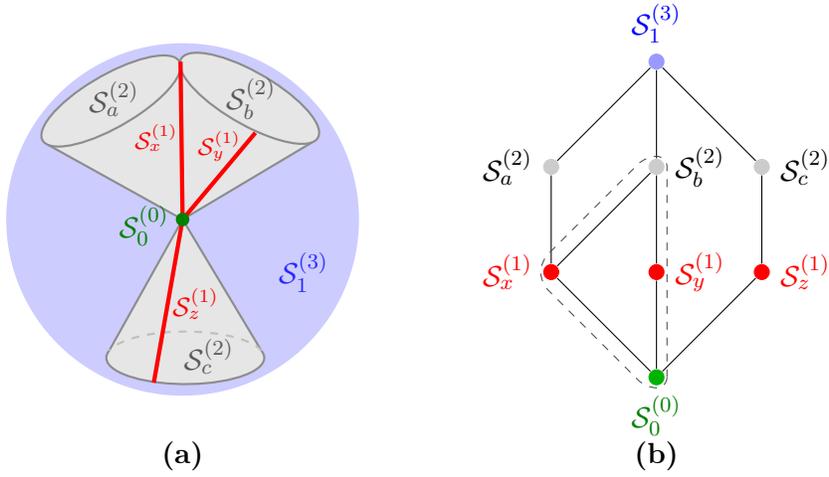

The pattern of inclusions among components will be central in the discussion that follows, so it will be useful to have a simple visualization of the inclusion relations.  This is given by the \emph{Hasse diagram} of the poset.  This is a graph in which each node corresponds to an element of the poset $P$.  Two nodes $i,j\in P$ are connected by an edge if and only if $i$ \emph{covers} $j$, that is, if $i>j$ and there is no $k\in P$ such that $i>k>j$.  Furthermore, if $i$ covers $j$ then we position the $i$ node higher than the $j$ node on the page.  Finally, we incorporate the grading by dimension by placing all nodes with the same dimension at the same height on the page.  See figure \ref{fig:hasse} for an example.  Note that we label the nodes of the Hasse diagram by the strata; the components are then the union of its associated stratum with all the strata less than it in the partial ordering.

Note that the stratification as we have defined it so far is not unique.  For instance, it does not preclude including among the strata regular submanifolds of the closure of a covering stratum.  In what follows we argue that there is a natural stratification in which each stratum\footnote{Except for any strata maximal with respect to the partial ordering.  With our assumptions, there is a unique such stratum, $\cS^{(r)} = \cCrg$.} consists of metric and/or complex structure singularities of the closure of their covering strata.  Furthermore, this stratification is unique.  In this stratification, each stratum has a special Kahler geometry or, in special cases, a slight weakening of the special Kahler structure, and their closures each have the properties of a Coulomb branch geometry (i.e., that of a singular special Kahler geometry with physically allowable singularities).  We thus call this stratification the \emph{special Kahler stratification} of the Coulomb branch.

We will find that the special Kahler stratification has the property that the closure, $\cSb^{(d)}_i$, of each stratum contains strata of one dimension less (except for the 0-dimensional strata, which are their own closures).  This means in particular that the grading by dimension in \eqref{stra2} can be taken to be $d_j =j$ for $r\ge j\ge 0$.  

We will also see that the special Kahler stratification is compatible with our simplifying assumption of a $\C^*$ homothety coming from spontaneously broken superconformal invariance on $\cC$.  This means that each $\cSb^{(d)}_i$ is invariant under the $\C^*$ action, so itself inherits a special Kahler geometry compatible with it being the scale-invariant Coulomb branch of a rank $d$ SCFT.  Since there is a single fixed point in $\cC$ of the $\C^*$ action, it follows that there is just a single dimension 0 stratum.  Also, the assumption that the Coulomb branch chiral ring is freely generated (which was equivalent to the assumption that $\cC$ has no complex structure singularities) implies that there is just a single maximal-dimension stratum with respect to the partial ordering, and it has dimension $r$.  Thus, these assumptions imply that the Hasse diagram of the stratification will have the general shape shown in figure \ref{fig:hasse}b: one node at the top covering all the dimension-$(r-1)$ nodes, one node at the bottom covered by all the dimension-$1$ nodes, and no intermediate nodes which are maximal or minimal with respect to the partial ordering.

Unlike the $\C^*$ homothety property, we will find that the regular complex structure property is \emph{not} inherited by the special Kahler structures of the components.  Thus, even if $\cC=\cSb^{(r)}$ is regular as a complex space, the $\cSb^{(d)}_i$ for $d<r$ need not have this property, i.e., they may develop complex singularities.  

In what follows, we develop the special Kahler stratification in two steps.  First we show that a rank $d$ special Kahler structure restricts to a rank $(d-1)$ special Kahler structure on each of its codimension one singular components, $\cSb^{(d-1)}_i$.  It then follows by induction with respect to the partial ordering that all components inherit special Kahler structures.  In the second step we show that this inherited special Kahler structure is unique as long as there is a unique maximal stratum in the poset.  In other words, we show that the restriction of the special Kahler structures on two different covering strata are the same.

\subsection{The special Kahler structure induced on a codimension one singularity}
\label{sec:SK}

\paragraph{Topology of the singularity.}

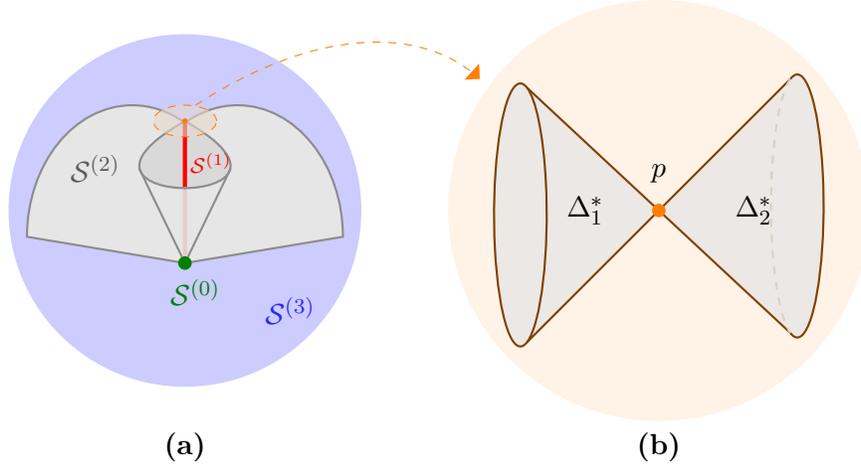
\begin{figure}[ht]
\centering
\begin{tikzpicture}
\begin{scope}[scale=.7,xshift=+0cm]
\fill[color=blue!20] (0,1) circle (3.35);
\draw[thick,draw=black!45,fill=black!10] (0,0) -- (3.0,0.5) arc (0:120:2.0cm and 2.5cm) -- cycle;
\draw[thick,draw=black!45,fill=black!10] (0,0) -- (-3.0,0.5) arc (180:60:2.0cm and 2.5cm) -- cycle;
\draw[thick,draw=black!45,fill=black!15] (0,2.7) .. controls (-3,1) and (3,1) .. cycle;
\draw[thick,draw=black!45,fill=black!10] (0,0) -- (0.87,1.8);
\draw[thick,draw=black!45,fill=black!10] (0,0) -- (-0.87,1.8);
\draw[ultra thick,draw=red] (0,1.45) -- (0,2.7);
\draw[ultra thick,draw=red!50!black!20] (0,0) -- (0,1.40);
\node[circle,fill=green!50!black,scale=.5] at (0,0) {};
\node[green!50!black] at (0.2,-.55) {$\cS^{(0)}$};
\node[red] at (.5,1.9) {\scriptsize $\cS^{(1)}$};
\node[black!65] at (-1.7,1.8) {$\cS^{(2)}$};
\node[blue!85] at (2.0,-.9) {$\cS^{(3)}$};
\draw[draw=orange, dashed, fill=orange!50!black!20, fill opacity=0.6] (0,2.7) ellipse (0.6 and 0.3);
\node[circle,fill=orange,scale=.2] at (0,2.7) {};
\draw[->,orange,dashed] (0.2,3.0) .. controls (2.5,4.5) and (4.5,4.5) .. (5.6,3.5);
\node at (0,-3.5) {\bf (a)};
\end{scope}
\begin{scope}[scale=.7,xshift=+9cm,yshift=+1cm]
\fill[color=orange!10] (0,0) circle (4.0);
\draw[rotate=90,thick,draw=orange!50!black,fill=orange!20!black!10] (0,0) -- (2.5,-2.5) arc (375:165:2.5cm and .5cm) -- cycle;
\draw[rotate=90,thick,dashed,orange!50!black!25] (2.5,-2.5) arc (15:165:2.5cm and .5cm);
\draw[rotate=270,thick,draw=orange!50!black,fill=orange!20!black!10] (0,0) -- (2.5,-2.5) arc (375:165:2.5cm and .5cm) -- cycle;
\draw[rotate=270,thick,orange!50!black] (2.5,-2.5) arc (15:165:2.5cm and .5cm);
\node[circle,fill=orange,scale=.5] at (0,0) {};
\node at (0,.7) {$p$};
\node at (-1.4,0) {$\D^*_1$};
\node at (1.8,0) {$\D^*_2$};
\node at (0,-4.5) {\bf (b)};
\end{scope}
\end{tikzpicture}
\caption{Figure (a)  shows a 3-dimensional scale-invariant Coulomb branch, where each real dimension in the figure represents 1 complex dimension and where (part of) a 2-dimensional component $\cSb^{(2)}$ self-intersects along a 1-dimensional stratum $\cS^{(1)}$.  Also shown is a neighborhood (orange disk) in a transverse slice in $\cSb^{(3)}$ through a point $p\in\cS^{(1)}$ (orange dot).  Figure (b) depicts the intersection of this neighborhood with $\cSb^{(2)}$, where now each real dimension in the figure represents 1 real dimension.  The intersection with $\cS^{(2)}$ is the disjoint union of two punctured disks, $\D^*_1 \coprod \D^*_2$, pictured as cones with the point $p$ as their common vertex.}
\label{stratumlink}
\end{figure}

Consider the codimension one singular locus $\cSb^{(d-1)} \subset \cSb^{(d)}_i$.  As we mentioned earlier, the Coulomb branch, like any moduli space of vacua, is a metrically complete space.  In particular, a smooth special Kahler metric on the $\cS^{(d)}_i$ stratum induces a non-degenerate distance function on $\cSb^{(d-1)}$. The open balls in this metric define the topology of $\cSb^{(d)}_i$.  Pick a point, $p$, in one of the $(d-1)$-dimensional strata, say,
\begin{align}\label{}
p \in \cS^{(d-1)}_j , \quad j\in I^{(d-1)} \quad \text{with} \quad j < i .
\end{align}
It then follows from the topological structure of the Whitney stratification \cite{Kirwan:2006} that there is a neighborhood $p\in U \subset \cSb^{(d)}_i$, open in $\cSb^{(d)}_i$, for which $U \cap \cS^{(d)}_i$ is homeomorphic to a finite disjoint union 
\begin{align}\label{}
U \cap \cS^{(d)}_i  &\simeq  \left(\coprod_{a=1}^n \D^*_a \right) \times \D^{d-1} 
\end{align}
where $\D := \{ z\in \C \, |\, |z|<1\}$ is the open unit disk and each $\D^*_a \simeq \D^* := \D \setminus \{0\}$ is the punctured unit disk. Viewing the punctured disk as an open interval times a circle, and since
\begin{align}\label{}
U \cap \cS^{(d-1)}_j \simeq \D^{d-1},
\end{align}
we can describe the topology of the neighborhood as a polydisk times a cone over $n$ distinct circles,
\begin{align}\label{indSK0}
U = \D^{d-1} \times C(\coprod_a S^1_a),
\end{align}
where the open cone of a space is defined by $C(L) := (L\times [0,1) )/(L \times \{0\})$.  The singular locus is thus the vertex of the cone times the polydisk.  This topology can be visualized as in figure \ref{stratumlink}a where the neighborhood $U$ would be the intersection of a ball centered on $p$ (the orange dot) with $\cSb^{(2)}$.

The common $\D^{d-1}$ dimensions correspond to the directions ``parallel'' to $\cS^{(d-1)}_j$, while the punctured disks are transverse to $\cS^{(d-1)}_j$ in $\cSb^{(d)}_i$.  A \emph{transverse slice} in $\cSb^{(d)}_i$ through the point $p$ is depicted in figure \ref{stratumlink}b as the bouquet of two open gray cones with $p$ their common vertex.  Such slices of a component transverse to a lower-dimensional stratum included in the component will play a central role in the discussion in later sections of the paper.  In a later subsection we will show that, like the strata, the transverse slices also inherit a special Kahler geometry.

\paragraph{Complex structure of the singularity.}

Now focus on the component of $U$ given by $\D^{d-1}\times C(S^1_a) := U_a$ for a single circle $S^1_a$.  It is relatively straightforward to characterize the complex structure of $U_a$.  By an argument given in more detail in \cite{Argyres:2018zay}, the special coordinates, $\s$, are good holomorphic coordinates on $\D^{d-1}\times (\text{wedge of }\D^*)$ where by ``wedge of $\D^*$'' we mean a subset of the punctured unit disk for which the argument of the coordinate $z$ lies in an interval of length less than $2\pi$.   (They are generally not single-valued on the whole punctured disk because they suffer electric-magnetic duality monodromies.)  And since the special coordinates are holomorphic and non-degenerate off of the singular locus $\cSb^{(d-1)}_j$ and do not diverge as they approach any point in $\cSb^{(d-1)}_j$ (by an argument given earlier), they have a definite limit as they approach $\cSb^{(d-1)}_j$ \cite{Forstneric:1992}.  It follows that the singular locus $\cS^{(d-1)}_j \cap U_a \simeq \D^{d-1}$ is itself a (regular) complex submanifold of $U_a$.  So we can pick a set of good complex coordinates on $U_a$, $(u_a^\perp,\bu_a^\parallel)$, such that $\cS^{(d-1)}_j$ is at $u_a^\perp=0$ and such that they are holomorphic in the special coordinates away from $u_a^\perp=0$ in any wedge domain.

It is important to note that although $\cS^{(d-1)}_j \cap U_a$ is a submanifold of $U_a$, it may still be a locus of complex singularities of $U_a$.  The possibility of such (non-normal) codimension one complex singularities occurring on a Coulomb branch was explored in \cite{Argyres:2017tmj}, where they could not be ruled out, but no examples among rank 1 Coulomb branches were found.  We will point out below when discussing rank 2 examples that such complex structure singularities in codimension one do commonly occur on 1-dimensional components of rank 2 stratifications.  We will call the regular $u_a^\perp$ coordinate chosen above the \emph{uniformizing parameter} of the transverse slice.  It is the transverse coordinate on the \emph{normalization} of the codimension one singularity.

\paragraph{Special geometry of the singularity.}

It is now relatively easy to see that $\cS^{(d-1)}_j \cap U_a$ inherits a special Kahler structure from that on $U_a$ essentially just by restriction.  This closely parallels and generalizes an argument given in the $\cN=3$ case \cite{Argyres:2019yyb}.  (The $\cN=3$ case was easier because there the special coordinates are flat so $\cS^{(d-1)}_j \cap U_a$ can be described by linear algebra, and because in $\cN=3$ the lattice spanned by the charges of states becoming massless on a codimension one stratum is a rank 2 sublattice of the full charge lattice.)  Also, most of the ingredients of the following argument were described in section 4.2 of \cite{Argyres:2018zay} in the $\cN=2$ context.

To show that $\cS^{(d-1)}_j \cap U_a$ inherits a special Kahler structure, we need to show that both the rank-$2d$ charge lattice and the rank-$2d$ vector bundle of special coordinates, $\s$, restrict appropriately to rank-$2(d-1)$ versions.  We saw that the holomorphic special coordinate section $\s$ has a well-defined holomorphic limit on $\cS^{(d-1)}_j \cap U_a$.  Recall that $\s$ takes values in the complex vector space
\begin{align}\label{}
\S_i := (\C \otimes \L_i)^*,
\end{align}
the complex linear dual of the complexification of the charge lattice, and inherits a symplectic inner product $\vev{\cdot,\cdot}$ from the Dirac pairing on $\L_i$.  (Here the $i$ subscript denotes the $\cSb^{(d)}_i$ component.)  $\s$ is a section of an $\Sp(2d,\Z)$ bundle over $\cS^{(d)}_i$ with fiber $\S_i$.  

Following the discussion of  subsection \ref{ssCplxAnal}, define
\begin{align}\label{indSK1}
\Phi_{j<i} &:= \text{set of charges of massive states on $\cS^{(d)}_i$ which become massless on $\cS^{(d-1)}_j$,}
\nn\\
\L_{j<i} &:= \text{lattice given by the integral span of $\Phi_{j<i}$.}
\end{align}
Though the \emph{subset} $\Phi_{j<i}$ might change discontinuously as one encircles $\cS^{(d-1)}_j \cap U_a$ in $U_a$ by crossing a wall of marginal stability, we showed in section \ref{ssCplxAnal} that the \emph{sublattice} $\L_{j<i}$ is constant along $\cS^{(d-1)}_j \cap U_a$.  Thus although $\Phi_{j<i}$ is not necessarily well-defined, $\L_{j<i}$ is.  Then in any wedge domain of $U_a$, $\cS^{(d-1)}_j \cap U_a$ is given in special coordinates by an equation $Z_Q(\s) = Q^T \s =0$ for any and all charges $Q \in \L_{j<i}$.  

Thus on the stratum we have that the following linear combinations of the special coordinates vanish:
\begin{align}\label{indSK2}
\L_{j<i}(\s) =0,
\end{align}
where by $\L(\S)$ we mean the dual pairing $Q^T\s$ for $Q\in\L$ and $\s\in\S$.  By the regularity of the metric induced from \eqref{SKmetric} on $\cS^{(d-1)}_j$ it follows that the tangential derivatives of $\s$ along $\cS^{(d-1)}_j$ span a rank-$2(d{-}1)$-dimensional vector subspace, $\S_j \subset \S_i$.  The special Kahler condition \eqref{SKconds} implies that $\S_j$ is a symplectic subspace.  Finally, by taking tangential derivatives of \eqref{indSK2}, it follows that
\begin{align}\label{indSK3}
\L_{j<i}(\S_j) =0.
\end{align}
Since $\S_j$ has dimension $2(d-1)$, it follows that
\begin{align}\label{indSK4}
\text{rank}(\L_{j<i}) \le 2 .
\end{align}
There are thus two possibilities: $\L_{j<i}$ has rank 2 or rank 1.\footnote{rank$(\L_{j<i})=0$ would mean that no charged states are becoming massless at $\cSb^{(d-1)}_j$. This would violate our physical assumption that metric singularities occur only when charged states become massless.  But there is the logical possibility that there could be a complex structure singularity at $\cSb^{(d-1)}_j$ while it remains metrically smooth.  In later sections we will see examples where precisely this occurs, albeit only in the case where $d=1$, so the singular stratum is a point, and the question of its induced special Kahler structure is empty.  It is an interesting question whether such metrically smooth complex singularities can occur along higher-dimensional strata.}  The construction of the induced special Kahler structure on $\cSb^{(d-1)}_j\cap U_a$ has a qualitatively different flavor in the two cases, so we discuss them separately.

\paragraph{Case when rank($\L_{i<j}$)=2.}

In this case since $\S_j$ is symplectic, \eqref{indSK3} implies that $\L_{i<j}$ is also a symplectic sublattice with respect to the Dirac pairing.  Furthermore, $\S_i$ and $\L_i$ and have symplectic-orthogonal decompositions
\begin{align}\label{indSK5}
\S_i &= \S_{j<i} \oplus \S_j &
&\text{and} &
\L_i &=\L_{j<i} \oplus \L_j &
&\text{with} &
\L_j(\S_{j<i}) &= \L_{j<i}(\S_j) = 0,
\end{align}
where $\S_{j<i}$ represents the symplectic complement of $\S_j$ in $\S_i$, and $\L_j$ the symplectic complement of $\L_{i<j}$ in $\L_i$.  Thus
\begin{align}\label{}
\S_j &= (\C\otimes \L_j)^* &
&\text{and} &
\S_{j<i} &= (\C\otimes\L_{j<i})^* .
\end{align}
These thus define a flat $\Sp(2d-2,\Z)$ bundle and a flat $\Sp(2,\Z)$ bundle, respectively, over $\cS^{(d-1)}_j\cap U_a$.  Related bundles also exist on $\cS^{(d)}_i$ as subbundles of the $\Sp(2d,\Z)$ bundle over $\cS^{(d)}_i$ with fiber $\S_i$.  But note that while the limit as one approaches the $\cS^{(d-1)}_j$ stratum of the $\S_j$ subbundle exists and gives the $\S_j$ bundle over $\cS^{(d-1)}_j$, the same is not true of the $\S_{j<i}$ subbundle which has no well-defined limit.

Denote the restriction of the special coordinate section, $\s$, to $\cS^{(d-1)}_j \cap U_a$ by $\s|_j$.  As discussed above, the components of $\s|_j$ exist and are holomorphic on $\cS^{(d-1)}_j \cap U_a$.  We now want to show that $\s|_j$ is a section of the $\S_j$ bundle over $\cS^{(d-1)}_j \cap U_a$.  Recall that the $\S_j$ bundle was defined by the subspace of $(\C\otimes \L_i)^*$ spanned by the derivatives of $\s$ tangent to $\cS^{(d-1)}_j$, which is characterized by \eqref{indSK3}.  On the other hand, \eqref{indSK2} implies that $\s|_j$ satisfies the same equation, $\L_{j<i}(\s|_j)=0$.  Since rank$(\L_{j<i})=2$ this implies that both $\s|_j$ and its derivatives along $\cS^{(d-1)}_j$ take values in $\S_j$, and so $\s|_j$ is a holomorphic section of the $\S_j$ bundle.

We have thus associated to $\cS^{(d-1)}_j$ a linear system of rank $2(d-1)$ charge lattices, $\L_j$, along with a dual $\Sp(2d-2,\Z)$ bundle with fiber the complex vector space $\S_j$, and its holomorphic section $\s|_j$.  The special Kahler condition \eqref{SKconds} on $d\s$ restricts to the same condition on $d\s|_j$.  These are all the defining ingredients of a special Kahler structure on $\cS^{(d-1)}_j$. 

This completes the induction step showing that a unique special Kahler structure is induced on a codimension one singular stratum, at least for each conical component of a neighborhood \eqref{indSK0} of the stratum.

\paragraph{Case when rank($\L_{i<j}$)=1.}
In this case the equation \eqref{indSK3} obeyed by $\S_j$ and $\L_{j<i}$ no longer uniquely characterizes them, and we no longer have the perfect decomposition \eqref{indSK5} of the charge lattice and dual vector space into complementary dual symplectic subspaces.  In particular, defining the symplectic subspace $\S_j$ as we did above \eqref{indSK3} as the space spanned by the tangential derivatives of $\s$, it still follows that it is a $2(d-1)$-dimensional complex symplectic space, and so there is a unique symplectic-orthogonal decomposition $\S_i = \S_j \oplus \S_{j<i}$, thus defining the 2-dimensional symplectic subspace $\S_{j<i}$.  

Now suppose we were to try to use \eqref{indSK3} to define a sublattice $\L'_{j<i}$ by
\begin{align}\label{indSK6}
\L'_{j<i}(\S_j)=0.
\end{align}
Depending on $\S_j$ this equation might have a rank-2 symplectic lattice, a rank-1 lattice, or a rank-0 lattice (i.e., just the origin) as its solution.  This is because the (dual) subspace $\S_j$ may or may not be commensurate with the rational lattice structure.  

In the present case we know that $\L'_{j<i} \supset \L_{j<i}$, so $\L'_{j<i}$ always contains a rank-1 sublattice.  If $\L'_{j<i}=\L_{j<i}$ (so has rank 1) then the $\S_j \oplus \S_{j<i}$ symplectic-orthogonal decomposition is not rational with respect to the integral (dual) symplectic form inherited from the Dirac pairing.  This means that when we perform the construction as in the rank 2 case of the $\S_j$ bundle and its restricted special coordinate section $\s|_j$, we find only that the $\S_j$ bundle is an $\Sp(2d-2,\R)$ bundle, and not an $\Sp(2d-2,\Z)$ bundle.  Relatedly, there is not a unique choice of symplectic rank-$2(d-1)$ charge lattice $\L_j$ orthogonal to $\L_{j<i}$ to associate to a special Kahler structure on $\cS^{(d-1)}_j$.

On the other hand, if $\L'_{j<i}$ has rank 2, then it is unique, and uniquely defines a symplectic-orthogonal decomposition of the charge lattice, and $\s|_j$ is a section of an $\Sp(2d-2,\Z)$ bundle with fiber $\S_j$.  Thus in this case $\cS^{(d-1)}_j\cap U_a$ inherits a unique special Kahler structure from $\cS^{(d)}_i$.

We do not know of an argument showing that $\L'_{j<i}$ must have rank 2. Therefore cannot prove that strata of this kind inherits a special Kahler stratification. But we have shown that they do inherit a weakened form of special Kahler geometry, which we will call \emph{loose} special Kahler geometries, in which the monodromies are allowed to be in $\Sp(2d,\R)$ rather than in $\Sp(2d,\Z)$.  We will get back to this point in section \ref{sec:In}.

\subsection{Uniqueness of the special Kahler stratification}

We have shown that a unique special Kahler structure on a stratum of dimension $d-1$ is inherited from a conical component of a neighborhood in an enclosing stratum of dimension $d$.  This can be used to show a unique special Kahler structure on every stratum only if the special Kahler structures induced from each conical neighborhood and each enclosing stratum all coincide.

A priori, there is no reason they should coincide.  Even the special Kahler structures induced from two different conical neighborhoods of the same enclosing stratum, as in figure \ref{stratumlink} need not be the same since the analytic continuation of the special coordinates within the enclosing stratum from one cone to the other need not be single-valued.

But with our assumption that the coordinate ring of the Coulomb branch of the SCFT in question is freely generated, it follows that there is a unique top-dimensional stratum which has no complex singularities.  Since the special coordinate section $\s|_j$ induced on lower-dimensional strata is by restriction of the values of that section on enclosing strata, it follows that if there is a single simple stratum at the top of the stratification, then all the induced sections will agree on a given stratum.  In other words, the answer one gets by restriction will be independent of the path through the Hasse diagram one follows to get to a given stratum, and also independent of any non-trivial paths one might follow within the enclosing stratum to arrive at the given stratum. 

This then shows that, at least for Coulomb branches, $\cC$, which are equivalent to $\C^r$ as complex manifolds, all strata of the stratification of the metric singularities of $\cC$ inherit unique (perhaps loose) Coulomb branch structures.  

\subsection{Transverse slices and the physical interpretation of the stratification}

The above construction of the special Kahler structure of a $(d-1)$-dimensional stratum not only produced the charge lattice and special coordinates of a rank-$(d-1)$ Coulomb branch, but also those for a rank-$1$ Coulomb branch --- to wit, the local system of rank-2 symplectic lattices $\L_{j<i}$, the restriction of the Dirac pairing to it, and its dual rank-2 complex vector bundle with fibers $\S_{j<i}$.  Indeed, through any point $p \in \cS^{(d-1)}_j$ one can, at least locally,  define a 1-complex-dimensional subspace of $\cSb^{(d)}_i$ for $i>j$ as the solution of the first order differential equations
\begin{align}\label{tslice1}
\L_j(d\s|_i)=0
\end{align}
going through the point $p$. The physical meaning of these equations will be explained momentarily. We call this subspace the \emph{elementary slice} to $\cS^{(d-1)}_j$ in $\cS^{(d)}_i$, and denote it by $\mT_{j<i}$.

$\mT_{j<i}$ thus has a special Kahler structure, and so the interpretation as a rank-1 Coulomb branch.  In fact, this interpretation has a simple physical origin.  The low energy theory of a rank-$r$ $\cN=2$ theory in the $\cS^{(d)}_i$ stratum has $d$ free $\U(1)$ gauge factors (times some rank-$(r{-}d)$ SCFT or IR-free field theory).  As one approaches the codimension-1 singularity $\cS^{(d-1)}_j$, states charged with respect to one of these $\U(1)$ factors become massless, while states charged with respect to the other $\U(1)^{d-1}$ factors remain massive.  Indeed, since $\L_{j<i}$ is the sublattice of charges of states becoming massless, this is precisely what we showed in section \ref{sec:SK} when we showed that rank$(\L_{j<i})\le 2$ and that $\L_{j<i}$ is a symplectic sublattice when it has rank 2.  This means that we can, by an electric-magnetic duality rotation, make that sublattice purely electric and/or magnetic with respect to a single $\U(1)$ and therefore \eqref{tslice1} follows.

Thus $\mT_{j<i}$ has the physical interpretation as the Coulomb branch of the rank-1 theory consisting of that $\U(1)$ gauge factor and the light states charged under it;  the other $\U(1)^{d-1}$ gauge factors remain free in the IR and so are decoupled from the $\mT_{j<i}$ theory.  Near $p\in \cS^{(d-1)}_j$ the $\mT_{j<i}$ theory will be the Coulomb branch of either an interacting rank-1 SCFT or an IR-free rank-1 scale-invariant theory.  So locally, such a Coulomb branch will look like a bouquet of 2-real-dimensional cones with common vertex $p$.  In fact, the two grey cones in figure \ref{stratumlink}b is a depiction of an elementary slice through $\cS^{(1)}$ in $\cS^{(2)}$.

Associated with such a rank-1 Coulomb branch are various invariants of the special Kahler geometry for each cone in the bouquet.  For instance, assuming that each cone has no complex singularity --- i.e., as a complex space each cone is a copy of $\C$, and the bouquet of cones is simply their transverse intersection at a point, $u=0$, for $u\in\C$ --- then the set of possible asymptotic forms of their special Kahler geometries as $u\to 0$, along with their most important invariants, are given in table \ref{tab:Kodaira}.  If we allow the cones to have complex singularities, then a more extensive list of possible asymptotic forms of their special Kahler geometries is allowed.  This list is given in table \ref{table:eps} in the next section, where it will be discussed in more detail.

\begin{table}
\centering
$\begin{array}{|c|l|c|c|c|c|c|}
\hline
\multicolumn{7}{|c|}{\text{\bf Possible scaling behaviors near singularities of a rank 1 Coulomb branch}}\\
\hline\hline
\text{Name} & \multicolumn{1}{c|}{\text{planar SW curve}} & \ \text{ord}_0(D_{x})\ \ &\ \D(u)\ \ & M_{0} & \text{deficit angle} 
& \t_0 \\
\hline
II^*   &\parbox[b][0.45cm]{4cm}{$\ y^2=x^3+u^5$}             
&10 &6 &ST &5\pi/3 & \ e^{i\pi/3}\ \\
III^*  &\ y^2=x^3+u^3x &9 &4 &S &3\pi/2 & i\\
IV^*  &\ y^2=x^3+u^4 &8 &3 &-(ST)^{-1} &4\pi/3 & e^{i\pi/3}\\
I_0^* &\ y^2=\prod_{i=1}^3\left(x-e_i(\t)\, u\right)
&6 &2 &-I &\pi & \t\\
IV &\ y^2=x^3+u^2 &4 &3/2 &-ST &2\pi/3 & e^{i\pi/3}\\
III &\ y^2=x^3+u x &3 &4/3 &S^{-1} &\pi/2 & i\\
II  &\ y^2=x^3+u &2 &6/5 &(ST)^{-1} &\pi/3 &e^{i\pi/3}\\
I_0 &\ y^2=\prod_{i=1}^3\left(x-e_i(\t)\, \right)
&0 &1 &I &0 & \t\\
\hline
\hline
I^*_n\ \ (n{>}0) &
\parbox[b][0.45cm]{5cm}{
$\ y^2=x^3+ux^2+\L^{-2n}u^{n+3}\ \ $}
& n+6 & 2 & {-T^n} & \pi
& i\infty\\
I_n\ \ (n{>}0)    &\ y^2=(x-1)(x^2+\L^{-n}u^n)  
& n     & 1 & {T^n} & 0 
& i\infty\\[0.5mm]
\hline
\end{array}$
\caption{\label{tab:Kodaira} Scaling forms of rank 1 planar special Kahler singularities, labeled by their Kodaira type (column 1), a representative family of elliptic curves with singularity at $u=0$ (column 2), order of vanishing of the discriminant of the curve at $u=0$ (column 3), mass dimension of $u$ (column 4), a representative of the $\SL(2,\Z)$ conjugacy class of the monodromy around $u=0$ (column 5), the deficit angle of the associated conical geometry (column 6), and the value of the low energy $\U(1)$ coupling at the singularity (column 7).  The first eight rows are scale invariant.  The last two rows give infinite series of singularities which have a further dimensionful parameter $\L$ so are not scale invariant; they can be interpreted as IR free theories since $\t_0=i\infty$.}
\end{table}

We focus on the the asymptotic geometry of the elementary slice near the singularity since that is the limit in which any massive charged states decouple from the interacting rank-1 theory in the IR.  In particular, there is no reason to suppose that the elementary slice defined by \eqref{tslice1} will be scale invariant, or even that the solution to \eqref{tslice1} will extend to a metrically complete space.  Thus we should think of the elementary slices as local geometric properties of a neighborhood of $\cS^{(d-1)}_j$ in $\cS^{(d)}_i$.  This is in marked contrast to the situation in $\cN=3$ Coulomb branches, or on Higgs branches, where the stratification is perfect, and the transverse slices extend to complete, scale-invariant geometries, due to the greater rigidity of triply special Kahler and hyperkahler spaces.

The notion of elementary slice easily generalizes to the more general \emph{transverse slice} to any stratum $\cS^{(d)}_\ell$ in $\cS^{(d')}_k$ as long as $\ell < k$ (thus $d'>d$).  We denote this slice by $\mT_{\ell <k}$.  It is, in a small enough neighborhood in $\cSb^{(d')}_k$ of a point $p\in\cS^{(s)}_\ell$, a $(d'{-}d)$-dimensional special Kahler space through $p$ transverse to $\cS^{(s)}_\ell$.  It clearly has the physical interpretation as the Coulomb branch of the rank-$(d'{-}d)$ IR SCFT describing the light charged states near $p$.  By our previous arguments, they are charged only under $d'{-}d$ $\U(1)$ gauge factors.

These transverse slices are transverse in the sense of \cite{slodowy1980}.  In the terminology of \cite{Bourget:2019aer, Grimminger:2020dmg}, two strata labelled by $\ell$ and $k$ in the stratification poset are called \emph{neighboring strata} if $k$ covers $\ell$, and the transverse slice between neighboring strata is an elementary slice.  Recall that neighboring strata are ones which are connected by an edge of the Hasse diagram of the stratification.  Thus to each edge corresponds an elementary slice.

This stratified structure is very reminiscent of the structure of nilpotent orbits of Lie algebras \cite{kraft1981minimal, kraft1982geometry, fu2017generic, Cabrera:2016vvv, Cabrera:2017njm} and more generally of symplectic singularities \cite{brieskorn1970singular, slodowy1980, beauville1999symplectic}, for which the existence of the stratification has been demonstrated in \cite{kaledin2006symplectic}. Symplectic singularities also play an important role in understanding Higgs branches of four dimensional $\cN=2$ theories and these have been recently studied, for instance, in \cite{Bourget:2019aer,Bourget:2020asf}.  A crucial difference between the stratification discussed here and the geometries studied in \cite{Bourget:2019aer,Bourget:2020asf}, is that in the general singular hyperkahler case considered there, the elementary slices can have arbitrarily large (quaternionic) dimension, whereas in the special Kahler case, they always have the minimal dimension allowed for a special Kahler space, i.e., one complex dimension. In other words, the elementary slices --- or equivalently the edges of the Hasse diagram --- correspond to rank 1 $\cN=2$ Coulomb branch geometries, which in turn are given by the Kodaira classification given in table \ref{tab:Kodaira} or its simple generalization to be described shortly in table \ref{table:eps}, providing a very constraining picture. On the other hand, the existence of the possibility of strata with loose special Kahler structures makes the Coulomb branch stratification richer than the one on the Higgs branch.

\begin{figure}[ht]
\centering
\begin{tikzpicture}
\begin{scope}[scale=.7,xshift=0cm]
	\node [circle,scale=2,color=blue!40,draw,fill,inner sep=1pt, 
	label={[blue]above:$\cS^{(3)}_1$}] 
	at (0,3) [] (3) {};
	\node [circle,scale=2,color=black!20,draw,fill,inner sep=1pt, 
	label=right:$\cS^{(2)}_c$] 
	at (+2,1) [] (2c) {};
	\node [circle,scale=2,color=black!20,draw,fill,inner sep=1pt, 
	label=right:$\cS^{(2)}_b$] 
	at (0,1) [] (2b) {};
	\node [circle,scale=2,color=black!20,draw,fill,inner sep=1pt, 
	label=left:$\cS^{(2)}_a$] 
	at (-2,1) [] (2a) {};
	\node [circle,scale=2,color=red,draw,fill,inner sep=1pt, 
	label={[red]right:$\cS^{(1)}_z$}] 
	at (+2,-1) [] (1z) {};
	\node [circle,scale=2,color=red,draw,fill,inner sep=1pt, 
	label={[red]right:$\cS^{(1)}_y$}] 
	at (0,-1) [] (1y) {};
	\node [circle,scale=2,color=red,draw,fill,inner sep=1pt, 
	label={[red]left:$\cS^{(1)}_x$}] 
	at (-2,-1) [] (1x) {};
	\node [circle,scale=2,color=green!70!black,draw,fill,inner sep=1pt, 
	label={[green!50!black]below:$\cS^{(0)}_0$}] 
	at (0,-3) [] (0) {};
	\draw (3) edge (2c);
	\draw (3) edge (2b);
	\draw (3) edge (2a);
	\draw (2c) edge (1z);
	\draw (2b) edge (1y);
	\draw (2b) edge (1x);
	\draw (2a) edge (1x);
	\draw (1z) edge (0);
	\draw (1y) edge (0);
	\draw (1x) edge (0);
	\draw[black!70, dashed, rounded corners] (0.2,-3.2) -- (0.2,1.2) -- (-0.2,1.2) -- (-2.2,-.8) -- (-2.2,-1.2) -- (-0.2,-3.2) -- cycle;
\node at (0,-4.5) {\bf (a)};
\end{scope}
\begin{scope}[scale=.7,xshift=+9cm]
	\node [circle,scale=2,color=blue!40,draw,fill,inner sep=1pt, 
	label={[blue]above:$\cT^{(0)}_1$}] 
	at (0,3) [] (3) {};
	\node [circle,scale=2,color=black!20,draw,fill,inner sep=1pt, 
	label=right:$\cT^{(1)}_c$] 
	at (+2,1) [] (2c) {};
	\node [circle,scale=2,color=black!20,draw,fill,inner sep=1pt, 
	label=right:$\cT^{(1)}_b$] 
	at (0,1) [] (2b) {};
	\node [circle,scale=2,color=black!20,draw,fill,inner sep=1pt, 
	label=left:$\cT^{(1)}_a$] 
	at (-2,1) [] (2a) {};
	\node [circle,scale=2,color=red,draw,fill,inner sep=1pt, 
	label={[red]right:$\cT^{(2)}_z$}] 
	at (+2,-1) [] (1z) {};
	\node [circle,scale=2,color=red,draw,fill,inner sep=1pt, 
	label={[red]right:$\cT^{(2)}_y$}] 
	at (0,-1) [] (1y) {};
	\node [circle,scale=2,color=red,draw,fill,inner sep=1pt, 
	label={[red]left:$\cT^{(2)}_x$}] 
	at (-2,-1) [] (1x) {};
	\node [circle,scale=2,color=green!70!black,draw,fill,inner sep=1pt, 
	label={[green!50!black]below:$\cT^{(3)}_0$}] 
	at (0,-3) [] (0) {};
	\draw (3) edge (2c);
	\draw (3) edge (2b);
	\draw (3) edge (2a);
	\draw (2c) edge (1z);
	\draw (2b) edge (1y);
	\draw (2b) edge (1x);
	\draw (2a) edge (1x);
	\draw (1z) edge (0);
	\draw (1y) edge (0);
	\draw (1x) edge (0);
	\draw[black!70, dashed, rounded corners] (-2.2,1.2) -- (-2.2,-1.2) -- (-1.8,-1.2) -- (0.2,0.8) -- (0.2,3.2) -- (-0.2,3.2) -- cycle;
	\node at (1.7,-2.3) {$\mT_{0<z}$};
        \node at (2.7,0) {$\mT_{z<c}$};
        \node at (0,-4.5) {\bf (b)};
\end{scope}
\end{tikzpicture}
\caption{Hasse diagram of a 3-dimensional scale-invariant Coulomb branch.  Figure (a) shows the nodes labelled by the strata $\cS^{(d)}_i$ of dimension $d$ and $i$ is a unique label.  $\cS^{(3)}_1$ is the manifold of non-singular points of the Coulomb branch, $\cS^{(0)}_0$ is the unique superconformal vacuum.  The union of the strata enclosed by the dashed line is the component $\cSb^{(2)}_b$ associated to the stratum $\cS^{(2)}_b$.  Figure (b) labels the nodes by the theories $\cT^{(r)}_i$ supported on the strata where $r$ is the rank of the theory and $i$ is the same unique stratum label.  $\cT^{(3)}_0$ is the whole theory, $\cT^{(0)}_1$ is the empty theory, and the nodes enclosed by the dashed line form the Hasse diagram of the Coulomb branch of the $\cT^{(2)}_x$ theory.  A few edges have also been labelled by their elementary slices.}
\label{hasse2}
\end{figure}
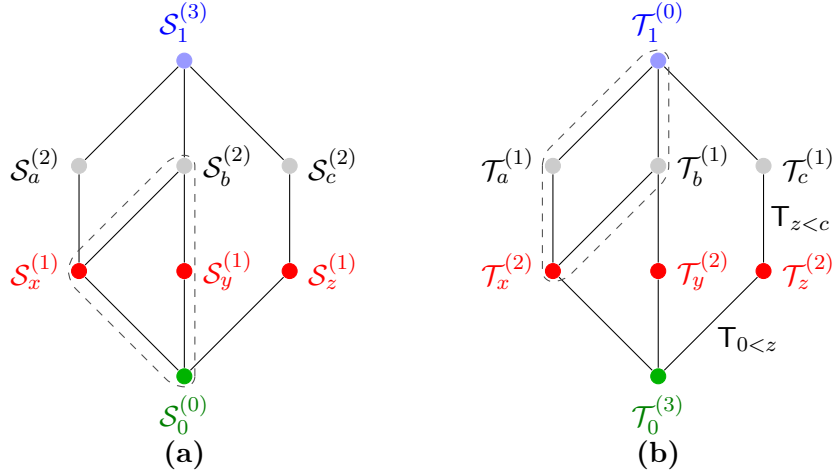

To each stratum there is associated a unique maximal transverse slice.  It is simply its transverse slice in the maximal stratum which dominates it in the poset.   With our assumption of a SCFT with freely-generated Coulomb branch chiral ring, there is a unique maximal stratum which dominates all other strata, thus this transverse slice is unique.  So if we are working with a rank-$r$ SCFT, to a stratum $\cS^{(d)}_i$ is associated a maximal transverse slice $\mT_i$ of dimension $r-d$.  The corresponding rank-$(r-d)$ SCFT will be denoted $\cT^{(r-d)}_i$.  We will call this SCFT the the \emph{theory supported on the stratum} $\cS^{(d)}_i$.  Thus we can equally well label the nodes of the Hasse diagram not only by the strata, but also by the theories that are supported on them.  For instance, the Hasse diagram of figure \ref{fig:hasse} can be labeled in the two ways shown in figure \ref{hasse2}.

\subsection{Revisiting the rank$\big(\L_{i<j}\big)$=1 case}\label{sec:In}

In our analysis, if the rank of the lattice of charges becoming massless along a stratum $\cS_i$ is 1, we could only conclusively establish that $\cS_i$ inherits the weaker version of a special Kahler structure, which we call a loose special Kahler structure.

Physically the rank$\big(\L_{i<j}\big)=1$ implies that the charges of the states becoming massless are all mutually local and so there exists an electric-magnetic duality basis in which they can all be thought as electric charges under a single $\U(1)$ gauge factor. In the case in which the stratum is complex co-dimensions one, this implies that the theory supported on the stratum $\cT^{(1)}_i$ (which in this case is a rank-1 theory) is a $\U(1)$ gauge theory with massless hypermultiplets. This case happens quite frequently in examples below. 

Irregardless of this fact, if the theory $\cT_i$ (we write no-superscript in this case as we are keeping the complex co-dimension of the stratum generic) supported on the stratum has a non-trivial Higgs branch then there is a physical argument to prove that the stratum $\cS_i$ does inherit a special Kahler structure (as opposed to a loose one). This is because turning on the Higgs branch moduli of $\cT_i$ we can move from $\cS_i$ into a mixed branch stratum $\cM_i$. In this case, the stratum $\cS_i$ is the intersection of the mixed branch with the Coulomb branch: $\cS_i:=\cM_i\cap\cC$ and it inherits the special Kahler structure from that of $\cM_i$. We will see how this works in examples when we discuss the Hasse diagram of the full moduli space. 

With this observation, and restricting to complex co-dimension one strata which will play a prominent role below, we have shown that the special Kahler stratification only reduces to its weaker form if the strata support $\U(1)$ gauge theories with trivial Higgs branch. These theories, which sometimes are called \emph{frozen $I_n$} type, will be labeled below as $[I_n,\varnothing]$.

\section{A closer look at the stratification of rank 2 Coulomb branches}\label{sec:rank2}

\begin{table}
\centering
$\begin{array}{|c|c|}
\multicolumn{2}{c}{{\text{\bf\quad Possible Coulomb branch scaling dimensions of rank-2 SCFTs\quad}}}\\
\hline\hline
&\\
\multirow{-2}{*}{$\quad \text{fractional}\quad\,$}&\multirow{-2}{*}{{\large 
$\ \frac{12}{11} \, ,\, \frac{10}{9} \, ,\, \frac{8}{7} \, ,\,
\frac{6}{5} \, ,\, \frac{5}{4} \, ,\, \frac{4}{3} \, ,\, 
\frac{10}{7} \, ,\, \frac{3}{2} \, ,\, \frac{8}{5} \, ,\, 
\frac{5}{3} \, ,\, \frac{12}{7} \, ,\, \frac{12}{5} \, ,\, 
\frac{5}{2} \, ,\, \frac{8}{3} \, ,\, \frac{10}{3} \ $}}\\
\hline
&\\
\multirow{-2}{*}{$\quad \text{integer}\quad\,$}&\multirow{-2}{*}{
$\ 1 \ ,\ 2 \ ,\ 3 \ ,\ 4 \ ,\  5 \ ,\  6 \ ,\  8 \ ,\  10 \ ,\ 12\ $}\\
\hline
\end{array}$
\caption{List of the allowed values of scaling dimensions of Coulomb branch operators for rank-2 $\cN=2$ SCFTs. This list relies on the assumption that the Coulomb branch chiral ring is freely generated.}
\label{ScaDimList} 
\end{table}

Let us now specialize the discussion above to rank-2 which is the simplest example where we can study the stratification of the Coulomb branch singular locus. We will call the globally defined coordinates of this two complex dimensional Coulomb branch $(u,v)$. We will indicate their scaling dimension as $[u]$ ($[v]$) and $\D_u$ ($\D_v$) interchangeably and we will choose the labeling of the coordinates such that $\D_u\leq\D_v$.

It is by now well known that the allowed scaling dimensions for Coulomb branch coordinates are strongly constrained by compatibility with special Kahler geometry \cite{Caorsi:2018zsq,Argyres:2018urp}. In particular at rank 2 there are only 24 allowed rational values \cite{Argyres:2018zay}, which are reported in table \ref{ScaDimList}. A more refined analysis of the structure of monodromies restricts also the allowed pairs of $(\D_u,\D_v)$ (see table 5 of \cite{Caorsi:2018zsq}).

At rank 2 the complex co-dimension one strata are the only strata that need to be identified and their topology has already been analyzed in \cite{Argyres:2018zay}. We will summarize these results below. Therefore we can write explicitly the stratification 
\beq\label{strar2}
\cC^{\rm rank=2}=\cCrg\cup\Big(\cSb_{\rm metric}\setminus \{0\}\Big)\cup\{0\}\equiv \cS^{(2)}\cup\left(\coprod_{i\in I^{(1)}}\cS^{(1)}_i\right)\cup\{0\}
\eeq
This also makes it obvious that to specify the full Hasse diagram of a rank-2 Coulomb branch geometry we only need to identify:
\begin{itemize}
\item[1-] The complex co-dimension 1 strata $\cS_i^{(1)}$.
\item[2-] Their complex dimension 1 transverse slices $\mT_i$.
\end{itemize}

Since in this case there are only strata of complex codimension 1 (as the complex co-dimension 0 stratum is $\cCrg$ and the co-dimension two stratum is the superconformal vacuum $\{0\}$), we will drop the superscript to lighten the notation: $\cS^{1}_i\mapsto \cS_i$. Also at rank 2 the component $\cSb_i$ can be straightforwardly obtain by adding a point, $\{0\}$, to their corresponding stratum $\cS_i$. Since the difference is minor we might at places mix them up.

In this section we will quickly review the topology of the possible strata and how to translate the topological information into the full special Kahler characterization of the stratum. In particular we will explain in detail the phenomenon already hinted above of the appearance of  elementary slices, called \emph{irregular geometries} and reported in table \ref{table:eps}, beyond the Kodaira classification \ref{tab:Kodaira}. These geometries were introduced and studied systematically in \cite{Argyres:2017tmj} and, among the various unconventional properties they posses, is the fact that they give rise to an apparent violation of unitarity along their Coulomb branches. 

\subsection{Topology of the embedding of the strata: unknots and $(p,q)$ torus links}\label{sec:knots}

Summarizing here the work in \cite{Argyres:2018zay}, at rank 2 there are three qualitatively different 1-dimensional strata.  These arise as topologically inequivalent orbits of the globally defined $\C^*$ action: 
\beq
\l\ \circ:\left(
\begin{array}{c}
u\\
v
\end{array}
\right)
\mapsto
\left(
\begin{array}{c}
\lambda^{\D_u}u\\
\lambda^{\D_v}v
\end{array}
\right),\quad
\l\in\C^*.
\eeq
We then obtain the following:
\begin{itemize}
\item (a): the orbit through the point $(u,v)=(1,0)$. This is the submanifold consisting of the $v=0$ plane minus the origin and will be indicated by $\cS_v$.
\item (b): the orbit through the point $(u,v)=(0,1)$. This is the submanifold consisting of the $u=0$ plane minus the origin and will be indicated by $\cS_u$.
\item (c): the orbit through a point $(u,v)=(\w,1)$ for $\w\neq0$. These orbits are the non-zero solutions to the equation $u^p+\w \, v^q=0$, $(p,q)\in\N$, for a given $\w\in\C^*$ and will be indicated by $\cSpq$. \label{orbitc} 
\end{itemize}
Notice that $\cSpq$, for $p\neq q\neq 1$, is singular as a complex variety at the origin and in general non-normal. Since $\cSpq$ is one complex dimensional, taking its normalization will also resolve the singularity. The fact that $(p,q)\in\N$ follows from the non-trivial analysis of \cite{Argyres:2018zay}, and of course by compatibility with the $\C^*$ action $q/p=\D_u/\D_v$. 
In \cite{Argyres:2018zay} we introduced the name of ``unknotted'' orbits for the types (a) or (b), and ``knotted" orbits for orbits of type (c).

It is possible to argue \cite{Argyres:2018urp} that if only unknotted orbits are present in $\cS_\w$, the Coulomb branch geometry factorizes into that of two decoupled rank 1 SCFTs which we interpret as strong evidence that the SCFT living at the origin is in fact a product of two rank 1 SCFTs. We enshrine this as\\
\begin{tcolorbox}
\begin{fact}\label{knot}
A four dimensional rank 2 $\cN=2$ SCFT which cannot be decomposed into the product of two rank-1 theories has at least one knotted stratum. \end{fact}
\end{tcolorbox}\vspace{0.5em}
\noindent This can be straightforwardly generalized to higher ranks \cite{Argyres:2018urp}.

The locus of metric singularities, $\cSb$, should then be identified with the closure of the disjoint union of a variety of possible strata $\cS_\w$. Therefore $\cSb$ can be described by the following reduced polynomial in $\C^2$:
\begin{align}\label{singularity}
\cSb = \left\{ u^{\ell_u}\cdot 
\prod_{j=1}^\ell (u^p+\w_j v^q)
\cdot v^{\ell_v} = 0 \right\} ,
\end{align}
where the $\w_j\in\C^*$ are all distinct. Here $\ell_u$ and $\ell_v$ are either 0 or 1, depending on which unknotted orbits are present, and $\ell$ is the number of knotted orbits in $\cSb$.  The zeros of \eqref{singularity} is defined to be the \emph{discriminant locus} of $\cC$, see \cite{Martone:2020nsy} for a more systematic discussion. In particular, $\cSb\setminus\{0\}$ is a smoothly embedded 1-dimensional complex submanifold of $\cC$ with $\ell_u+\ell+\ell_v$ disconnected components.\footnote{While \eqref{singularity} characterizes $\cSb$ topologically, the extra structure arising from special Kahler geometry can almost fully be characterized by introducing an extra positive integer for each component corresponding to the degree of vanishing of the \emph{quantum discriminant} of $\cC$. Using the fact that all genus 2 curves are hyperelliptic, the quantum discriminant can be defined in terms of the Seiberg Witten curve for all rank-2 geometries. See again \cite{Martone:2020nsy} for more details.}

To understand the (point set) topology of how $\cSb$ is embedded in $\cC$, we intersect $\cSb$ with a family of topological 3-spheres,  $S^3_\r$ for $\r\in\R$, foliating $\C^2\setminus\{0\}$ and notice that $\cSb \cap S^3_{\r=0}$ is a ``deformation retract'' of $\cSb \setminus \{ 0\}$ in $\C^2$.  Therefore $\pi_1 (\C^2\setminus \cSb ) \simeq \pi_1 (S^3_0 \setminus \cSb \cap S^3_0 )$. We find that these intersections with the various components (a)-(c) in \eqref{singularity}, give rise to the following:
\begin{align}\label{knots}\nonumber
\cSb_u\cap S^3_\r &= \left\{(u,v)\in\C^2 \ |\  
u = 0, \,
\,\qquad v = 2^{1/q} e^{i\f}
\quad\text{with}\quad \f\in\R \mod 2\pi\right\},\\ 
\cSb_v\cap S^3_\r &= \left\{(u,v)\in\C^2 \ |\  
u = 2^{1/p} e^{i\th},
v = 0 
\quad\qquad\, \ \text{with}\quad \th\in\R \mod 2\pi\right\},\\ \nonumber
\cSbpq\cap S^3_\r &= \left\{(u,v)\in\C^2 \ |\  
u = e^{i\th}, 
\quad\ \ v = e^{i\f}
\quad \quad\ \ \text{with}\quad p\th = q\f \mod 2\pi \right\}.
\end{align}
The first two are easily seen to be embedded as unknotted circles in $S^3_\r$, while the last is embedded as a $(p,q)$ torus knot.  In \cite{Argyres:2018zay} we introduced the following notation to denote the topology of $\cSb$ given by the total link consisting of a torus link together with unknots by
\begin{align}\label{genlink}
L_{(p,q)}(\ell_u,\ell_v,\ell) := \Big(\cSb_u^{\ell_u} \cup \cSb_v^{\ell_v}\cup \cSbpq^\ell\Big)\cap S^3_0.
\end{align}
Here we are using a notation where $\cS_{u,v}^{\ell_{u,v}} := \cS_{u,v}$ if $\ell_{u,v}= 1$, and $:=\varnothing$ if $\ell_{u,v}=0$.  Similarly, $\ell=0$ means that there is no torus link component.

\begin{figure}[ht]
\centering
\includegraphics[width=.50\textwidth]{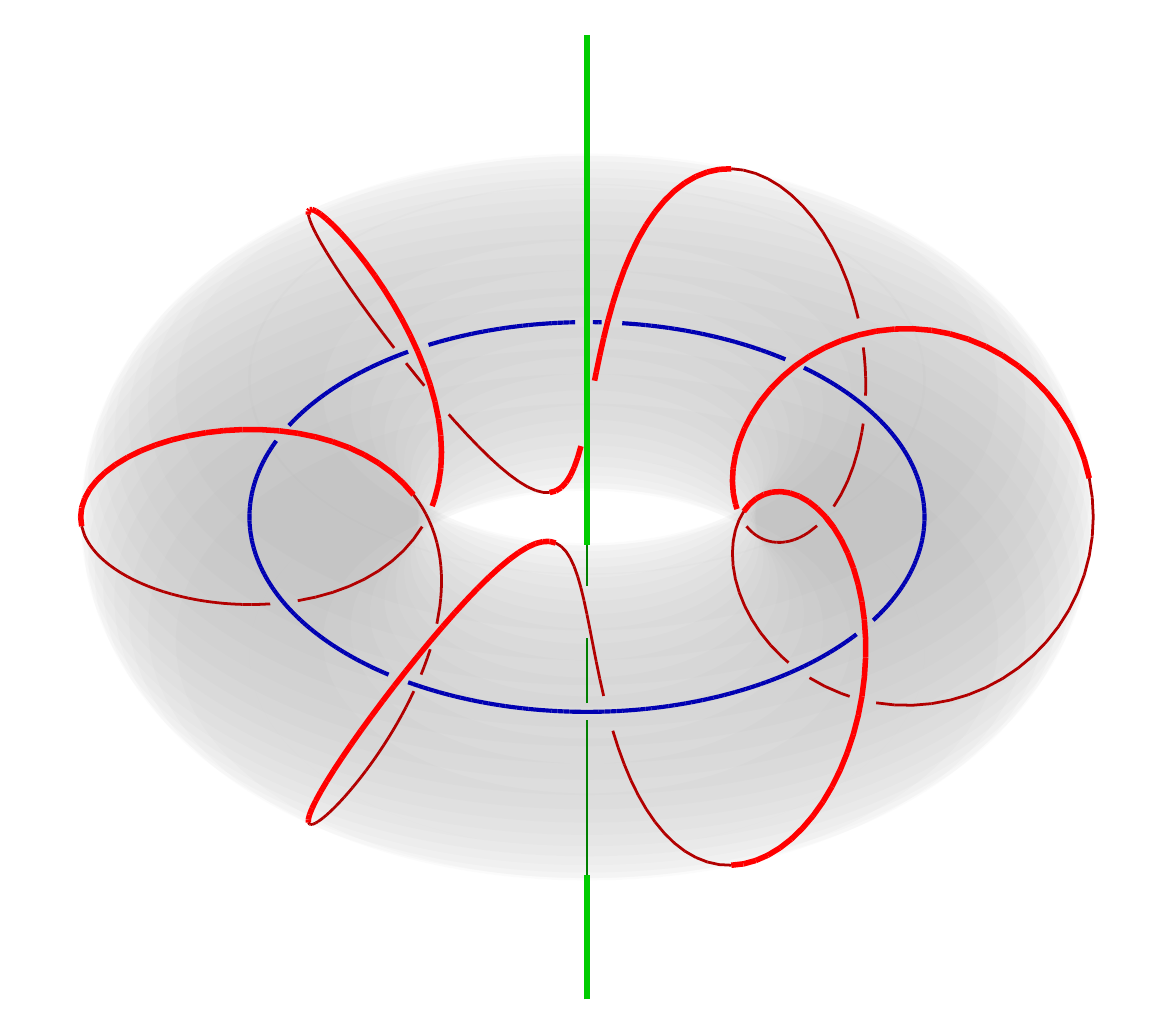}
\caption{Depiction of an $L_{(1,6)}(1,1,1)$ link consisting of the blue, red, and green circles.  The solid gray torus is there for visualization purposes.}
\label{knot1}
\end{figure}

These links are relatively easy to visualize.  For example, figure \ref{knot1} depicts an $L_{(1,6)}(1, 1, 1)$ link with the $(1,6)$ knot in red on the surface of a solid gray torus (the torus is present purely for visualization), unknot threading the interior of the torus in blue, and the other unknot as the ``z-axis" in green. The three dimensions are the stereographic projection of $S^3_0$ to $\R^3$ with the point at infinity being $(u,v)=(-2^{1/p},0)$ and origin being $(u,v)=(+2^{1/p},0)$.   Thus the green line goes through the point at infinity, so is topologically a circle. This clarifies the name unknotted (knotted) for orbit (a) and (b) (orbit (c)).

\subsection{Special Kahler structure of the strata and occurrence of irregular geometries}

Thus far we have discussed only the topology of the components of $\cSb$. Here we will discuss how to characterize the full special Kahler structure and therefore see each component of $\cSb$ as a rank-1 Coulomb branch geometry in its own right. This can be done quite straightforwardly by applying the analysis of section \ref{sec:SK} to the rank-2 case. We will attempt to avoid technical details and instead concentrate on explaining the calculations needed to carry out this characterization.

Consider first the decomposition \eqref{indSK5} near $\cSb_v$. In order to implement our analysis it is enough to analyze only the $r$ $\boldsymbol{a}$ components of the special coordinates $\s$ in \eqref{specCo} near $\cSb_v$. By scale invariance, $[a_\perp]=[a_\parallel]=1$, and imposing that $a_\perp\to0$ on $\cSb_v$, we obtain \cite{Caorsi:2018zsq}
\begin{align}\label{aXu}
a_\parallel \sim u^{\frac1{\D_u}},
\qquad
a_\perp\sim v\, u^{\frac{1-\D_v}{\D_u}},
\end{align}
where these expression are valid for $|v/u|\ll1$.  Consider the loop 
\begin{align}\label{gu}
\g_u := \biggl\{
v = 0 , \quad
u =u_* e^{i\vf}, 
\quad
\vf \in \left[0,2\pi\right) 
\biggr\} \, ,
\end{align}
which lies on $\cS_v$ and encircles $\{0\}$.  Then \eqref{aXu} allows one to directly calculate the monodromy induced by $\g_u$. In the notation used in \cite{Argyres:2018zay}, it is in the $\Sp(4,\Z)$ conjugacy class $[1^2\cdot \D_u]$.\footnote{The monodromies arising from considering loops of this kind, we called $U(1)_R$ monodromies in \cite{Argyres:2018zay,Argyres:2018urp}, and this is because the loops associated to these monodromies, like \eqref{gu}, are orbits of the $U(1)$ component of the R-symmetry.}  This readily implies that the stratum associated with the $\cSb_v$ sublocus has $\SL(2,\Z)$ monodromy in the $[\D_u]$ elliptic conjugacy class, or in other words is of Kodaira type $II^*$, $III^*$, $IV^*$, $I_0^*$, $IV$, $III$, $II$ if $\D_u= 6$, 4, 3, 2, $\frac32$, $\frac43$, $\frac65$, respectively.\footnote{Note that the $I_0$ type corresponding to $\D_u=1$ is not allowed. This is because we are considering the degeneration along unknotted singularities.}  In particular we notice that $\D_u$ can only have one of the Coulomb branch scaling dimensions which are allowed at rank 1.

This analysis can be replicated identically for $\cSb_u$ leading to the following fact which further restricts the type of singularities which are allowed:\\
\begin{tcolorbox}
\begin{fact}\label{scaling}
A two complex dimensional Coulomb branch $\cC$, parametrized by Coulomb branch coordinates $(u,v)$, only admits a stratum corresponding to the $\C^*$ orbit $\cS_v$ ($\cS_u$) if $\D_u$ ($\D_v$) is a scaling dimension allowed at rank 1.
\end{fact}
\end{tcolorbox}\vspace{0.5em}
\noindent This statement also generalizes straightforwardly to theories of rank $r$.

To perform a similar calculation for $\cSpq$ we first need to introduce the concept of the uniformizing parameter. A knotted component in general has a non-normal complex singularity at the origin.  Its stratum $\cS$ is a connected 1-complex-dimensional manifold with a non-trivial holomorphic $\C^*$ action without a fixed point, so is holomorphically equivalent to $\C^*$ itself \cite{Akhiezer:1995}.  We can then coordinatize $\cS\simeq \C^*$ as the complex $t$-plane minus the point $t=0$, and take the $\C^*$ action to be simply $t \mapsto \l^{\D_t} t$ for $\l\in\C^*$.   This complex $t$ coordinate is the \emph{uniformizing parameter} for $\cS$, or $\cSb$ if we include the origin.

What is the relation between $t$ and $(u,v)\in \C^2$, the coordinates of the rank-2 Coulomb branch $\cC$?  For knotted strata given by $u^p=\w v^q$ we can proceed as follows. First observe that by scale invariance and up to a choice of normalization $t= u^{p\g} = \omega^\g v^{q\g}$ for some real $\g$. Then we can fix $\g$ by the requirement that $t$ is a good (i.e., single-valued) coordinate near the origin. In particular, upon looping around the origin, $t\to e^{2\pi i}t$, $(u,v)$ transform to $(e^{2\pi i/p\g}u, e^{2\pi i/q\g}v)$.  Their single-valuedness then implies that $\g=1/pq$, so
\beq\label{unipq}
t = u^{\frac1q} = \omega^{\frac 1 {pq}} v^{\frac1p},
\eeq 
and thus the scaling dimension of the uniformizing parameter is
\beq\label{scapq}
\D_t=\frac1q \D_u=\frac1p \D_v.
\eeq
Notice that \eqref{scapq} can be less than one even though $\D_v\geq\D_u>1$. This apparent violation of the unitarity bound, which was already noticed in \cite{Argyres:2017tmj}, actually takes place in concrete examples as we will see below. In fact there is a special set of Coulomb branch geometries at rank-1 which admit a consistent special Kahler structure while having uniformizing parameter of scaling dimension less than 1. These are the irregular geometries which we have mentioned a few times already and are reported in table \ref{table:eps}. For more details on their full Seiberg-Witten geometry see \cite{Argyres:2017tmj}. We can also define the uniformizing parameter for unknotted strata. In this case the uniformizing parameter is obvious; it is simply given by non-vanishing Coulomb branch global coordinate. That is by $u$ for $\cS_v$ and $v$ for $\cS_u$.

\begin{table}
\centering
$\begin{array}{|c|c|c|c|c|}
\hline
\multicolumn{5}{|c|}{\text{\bf Irregular rank 1 geometries $K^{(m)}$}}\\
\hline\hline
\ \text{name}\ \,&\ \text{opening angle}\ \,&\ \D_t^{(m)}\ \, & M_0               & \t_0  \\
\hline
II^{*(m)} 
& 2\pi (m+\frac16) & \frac6{1+6m} & ST & e^{i\pi/3}  \\
III^{*(m)} 
& 2\pi (m+\frac14) & \frac4{1+4m} & S   & i                    \\
IV^{*(m)} 
& 2\pi (m+\frac13) & \frac3{1+3m} & -(ST)^{-1} & e^{2i\pi/3}\\
I_0^{*(m)} 
& 2\pi (m+\frac12) & \frac2{1+2m} & -I   &\ \text{any}\ \t\ \, \\
IV^{(m)} 
& 2\pi (m+\frac23) & \frac3{2+3m} & -ST & e^{2i\pi/3}     \\
III^{(m)} 
& 2\pi (m+\frac34) & \frac4{3+4m} & S^{-1}  & i               \\
II^{(m)} 
& 2\pi (m+\frac56)& \frac6{5+6m} & (ST)^{-1} & e^{i\pi/3} \\
I_0^{(m)} 
& 2\pi (m+1)        & \frac1{1+m}  & I    & \text{any}\ \t \\[1mm]
\hdashline
I_n^{*(m)} 
& 2\pi (m+\frac12)\ & \frac2{1+2m} & -T^n  & i\infty         \\
I_n^{(m)} 
& 2\pi (m+1)\ & \frac1{1+m}  & T^n    & i\infty   \\[1mm]
\hline
\end{array}$
\caption{All geometries of 1-dimensional special Kahler cones.  $m\ge0$ is a non-negative integer, and $n>0$ is a positive integer.  For $m=0$ we reproduce the standard Kodaira classification, if $m>0$ we instead obtain the irregular geometries. Shown are the allowed values of the opening angle of the cone, of the scaling dimension of the uniformizing parameter $\D^{(m)}_t$, of the EM duality monodromy conjugacy class about the tip of the cone $M_0$, and of the value of the complex modulus of the fiber at the tip $\t_0$. \label{table:eps}}
\end{table} 

We are now ready to perform the monodromy analysis also for knotted strata. Labeling as $\tilde t$ a possible choice of the transverse coordinate to $\cSpq$, such that $\cSpq$ is located at $\tilde t=0$, say,
\beq
\til t := \left(\frac12\left(u^p+\omega v^q\right)\right)^{\frac 1 {pq}},
\eeq
we can write the expression for $\boldsymbol{a}=(a_1,a_2)$ arbitrarily near $\cSpq$, generalizing \eqref{aXu}, as
\begin{align}\label{aXpq}
a_\parallel\sim t^{\D_t},
\qquad
a_\perp\sim \til t\, t^{\frac{1-\D_t}{\D_t}},
\end{align}
valid for $|t/\tilde t|\ll1$.  This then implies that monodromy induced along the loop
\begin{align}\label{gpq}
\g_{p,q} := \biggl\{
t = t_* e^{i\vf},  , \quad
\tilde t = 0,
\quad
\vf \in \left[0,2\pi\right) 
\biggr\} \, ,
\end{align}
belongs to the $\Sp(4,\Z)$ conjugacy class $[1^2\D_t]$ and therefore the corresponding $\SL(2,\Z)$ elliptic conjugacy class is $[\D_t]$. In this case, as mentioned multiple times, there is no unitarity bound providing a lower bound for it, $\D_t$ is not bounded to take the ``standard Kodaira'' values, and the corresponding knotted strata can belong to the enlarged class in table \ref{table:eps}.  We will label this enlarged set of geometries as $K^{(m)}$ ($K$ for ``Kodaira'') and the scaling dimension of the corresponding uniformizing parameter as $\D_t^{(m)}$, where $\D_t^{(0)}$ correspond to the allowed rank-1 Coulomb branch scaling dimensions.  Thus
\beq
\D_t^{(m)}:=\frac{\D_{t,n}}{\D_{t,d}+m\, \D_{t,n}}\quad {\rm where}\quad \frac{\D_{t,n}}{\D_{t,d}}:=\D^{(0)}_t\in\left\{6,4,3,2,\frac32,\frac43,\frac65,1\right\} .
\eeq
Because the elliptic eigenvalues are roots of unity, it is also easy to see that  $\bigl[\D_t^{(m)}\bigr]$ corresponds to the same $\SL(2,\Z)$ conjugacy class as $\bigl[\D_t^{(0)}\bigr]$ which is in turn consistent with the results reported in table \ref{table:eps}.

Because of the apparent violation of the unitarity bound, which happens for all irregular geometries, $K^{(m)}$ for $m>0$, the theories $\cT_i$ supported over a stratum $\cS_i\cong K^{(m)}$ are strongly constrained. In particular we have the following:\\
\begin{tcolorbox}
\begin{fact}\label{NoHB}
A theory $\cT_i$ supported on a stratum $\cS_{i}\cong K^{(m)}$ necessarily has $\cH=\{\varnothing\}$, where $\cH$ denotes the Higgs branch of $\cT_i$.
\end{fact}
\end{tcolorbox}\vspace{0.5em}

This concludes our analysis of the special Kahler structure of each stratum and its associated component. In summary:\\
\begin{tcolorbox}
\begin{center}
\textbf{Conditions on special Kahler stratifications at rank 2}
\end{center}\vspace{-.5em}
\begin{itemize}[leftmargin=*]
\item Here we are only interested in rank-2 theories which are not product of rank-1 theories. Then by fact \ref{knot}, there will always be at least one knotted component which can be any of the special Kahler geometries in table \ref{table:eps}.
\item If the knotted strata $\cSpq\cong K^{(m)}$, $m>0$, then $\cT_{(p,q)}$ has a trivial Higgs branch.
\item An unknotted component algebraically embedded in $\cC$ as $u=0$ ($v=0$), will have a rank 1 special Kahler characterized by the $\SL(2,\Z)$ monodromy $[\D_v]$ ($[\D_u]$).
\item If $\D_u$ ($\D_v$) is not an allowed rank 1 scaling dimension, by fact \ref{scaling} no unknotted component algebraically embedded as $v=0$ ($u=0$) is present.
\end{itemize}
\end{tcolorbox}\vspace{0.5em}

\subsection{Transverse slices and central charge formulae}

The last thing that is left to identify are the transverse slices. This is done by using the formulae presented in \cite{Martone:2020nsy} (summarized below) which leverage information about the central charges of the theory.

Generalizing the results of Shapere and Tachikawa \cite{Shapere:2008zf}, it is possible to show that the $(a,c)$ conformal central charges and $k_a$ flavor symmetry central charges (where $a$ labels the simple factors of the flavor symmetry of the SCFT) depend in a precise way on the central charges of the rank-1 theories supported on the complex co-dimension one strata $\cS^{(r-1)}_i$. We have re-introduced the superscript $^{(r-1)}$ labeling the complex dimension of the stratum, because this is a result which applies to arbitrary ranks, even though we will only use it below at rank 2.  Via these formulae we will be able to determine the theories $\cT_i^{(1)}$ and therefore the transverse slice $\mT_i$ by exploiting the local identification between $\mT_i$ and the Coulomb branch of $\cT_i^{(1)}$.

Now we restrict to rank-2, so we will henceforth drop the superscripts. The main formulae are
\begin{subequations}
\begin{align}\label{acc}
24a&=10+h+6(\D_u+\D_v-2)+\sum_{i\in I} b_i\cdot \Dsi_i,\\\label{ccc}
12c&=4+h+\sum_{i\in I} b_i\cdot \Dsi_i,\\\label{kcc}
k_a&=\sum_{i\in I_{\ff_a}}\left(\frac{\Dsi_i}{d^a_i\D_i}\left(k_a^i-T({\bf2}\bh_i)\right)\right)+T({\bf 2}\bh),
\end{align}
\end{subequations}
where the sum in \eqref{acc} and \eqref{ccc} is done over all the strata of the singular locus while $I_{\ff_a}\subset I$ in \eqref{kcc} restricts the sum over the strata for which the $\cT_i$ flavor symmetry $\ff^i$ contains an $\ff_a$ factor, though there might be some IR flavor symmetry enhancement. For a systematic explanation and derivation see \cite{Martone:2020nsy}. Let us here explain the various factors that enter the central charge:
\begin{itemize}
\item $h$ is the quaternionic dimension of the extended Coulomb branch (ECB) of the theory \cite{Argyres:2016xmc}.
\item $\D_u$ and $\D_v$ are the globally defined Coulomb branch scaling dimensions of the coordinates on $\cC$.
\item The $b_i$, which were initially defined in \cite{Argyres:2016xmc},\footnote{Notice that between the $b_i$ in \cite{Argyres:2016xmc} and \eqref{bi} there is an inessential difference of a factor 2, this is due to a slightly different convention between \cite{Argyres:2016xmc} and \cite{Martone:2020nsy} for the measure factor of the partition function of the topologically twisted version of the SCFT which is used to derive the central charge formulae.} are integers which characterize the physics of the theory $\cT_i$ supported on the $\cS_i$ stratum, and are given by
\beq\label{bi}
b_i := \frac{12c_i-2-h_i}{\D_i}
\eeq
where $c_i$, $h_i$ and $\D_i$ are, respectively, the $c$ central charge, the quaternionic dimension of the ECB and the scaling dimension of the Coulomb branch parameter of the rank-1 theory $\cT_i$.
\item $\Dsi_i$ is the scaling dimension of the polynomial identifying $\cSb_i$ as a hypersurface in $\cC$. Explicitly $\Dsi_i=(\D_u,\D_v,p\D_u\equiv q\D_v)$ for, respectively $\cSb_i\cong(\cSb_u,\cSb_v,\cSbpq)$.
\item $k_a$ is the central charge of the $\ff_a$ flavor factor of the SCFT's flavor symmetry and $T({\bf 2}\bh)$ is the Dynkin index of the $\ff_a$ representation of the ECB, if there is one. $\ff_a$ is assumed to be simple.

\item $k_a^i$ is the corresponding quantity applied to the $i$-th stratum $\cS_i$. That is $k_a^i$ is the central charge of the flavor factor $\ff_a^i$ of the rank-1 theory $\cT_i$ which contains $\ff_a$ and $d^a_i$ is the index of embedding of $\ff_a$ in $\ff_a^i$. Since the sum in \eqref{kcc} is over $I_{\ff_a}$, the flavor symmetry of $\cT_i$ contains a $\ff_a$ factor.  $T({\bf2}\bh_i)$ is the Dynkin index of the $\ff^i_a$ representation of the ECB, if there is any, for the theory $\cT_i$.
\end{itemize}
We will now see how this works in concrete physical examples.

\section{Examples of rank 2 Coulomb branch stratifications}\label{sec:examp}

In this section we will go through a large series of examples where we can highlight how this entire structure comes together.  We will be fairly schematic by reporting only a few pieces of information for each theory, but we will record their complete Coulomb branch Hasse diagrams.  Apart from a few exceptions, the stratification of $\cC$ will have multiple disconnected components will be depicted schematically as:

\begin{tcolorbox}
\begin{center}
\textbf{Hasse diagrams of rank-2 Coulomb branches}\\[0.5 em]
\begin{tikzpicture}[decoration={markings,
mark=at position .5 with {\arrow{>}}}]
\begin{scope}[scale=1.5]
\node[bbc,scale=.5] (p0a) at (0,0) {};
\node[bbc,scale=.5] (p0b) at (0,-2) {};
\node[scale=.8] (t0a) at (0,.3) {$\cC$};
\node[scale=.8] (t0b) at (0,-2.3) {0};
\node[scale=.8] (p1) at (-.5,-1) {$\cT_1$};
\node[scale=.8] (p2) at (.5,-1) {$[I_n,\varnothing]$};
\node[scale=.8] (t1a) at (-.5,-.5) {$\mT_1$};
\node[scale=.8] (t1b) at (-.5,-1.5) {$\cS_1$};
\node[scale=.8] (t1a) at (.5,-.5) {$\mT_\nf$};
\node[scale=.8] (t1b) at (.5,-1.5) {$\cS_\nf$};
\node[scale=.8] (t2) at (0,-1) {$\ldots$};
\draw[red] (p0a) -- (p1);
\draw[red] (p0a) -- (p2);
\draw[red] (p1) -- (p0b);
\draw[red,dashed] (p2) -- (p0b);
\end{scope}
\end{tikzpicture}
\end{center}\vspace{-1em}
By \eqref{strar2}, the lowest and highest strata will be common to all Hasse diagrams and are respectively the origin $\{0\}$ and the entire Coulomb branch.  We label each intermediate strata with the theory supported on the given stratum $\cS_i$ as $\cT_i$.  The transverse slice to each stratum, $\mT_i$, is naturally identified with the Coulomb branch of $\cT_i$ and we use the corresponding entry in the Kodaira list (see table \ref{tab:Kodaira}) to label it. If the theory has an ECB we will write explicitly the extra free-hypermultiplets and we will label them as multiple copies of the quaternion $\H^\ell:=\H^{\otimes^\ell}$.\footnote{We thank J. Grimminger for suggesting this notation which considerably clarifies our Hasse diagrams.} The transverse slice to $0$ inside the component $\cSb_i$ is $\cSb_i$ itself, so we label the elementary slices emanating from the $0$ node by the corresponding enclosing strata $\cS_i$.  As we described at length above, their closures can be one of the entries in \ref{table:eps}. The dots indicate that the singular locus is in general the union of many strata. When a stratum supports an $[I_n,\varnothing]$ theory, so only a loose special Kahler stratification applies (see sec \ref{sec:In}), we indicate the enclosed elementary slice by a dashed edge of the Hasse diagram.
\end{tcolorbox}


Notice that to reconstruct the topology of $\cSb$ from the information provided by the Hasse diagram we need the extra information of the scaling dimension of the globally defined coordinates. For each example below, we report $(\D_u,\D_v)$, as well as other relevant SCFT data, in an adjacent table. Finally, the list of possible theories supported on the various strata, along with the information needed to check the matching of the central charge formula, is reported in table \ref{tab:theories} for ease of use. In the square bracket notation for naming these theories we only report the non-abelian part of the flavor symmetry of the corresponding theory.

\begin{table}
\centering
\renewcommand{\arraystretch}{1.3}
$\begin{array}{|c|c|c|c|c|c|c|c|cc}
\cline{1-8}
\multicolumn{8}{|c|}{\text{\bf Summary of rank-1 theories $\cT_i$ supported on $\cS_i$}}\\
\hhline{========~}
\text{Name} & \multicolumn{1}{c|}{\ \ 12\, c\ \ } &\quad \D_u\quad{} &\quad h\quad{}&\, {\boldsymbol R_{2h}}\,{}&\quad b\quad{} &\quad \ff\quad{} &\quad k_\ff\quad{}&\, {}&\quad{} \\
\cline{1-9}
\ \boldsymbol{ [II^*,E_8]}\ &62&6&0&{\bf1}&10&E_8&12&\multicolumn{2}{c}{\parbox[t]{2mm}{\multirow{6}{*}{\rotatebox[origin=c]{90}{SCFTs}}}}\\
\ \boldsymbol{[II^*,C_5]}\ &49&6&5&{\bf10}&7&C_5&7\\
\ \boldsymbol{[III^*,E_7]}\ &38&4&0&{\bf1}&9&E_7&8\\
\ \boldsymbol{[III^*,C_3A_1]}\ &29&4&3&({\bf6,1})&6&C_3\times A_1&(5,8)\\
\ \boldsymbol{[IV^*,C_2]}\ &19&3&2&{\bf4}_0&5&C_2\times U_1&(4,?)\\
\ \boldsymbol{[I_0^*,C_1]}\ &9&2&1&{\bf2}&3&C_1&3&\\
\hhline{=========~}
\ \boldsymbol{[I_1,\varnothing]}\ &3&1&0&{\bf 1}&1&U_1&1&\multicolumn{2}{c}{\parbox[t]{2mm}{\multirow{4}{*}{\rotatebox[origin=c]{90}{IR-free}}}}\\
\ \boldsymbol{[I_n^*,C_{\frac{n+4}{4}}]}\ &9+\frac34n&2&1+\frac{n}{4}&{\bf\frac{n+4}2}&3+\frac{n}4&C_{\frac{n+4}4}&3\\
\ \boldsymbol{[I_n,A_{n-1}]}\ &n+2&1&0&0&n&A_{n-1}&2\\
\ \boldsymbol{[I_n,A_1]}\ &5&1&0&0&3&A_{1}&2\\
\hhline{=========~}
\boldsymbol{[I^*_n,C_n]_{\Z_2}}\ & 2n+4&\red{2}&0&0&\red{n}&C_n&2\\
\cline{1-9}
\end{array}$
\caption{\label{tab:theories} For the convenience of the reader, we summarize the properties of the rank-1 theories which will appear on complex co-dimension one loci of the rank-2 theories which we analyze below. In applying the central charge formula for discretely gauged theories, some of the quantities refer to the ``parent'' theories;  we show these entries in red.}
\end{table}

\subsection{Lagrangian theories}

The moduli space of vacua of $\cN=2$ four dimensional gauge theories has been studied for decades in great detail, see, e.g., \cite{Argyres:1996eh}.  This section will add little to the common lore and should be seen as a warm-up to set up our notations and see how things work in well-known examples.

\subsubsection{$\SU(3)$ $\cN=4$}

Let us start from the simplest example, $\suf(3)$ $\cN=4$. This is an $\suf(3)$ $\cN=2$ gauge theory with a single hypermultiplet tranforming in the ${\bf 8}$. The Coulomb branch in this case is an orbifold space, $\cC=\C^2/S_3$, where $S_3$ is the symmetric group of degree 3 which is the Weyl group of $\suf(3)$. In this case a lagrangian description is available which allows the identification of the globally defined Coulomb branch coordinates with the vevs of gauge invariant combinations of free fields. Then $u=\left\langle {\rm Tr}\left(\Phi^2\right)\right\rangle$ and $v=\left\langle {\rm Tr}\left(\Phi^3\right)\right\rangle$, where $\Phi$ is the complex scalar component of the $\cN=2$ vector multiplet. It follows that $\D_u=2$ and $\D_v=3$. In this case the singular locus has a single connected component. This can be seen in two ways, either by looking at the fixed set of the orbifold action of $S_3$ on $\C^2$, or by analyzing the low energy physics. Let's take the second route. 

\begin{figure}[h!]
\ffigbox{
\begin{subfloatrow}
\ffigbox[7cm][]{\begin{tikzpicture}[decoration={markings,
mark=at position .5 with {\arrow{>}}}]
\begin{scope}[scale=1.5]
\node[bbc,scale=.5] (p0a) at (0,0) {};
\node[bbc,scale=.5] (p0b) at (0,-2) {};
\node[scale=.8] (t0a) at (0,.3) {$\cC$};
\node[scale=.8] (t0b) at (0,-2.3) {0};
\node[scale=.8] (p1) at (-0,-1) {$\boldsymbol{ [I_0^*,C_1]}{\times}\H$};
\node[scale=.8] (t1a) at (-0.3,-.5) {$I_0^*$};
\node[scale=.8] (t1b) at (-.3,-1.5) {$I_0$};
\draw[red] (p0a) -- (p1);
\draw[red] (p1) -- (p0b);
\end{scope}
\end{tikzpicture}
}
{\caption{\label{CBSU38}The Hasse diagram for the Coulomb branch of the $\cN=4$ $\suf(3)$ theory.}}
\end{subfloatrow}\hspace{1cm}%
\begin{subfloatrow}
\capbtabbox[7cm]{%
  \renewcommand{\arraystretch}{1.1}
  \begin{tabular}{|c|c|} 
  \hline
  \multicolumn{2}{|c|}{$\cN=4\ \suf(3)$}\\
  \hline\hline
  $(\D_u,\D_v)$  &\quad (2,3)\quad{} \\
  $24a$ &  48\\  
  $12c$ & 24 \\
$\ff_k$ & $\spf(1)_8$ \\ 
$h$&2\\
$T({\bf2}\bh)$&2\\
\hline\hline
$\cSb$&$L_{(2,3)}(0,0,1)$\\
\hline
  \end{tabular}
}{%
  \caption{\label{CcSU38}Central charges, flavor level, Coulomb branch parameters and ECB dimension.}%
}
\end{subfloatrow}}{\caption{\label{TotSU38} Information about the $\suf(3)$ $\cN=4$ theory.}}
\end{figure}

On a generic point of the Coulomb branch, there is a $\U(1)^2$ unbroken gauge symmetry and all components of the hypermultiplet are massive. Extra massless states appear on special loci where either there is an enhancement of the unbroken gauge group, or some component of the hypermultiplet become massless. There is certainly a locus where the former phenomenon happens, specifically  $\U(1)^2\to\suf(2)\times \U(1)$. This special locus coincides with the hypersurface $u^3=\omega v^2$. $\omega\in \C^*$ and it depends on the normalization of the Coulomb branch parameters. From the discussion of section \ref{sec:knots}, this is topologically a $(2,3)$ torus knot. This locus is also where some components of the hypermultiplet, which transform in the adjoint of the unbroken $\suf(2)$, become massless. This then implies that the theory on the generic point of the singular locus is an $\cN=4$ $\suf(2)$ which is itself a SCFT, which we label as $[I_0^*,C_1]$. This is the only locus of gauge symmetry enhancement and/or where hypermultiplet components become massless therefore the Hasse diagram will have a single stratum $\cS_1$.  Let's spell out in detail how to characterize its special Kahler structure. 

The simplest way is to compute the scaling dimension of the uniformizing parameter of the closure of $\cS_1$. From \eqref{unipq}
\beq
t_{\cS_1}= (u^3)^{\frac16} = (\omega v^2)^{\frac16}
\quad \Rightarrow\quad 
\D_t=1,
\eeq
or in other words $\cSb_1\equiv I_0$. We can, of course, reach an analogous conclusion performing a mononodromy analysis along a loop like in \eqref{gpq}. To finish our characterization of the Hasse diagram, which is shown in figure \ref{CBSU38}, we recall that the transverse slice is identified with the Coulomb branch of the (rank-1) theory supported on the $\cS_1$ and therefore $\mT_1\equiv I_0^*$.

A last check is that we are able to reproduce the central charges, reported in table \ref{CcSU38}, of this rank-2 theory from the stratification of the Coulomb branch locus using \eqref{acc}-\eqref{kcc}. Again as a warm-up we will spell out this step in detail. 

Since there is a single stratum the sum only involves a single term. $b_1$ will be equal to the entry in table \ref{tab:theories} corresponding to $[I_0^*,C_1]$, that is $b_1=3$, while $\Dsi_1=[u^3+ \omega v^2]=6$. Since this theory has a two quatenionic dimensional ECB, $h=2$, and plugging the appropriate scaling dimensions for $(u,v)$ in \eqref{acc}-\eqref{ccc}, we get $24a=48$ and $12c=24$ precisely matching the values in table \ref{CcSU38}. 

Finally let's match the central charge of the flavor symmetry. For that we need to concentrate on the theory supported on $\cS_1$.   Again we can read off all the information we need by looking up the entries corresponding to $[I_0^*,C_1]$ in table \ref{tab:theories}: $(\D_1,k_{C_1},T({\bf 2}\bh_1))=(2,3,1)$. The only quantity left is $T({\bf2}\bh)$, that is the Dynkin index of the representation of the ECB of the rank-2 theory. This value can again be found in table \ref{tab:theories} but let's explain how this works. Notice that the $h=2$ ECB arises from the two components of the adjoint hypermultiplet corresponding to the two Cartan generators of $\suf(3)$. Since the flavor symmetry commutes with the gauge symmetry, it acts by transforming the two half-hypermultiplet components into one another without acting on the gauge index, therefore ${\bf R}_h={\bf 2}\oplus {\bf2}$ and therefore $T({\bf2}\bh)=2$. Plugging everything in \eqref{kcc} gives precisely the expected result $k_{C_1}^{\suf(3)}=8$.

\subsubsection{$\SU(3)$ with $N_f=6$}

We now consider the Coulomb branch of another theory which has been analyzed extensively, the $\cN=2$ $\suf(3)$ gauge theory with 6 hypermultiplets in the ${\bf 3}$. The discussion here will be considerably less detailed than the previous one. Since the gauge algebra is still $\suf(3)$, $(\D_u,\D_v)=(2,3)$ and $u^3=\omega v^2$ will still semi-classically identify the only locus where there is a gauge enhancement. In this case no component of the hypermultiplets becomes massless there. It is well-known that the Coulomb branch of the $\cN=2$ pure $\suf(2)$ theory has two singular loci where the theory is effectively $[I_1,\varnothing]$ and it is therefore this phenomenon which gives rise to the splitting of the knotted singularity into two knots. Finally at $v=0$ one component of each hypermultiplet, which can be made charged under a single $\U(1)$, becomes massless. This gives rise to an unknotted singularity with an effective $[I_6,\suf(5)]$ supported over it and concludes our discussion of the singular locus which is topologically the link $L_{(2,3)}(0,1,2)$. 

\begin{figure}[h!]
\ffigbox{
\begin{subfloatrow}
\ffigbox[7cm][]{
\begin{tikzpicture}[decoration={markings,
mark=at position .5 with {\arrow{>}}}]
\begin{scope}[scale=1.5]
\node[bbc,scale=.5] (p0a) at (0,0) {};
\node[bbc,scale=.5] (p0b) at (0,-2) {};
\node[scale=.8] (t0a) at (0,.3) {$\cC$};
\node[scale=.8] (t0b) at (0,-2.3) {0};
\node[scale=.8] (p1) at (-.8,-1) {$\boldsymbol{[I_1,\varnothing]}$};
\node[scale=.8] (p2) at (.8,-1) {$\boldsymbol{[I_6,A_5]}$};
\node[scale=.8] (p3) at (0,-1) {$\boldsymbol{[I_1,\varnothing]}$};
\node[scale=.8] (t1a) at (-.6,-.4) {$I_1$};
\node[scale=.8] (t1b) at (-.6,-1.6) {$I_0$};
\node[scale=.8] (t2a) at (-.2,-.6) {$I_1$};
\node[scale=.8] (t2b) at (-.2,-1.5) {$I_0$};
\node[scale=.8] (t3a) at (.6,-.4) {$I_6$};
\node[scale=.8] (t3b) at (.6,-1.6) {$I_0^*$};
\draw[red] (p0a) -- (p1);
\draw[red] (p0a) -- (p2);
\draw[red] (p0a) -- (p3);
\draw[red,dashed] (p1) -- (p0b);
\draw[red] (p2) -- (p0b);
\draw[red,dashed] (p3) -- (p0b);
\end{scope}
\end{tikzpicture}}
{\caption{\label{CBSU363}The Hasse diagram for the Coulomb branch of the $\suf(3)$ gauge theory with 6 ${\bf 3}$s.}}
\end{subfloatrow}\hspace{1cm}%
\begin{subfloatrow}
\capbtabbox[7cm]{%
  \renewcommand{\arraystretch}{1.1}
  \begin{tabular}{|c|c|} 
  \hline
  \multicolumn{2}{|c|}{$\suf(3)$ {\rm w/} $N_f=6$}\\
  \hline\hline
  $(\D_u,\D_v)$  &\quad (2,3)\quad{} \\
  $24a$ &  58\\  
  $12c$ & 34 \\
$\ff_k$ & $\suf(6)_6$ \\ 
$h$&0\\
$T({\bf2}\bh)$&0\\
\hline\hline
$\cSb$&$L_{(2,3)}(0,1,2)$\\
\hline
  \end{tabular}
}{%
  \caption{\label{CcSU363}Central charges, Coulomb branch parameters and ECB dimension.}%
}
\end{subfloatrow}}{\caption{\label{TotSU363} Information about the $\suf(3)$ $\cN=2$ theory with 6 hypermultiplets in the ${\bf 3}$.}}
\end{figure}

Performing the special Kahler characterization as before we realize that the two knotted strata are again an $I_0$ while the unknotted one is the scale invariant special Kahler geometry parametrized by a scaling dimension 2 Coulomb branch parameter, that is an $I_0^*$. The complete Hasse diagram of the corresponding Coulomb branch is shown in figure \ref{CBSU363}.

Finally we can check that the central charges of the rank-2 theory can be reproduced from the stratification analysis. Since now the singular locus has three components, the sum has three terms. All the information needed to performing the computation are either in table \ref{tab:theories} or in table \ref{CcSU363}.  We leave it up to the reader to check that indeed all the information are perfectly reproduced.

\subsubsection{$\SU(3)$ with $1\, ({\bf 3})\oplus1\, ({\bf 6})$}

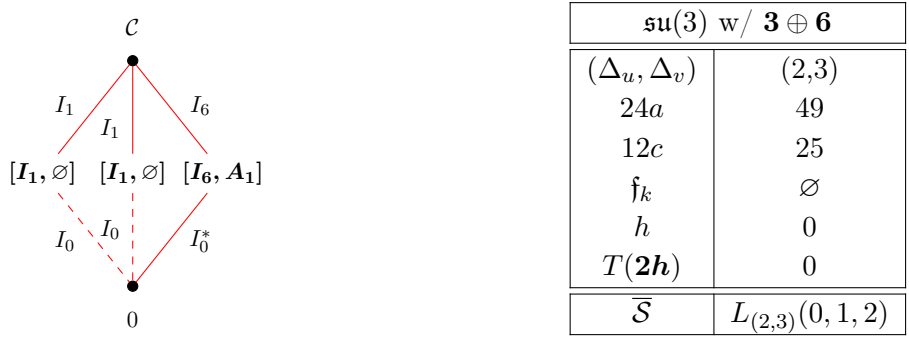
\begin{figure}[h!]
\ffigbox{
\begin{subfloatrow}
\ffigbox[7cm][]{
\begin{tikzpicture}[decoration={markings,
mark=at position .5 with {\arrow{>}}}]
\begin{scope}[scale=1.5]
\node[bbc,scale=.5] (p0a) at (0,0) {};
\node[bbc,scale=.5] (p0b) at (0,-2) {};
\node[scale=.8] (t0a) at (0,.3) {$\cC$};
\node[scale=.8] (t0b) at (0,-2.3) {0};
\node[scale=.8] (p1) at (-.8,-1) {$\boldsymbol{[I_1,\varnothing]}$};
\node[scale=.8] (p2) at (.8,-1) {$\boldsymbol{[I_6,A_1]}$};
\node[scale=.8] (p3) at (0,-1) {$\boldsymbol{[I_1,\varnothing]}$};
\node[scale=.8] (t1a) at (-.6,-.4) {$I_1$};
\node[scale=.8] (t1b) at (-.6,-1.6) {$I_0$};
\node[scale=.8] (t2a) at (-.2,-.6) {$I_1$};
\node[scale=.8] (t2b) at (-.2,-1.5) {$I_0$};
\node[scale=.8] (t3a) at (.6,-.4) {$I_6$};
\node[scale=.8] (t3b) at (.6,-1.6) {$I_0^*$};
\draw[red] (p0a) -- (p1);
\draw[red] (p0a) -- (p2);
\draw[red] (p0a) -- (p3);
\draw[red,dashed] (p1) -- (p0b);
\draw[red] (p2) -- (p0b);
\draw[red,dashed] (p3) -- (p0b);
\end{scope}
\end{tikzpicture}}
{\caption{\label{CBSU336}The Hasse diagram for the Coulomb branch of the $\suf(3)$ gauge theory with one ${\bf 3}$ and one ${\bf 6}$.}}
\end{subfloatrow}\hspace{1cm}%
\begin{subfloatrow}
\capbtabbox[7cm]{%
  \renewcommand{\arraystretch}{1.1}
  \begin{tabular}{|c|c|} 
  \hline
  \multicolumn{2}{|c|}{$\suf(3)$ {\rm w/} ${\bf 3}\oplus {\bf 6}$}\\
  \hline\hline
  $(\D_u,\D_v)$  &\quad (2,3)\quad{} \\
  $24a$ &  49\\  
  $12c$ & 25 \\
$\ff_k$ & $\varnothing$ \\ 
$h$&0\\
$T({\bf2}\bh)$&0\\
\hline\hline
$\cSb$&$L_{(2,3)}(0,1,2)$\\
\hline
  \end{tabular}
}{%
  \caption{\label{CcSU336}Central charges, Coulomb branch parameters and ECB dimension.}%
}
\end{subfloatrow}}{\caption{\label{TotSU336}Information about the $\suf(3)$ $\cN=2$ theory with one hypermultiplet in the ${\bf 3}$ and one in the ${\bf 6}$.}}
\end{figure}

Let's continue our analysis of Lagrangian theories with the $\cN=2$ $\suf(3)$ theory with one hypermultiplet in the ${\bf 3}$ and one in the ${\bf 6}$. The analysis of this case is closely analogous to the previous one therefore we will focus on what differentiates the two: the theory supported on the unknotted stratum. The Hasse diagram of the corresponding Coulomb branch is shown in figure \ref{CBSU336}.

Along $v=0$, the hypermultiplet in the fundamental contributes a single massless component and its corresponding $\U(1)$ charge can be made purely electric and normalized to 1. In the same basis and with the same normalization, the hypermultiplet in the ${\bf 6}$ contributes instead both a hypermultiplet with electric charge 1 and one with electric charge 2. The theory supported on the unknotted singularity is a $\U(1)$ theory with two hypermultiplets with charge one and one with charge two. The total contribution to the beta function of the $\U(1)$ gauge coupling is six as before and therefore the closure of the transverse slice will be an $I_6$. Yet the low energy theory is different and it is this difference that accounts for the different value for the $a$ and $c$ central charges between the two rank-2 theories.

One last remark is in order. Using the geometric information called the \emph{deformation pattern}, which played a central role in the rank-1 classification performed in \cite{Argyres:2015ffa,Argyres:2015gha,Argyres:2016xua,Argyres:2016xmc}, the two low-energy theories along the unknotted singularities can be written as
\begin{subequations}
\begin{align}\label{def1}
\suf(3)\ {\rm w/}\ N_f=6\quad&:\quad I_6\to\{I_1^6\},\\\label{def2}
\suf(3)\ {\rm w/}\ {\bf 3}\ \oplus{\bf 6}\quad&:\quad I_6\to\{I_1^2,I_4\}.
\end{align}
\end{subequations}
Without wanting to delve into the analysis of mass deformation of the rank-2 geometries, which is still largely beyond reach, we nevertheless notice that combining the existence of the two knotted $I_1$ singularities with the Dirac quantization condition \cite{Argyres:2015ffa} is enough to make \eqref{def1}-\eqref{def2} the only two allowed deformation pattern of the Hasse diagram in figure \ref{CBSU336}-\ref{CBSU363}.  For a slightly more detailed discussion, see \cite{Martone:2020nsy}.  This observation provides an encouraging piece of information that a complete analysis of rank-2 theories, in the style of \cite{Argyres:2015ffa,Argyres:2015gha,Argyres:2016xua,Argyres:2016xmc}, is possible.

\subsubsection{$G_2$ with $4\, ({\bf 7})$}\label{sec:g2}

We conclude the analysis of the lagrangian theories with a slightly less thoroughly studied example, an $\cN=2$ SCFT with $\gf_2$ gauge symmetry and four hypermultiplets in the ${\bf 7}$. The exceptional gauge symmetry does not change the analysis considerably. Since the Weyl symmetry of $\gf_2$ contains an extra $\Z_2$ factor, $(\D_u,\D_v)=(2,6)$. This implies that the knotted singularity is now a (3,1) knot.  And the Hasse diagram of the corresponding Coulomb branch is shown in figure \ref{CBG247}.

\begin{figure}[h!]
\ffigbox{
\begin{subfloatrow}
\ffigbox[7cm][]{
\begin{tikzpicture}[decoration={markings,
mark=at position .5 with {\arrow{>}}}]
\begin{scope}[scale=1.5]
\node[bbc,scale=.5] (p0a) at (0,0) {};
\node[bbc,scale=.5] (p0b) at (0,-2) {};
\node[scale=.8] (t0a) at (0,.3) {$\cC$};
\node[scale=.8] (t0b) at (0,-2.3) {0};
\node[scale=.7] (p1) at (-.8,-1) {$\boldsymbol{[I_1,\varnothing]}{\times}\H^4$\quad\ \ };
\node[scale=.7] (p2) at (.8,-1) {\quad\ $\boldsymbol{[I^*_{12},C_4]}$};
\node[scale=.7] (p3) at (0,-1) {\ \ $\boldsymbol{[I_1,\varnothing]}{\times}\H^4$};
\node[scale=.8] (t1a) at (-.6,-.4) {$I_1$};
\node[scale=.8] (t1b) at (-.6,-1.6) {$I^*_0$};
\node[scale=.8] (t2a) at (-.2,-.6) {$I_1$};
\node[scale=.8] (t2b) at (-.2,-1.5) {$I^*_0$};
\node[scale=.8] (t3a) at (.6,-.4) {$I^*_{12}$};
\node[scale=.8] (t3b) at (.6,-1.6) {$I_0^*$};
\draw[red] (p0a) -- (p1);
\draw[red] (p0a) -- (p2);
\draw[red] (p0a) -- (p3);
\draw[red,dashed] (p1) -- (p0b);
\draw[red] (p2) -- (p0b);
\draw[red,dashed] (p3) -- (p0b);
\end{scope}
\end{tikzpicture}}
{\caption{\label{CBG247}The Hasse diagram for the Coulomb branch of the $G_2$ gauge theory with 4 ${\bf 7}$s.}}
\end{subfloatrow}\hspace{1cm}%
\begin{subfloatrow}
\capbtabbox[7cm]{%
  \renewcommand{\arraystretch}{1.1}
  \begin{tabular}{|c|c|} 
  \hline
  \multicolumn{2}{|c|}{$\gf_2$ {\rm w/} 4\,${\bf {\large 7}}$}\\
  \hline\hline
  $(\D_u,\D_v)$  &\quad (2,6)\quad{} \\
  $24a$ &  98\\  
  $12c$ & 56 \\
$\ff_k$ & $\spf(4)_{7}$ \\ 
$h$&4\\
$T({\bf2}\bh)$&1\\
\hline\hline
$\cSb$&$L_{(1,3)}(0,1,2)$\\
\hline
  \end{tabular}
}{%
  \caption{\label{CcG247}Central charges, Coulomb branch parameters and ECB dimension.}%
}
\end{subfloatrow}}{\caption{\label{TotG247}Information about the $\gf_2$ $\cN=2$ theory with 4 hypermultiplets in the ${\bf 7}$.}}
\end{figure}

This theory posses a four quaternionic dimensional ECB which comes from the weight zero component of each hypermultiplet and therefore $T({\bf2}\bh)=1$ which allows to perfectly reproduce all the proprieties of the theory using the central charge formulae. 

\subsection{Argyres-Douglas theories}

Here we will perform the analysis of a subset of Argyres-Douglas (AD) theories \cite{Argyres:1995jj,Argyres:1995xn}, by which we mean any theory which has at least one Coulomb branch parameter of fractional scaling dimension and, if $\D<2$, its corresponding chiral deformation. This is an interesting case study which presents the interesting phenomenon discussed earlier of \emph{apparent violation of unitarity} and have strata described by the irregular special Kahler geometries shown in table \ref{table:eps}.

There is by now a large zoo of AD theories \cite{Xie:2012hs}; here we will focus on those AD theories which can be ``geometrically engineered'' by compactifying type IIB string theory on a Calabi-Yau threefold hypersurface singularity and which are labeled by a pair of ADE Dynkin diagrams ($G,G'$). The corresponding $\cN=2$ SCFT is then labeled with the same ($G,G'$) \cite{Cecotti:2010fi}. Analyzing the deformations of the singularity gives a lot of information about the $\cN=2$ SCFT, e.g., the rank of the theory's flavor symmetry, as well as the scaling dimensions of the Coulomb branch parameters and therefore the overall rank of the theory \cite{Shapere:1999xr}. We of course restrict our analysis to the theories of rank 2.

\subsubsection{$(A_1,A_4)$}

This is the rank-2 theory of an infinite series of rank-$n$ AD theories with trivial Higgs branch: $(A_1,A_{2n})$. The stratification of the Coulomb branch can be read off directly from the discriminant of the curve presented in \cite{Argyres:2005wx}. As discussed in \cite{Martone:2020nsy}, if a form of the SW curve is known where the curve is written as a fibration of an hyperelliptic curve over $\cC$, we can almost characterize the entire Hasse diagram, not just $\cSb$ by considering the quantum discriminant of the geometry. For the $(A_1,A_4)$ we have:
\beq
{\rm SW\ curve}:y^2=x^5 + u x + v\quad \Rightarrow\quad D^\L_x=256 u^5+3125 v^4,
\eeq
and therefore we readily conclude that the theory has a single (4,5) knotted intermediate stratum and that the theory supported on it is an $\cN=2$ $\U(1)$ gauge theory with a single hypermultiplet with charge 1. This in turns constrains $\mT_1$ to be an $I_1$. 

\begin{figure}[h!]
\ffigbox{
\begin{subfloatrow}
\ffigbox[7cm][]{
\begin{tikzpicture}[decoration={markings,
mark=at position .5 with {\arrow{>}}}]
\begin{scope}[scale=1.5]
\node[bbc,scale=.5] (p0a) at (0,0) {};
\node[bbc,scale=.5] (p0b) at (0,-2) {};
\node[scale=.8] (t0a) at (0,.3) {$\cC$};
\node[scale=.8] (t0b) at (0,-2.3) {0};
\node[scale=.8] (p1) at (-0,-1) {$\boldsymbol{ [I_1,\varnothing]}$};
\node[scale=.8] (t1a) at (-0.3,-.5) {$I_1$};
\node[scale=.8] (t1b) at (-.3,-1.5) {$I_0^{*(3)}$};
\draw[red] (p0a) -- (p1);
\draw[red,dashed] (p1) -- (p0b);
\end{scope}
\end{tikzpicture}}
{\caption{\label{CBA1A4}The Hasse diagram for the Coulomb branch of the $(A_1,A_4)$ AD theory.}}
\end{subfloatrow}\hspace{1cm}%
\begin{subfloatrow}
\capbtabbox[7cm]{%
  \renewcommand{\arraystretch}{1.1}
  \begin{tabular}{|c|c|} 
  \hline
  \multicolumn{2}{|c|}{$(A_1,A_4)$}\\
  \hline\hline
  $(\D_u,\D_v)$  &\quad $\left(\frac87,\frac{10}7\right)$\quad{} \\
  $24a$ &  $\frac{134}7$\\  
  $12c$ &$\frac{68}7 $\\
$\ff_k$ & $\varnothing$ \\ 
$h$&0\\
$T({\bf2}\bh)$&0\\
\hline\hline
$\cSb$&$L_{(4,5)}(0,0,1)$\\
\hline
  \end{tabular}
}{%
  \caption{\label{CcA1A4}Central charges, Coulomb branch parameters and ECB dimension.}%
}
\end{subfloatrow}}{\caption{\label{TotA1A4} Information about the $(A_1,A_4)$ AD theory.}}
\end{figure}
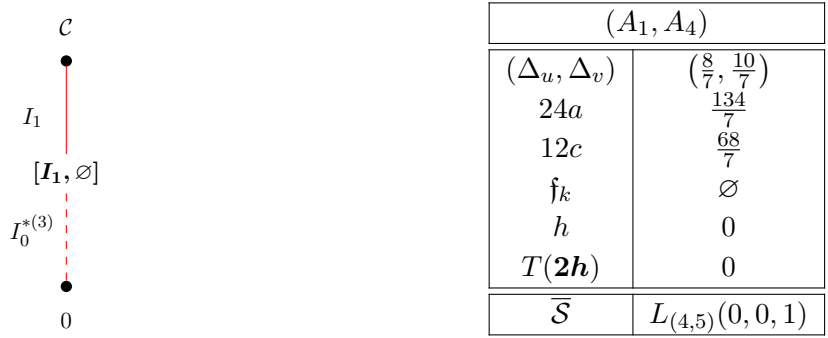

Let us analyze more closely the special Kahler structure of the knotted singularity. The uniformizing parameter for this hypersurface is
\beq
t_{(A_1,A_4)} \sim (u^5)^{\frac1{20}} \sim (v^4)^{\frac1{20}}
\quad \Rightarrow\quad 
\D_t=\frac27.
\eeq
As previously announced, we find indeed that $\D_t<1$. From the table \ref{tab:theories} we immediately identify $\bar{\cS_1}\equiv I_0^{*(3)}$. As an extra check of the consistency of this construction, we can compute the monodromy associated to the stratum following the discussion around \eqref{gpq} and find indeed $-\I$. Notice that, in line with conjecture \ref{NoHB}, the theory supported on this strata has a trivial Higgs branch. We leave it up to reader to check that central charges for this theory can be correctly reproduced from the complete Hasse diagram shown in figure \ref{CBA1A4} and the formulae \eqref{acc}-\eqref{kcc}.

\subsubsection{$(A_1,D_5)$}

This theory belongs instead to an infinite series of AD theory with a $\C^2/\Z_2$ Higgs branch: $(A_1,D_{2n+1})$. The stratification of the Coulomb branch singular locus can be again read off straightforwardly from the expression of the SW curve reported in \cite{Argyres:2005wx},
\beq\label{disA1D5}
{\rm SW\ curve}:y^2=x^5 +x( u x + v)\quad \Rightarrow\quad D_x^\L=v^2\left(27 u^4-256 v^3\right).
\eeq
This implies that there is a (3,4) knotted stratum ($\cS_1$) as well as an unknotted stratum of type $II$ ($\cS_2$).

\begin{figure}[h!]
\ffigbox{
\begin{subfloatrow}
\ffigbox[7cm][]{
\begin{tikzpicture}[decoration={markings,
mark=at position .5 with {\arrow{>}}}]
\begin{scope}[scale=1.5]
\node[bbc,scale=.5] (p0a) at (0,0) {};
\node[bbc,scale=.5] (p0b) at (0,-2) {};
\node[scale=.8] (t0a) at (0,.3) {$\cC$};
\node[scale=.8] (t0b) at (0,-2.3) {0};
\node[scale=.8] (p1) at (-.5,-1) {$\boldsymbol{[I_1,\varnothing]}$};
\node[scale=.8] (p2) at (.5,-1) {$\boldsymbol{[I_2,A_1]}$};
\node[scale=.8] (t1a) at (-.5,-.4) {$I_1$};
\node[scale=.8] (t1b) at (-.5,-1.6) {$I_0^{*(2)}$};
\node[scale=.8] (t2a) at (.5,-.4) {$I_2$};
\node[scale=.8] (t2b) at (.5,-1.6) {$II$};
\draw[red] (p0a) -- (p1);
\draw[red] (p0a) -- (p2);
\draw[red,dashed] (p1) -- (p0b);
\draw[red] (p2) -- (p0b);
\end{scope}
\end{tikzpicture}}
{\caption{\label{CBA1D5}The Hasse diagram for the Coulomb branch of the $(A_1,D_5)$ AD theory.}}
\end{subfloatrow}\hspace{1cm}%
\begin{subfloatrow}
\capbtabbox[7cm]{%
  \renewcommand{\arraystretch}{1.1}
  \begin{tabular}{|c|c|} 
  \hline
  \multicolumn{2}{|c|}{$(A_1,D_5)$}\\
  \hline\hline
  $(\D_u,\D_v)$  &$\quad \left(\frac65,\frac85\right)\quad{}$ \\
  $24a$ & $ \frac{114}5$\\  
  $12c$ & 12 \\
$\ff_k$ & $\suf(2)_{\frac{16}5}$ \\ 
$h$&0\\
$T({\bf2}\bh)$&0\\
\hline\hline
$\cSb$&$L_{(3,4)}(0,1,1)$\\
\hline
  \end{tabular}
}{%
  \caption{\label{CcA1D5}Central charges, Coulomb branch parameters and ECB dimension.}%
}
\end{subfloatrow}}{\caption{\label{TotA1D5}Information about the $(A_1,D_5)$ AD theory.}}
\end{figure}
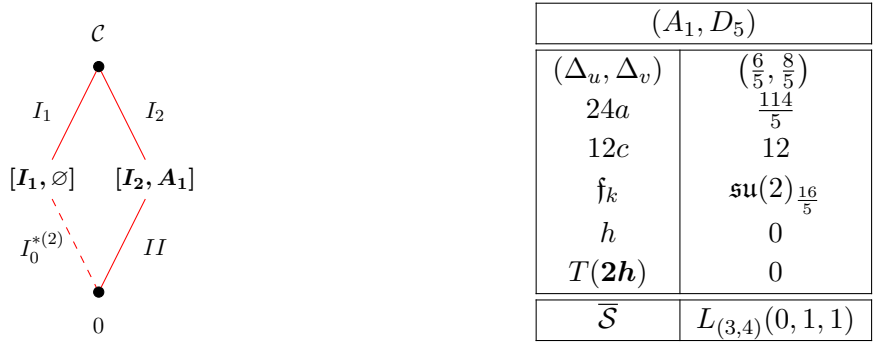

Again we can use the extra information which is provided from the discriminant. As before we infer that $\mT_1\equiv I_1$, again compatible with conjecture \ref{NoHB}. There is now an ambiguity in identifying $\mT_2$ as both an $I_2$ and a $II$ would be compatible with the discriminant \eqref{disA1D5}. There are two ways to lift this ambiguity:
\begin{itemize}
\item[1-] Using \eqref{acc} and \eqref{ccc} to match the central charge of the theory, we readily derive that $b_2=2$. Since the only allowed deformation pattern of the $II$ has $b=3$: $II\to\{I_1^3\}$, we immediately conclude that the theory supported on the stratum is a $\cN=2$ $U(1)$ gauge theory with two hypermultiplets of charge $Q=1$. This interpretation is in fact associated with the deformation pattern $I_2\to\{I_1^2\}$ which has indeed the claimed $b=2$.

\item[2-] We know that the $(A_1,D_5)$ theory has a one quaternionic dimensional Higgs branch and a non-trivial flavor symmetry. If the transverse slice $\mT_2=II$, the theory supported on $\cS_2$ must be the $(A_1,A_2)$ theory which has no flavor symmetry nor Higgs branch. This would imply that the $\suf(2)$ flavor symmetry of the rank-2 theory is not realized by any rank-1 theories supported on the singular locus of the Coulomb branch contradicting the $\cN=2$ UV-IR simple flavor condition in \cite{Martone:2020nsy}. The $I_2$ interpretation would instead carry a $\suf(2)$ flavor symmetry which could be identified with the flavor symmetry of the rank-2 SCFT and also perfectly reproducing the level $\frac{16}5$. This also implies that the entire Higgs branch of the theory is indeed a mixed branch.

\end{itemize}
We can summarize the arguments above in the Hasse diagram of the corresponding Coulomb branch shown in figure \ref{CBA1D5}. We observe that this Coulomb branch stratification has two nice implications consistent with known facts about this theory \cite{Beem:2019tfp}. First, the entire Higgs branch of the ($A_1,D_5$) extends over its Coulomb branch and it is therefore a mixed branch, and secondly the low-energy theory on the generic point of the Higgs branch of the theory is the rank-1 ($A_1,A_2$).\footnote{The fact that this theory is a rank-1 rather than a rank-2 theory is what makes the extension of the Higgs branch over the Coulomb branch a mixed branch rather than a ECB.}

\subsubsection{$(A_1,A_5)$}

This theory belongs instead to an infinite series of rank-$n$ AD theory with Higgs branch equal to $\C^2/\Z_{n+1}$: $(A_1,A_{2n+1})$. The stratification of the Coulomb branch singular locus can be again read off straightforwardly from the expression of the SW curve reported in \cite{Argyres:2005pp}:
\beq\label{disA1A5}
{\rm SW\ curve}:y^2=x^6 + u x + v\quad \Rightarrow\quad D^\L_x=\left(3125 u^6-46656 v^5\right),
\eeq
and again it implies that singular locus has topology $L_{(5,6)}(0,0,1)$. Performing an analysis analogous to the one above, the rest of the Hasse diagram, shown in figure \ref{CBA1A5}, can be completely characterized and the central $(a,c)$ correctly reproduced. Compatibly with our conjecture \ref{NoHB}, the theory supported on the irregular stratum, i.e., the single connected component of $\cSb$, has no Higgs branch yet the rank-2 theory at the origin has a non-trivial Higgs branch.  This in turn implies that the low-energy theory on the generic point of the Higgs branch is trivial.

\begin{figure}[h!]
\ffigbox{
\begin{subfloatrow}
\ffigbox[7cm][]{
\begin{tikzpicture}[decoration={markings,
mark=at position .5 with {\arrow{>}}}]
\begin{scope}[scale=1.5]
\node[bbc,scale=.5] (p0a) at (0,0) {};
\node[bbc,scale=.5] (p0b) at (0,-2) {};
\node[scale=.8] (t0a) at (0,.3) {$\cC$};
\node[scale=.8] (t0b) at (0,-2.3) {0};
\node[scale=.8] (p1) at (-0,-1) {$\boldsymbol{ [I_1,\varnothing]}$};
\node[scale=.8] (t1a) at (-0.3,-.5) {$I_1$};
\node[scale=.8] (t1b) at (-.3,-1.5) {$I^{(3)}_0$};
\draw[red] (p0a) -- (p1);
\draw[red,dashed] (p1) -- (p0b);
\end{scope}
\end{tikzpicture}}
{\caption{\label{CBA1A5}The Hasse diagram for the Coulomb branch of the $(A_1,A_5)$ AD theory.}}
\end{subfloatrow}\hspace{1cm}%
\begin{subfloatrow}
\capbtabbox[7cm]{%
  \renewcommand{\arraystretch}{1.1}
  \begin{tabular}{|c|c|} 
  \hline
  \multicolumn{2}{|c|}{$(A_1,A_5)$}\\
  \hline\hline
  $(\D_u,\D_v)$  &\quad $\left(\frac54,\frac32\right)$\quad{} \\
  $24a$ &  22\\  
  $12c$ & $\frac{23}2$ \\
$\ff_k$ &$ \U(1) $\\ 
$h$&0\\
$T({\bf2}\bh)$&0\\
\hline\hline
$\cSb$&$L_{(5,6)}(0,0,1)$\\
\hline
  \end{tabular}
}{%
  \caption{\label{CcA1A5}Central charges, Coulomb branch parameters and ECB dimension.}%
}
\end{subfloatrow}}{\caption{\label{TotA1A5}Information about the $(A_1,A_5)$ AD theory.}}
\end{figure}

\subsubsection{$(A_1,D_6)$}

This theory belongs instead to an infinite series of rank-$n$ AD theory with a more complicated Higgs branch (quickly described below): $(A_1,D_{2n+2})$. The stratification of the Coulomb branch singular locus can again be read off straightforwardly from the expression of the SW curve reported in \cite{Argyres:2005pp}:
\beq\label{disA1D6}
{\rm SW\ curve}:y^2=x^6 +x( u x + v)\quad \Rightarrow\quad D^\L_x=v^2\left(256 u^5+3125 v^4\right),
\eeq
and then we conclude that there is a (4,5) knotted singularity ($\cS_1$) as well as an unknotted singularity of type $III$ ($\cS_2$). The analysis which leads to the reproduction of the central charges as well as the full characterization of the Hasse diagram in figure \ref{CBA1D6}, is largely similar to the one above therefore we won't discuss it and instead focus on discussing the Higgs branch of this theory.

\begin{figure}[h!]
\ffigbox{
\begin{subfloatrow}
\ffigbox[7cm][]{
\begin{tikzpicture}[decoration={markings,
mark=at position .5 with {\arrow{>}}}]
\begin{scope}[scale=1.5]
\node[bbc,scale=.5] (p0a) at (0,0) {};
\node[bbc,scale=.5] (p0b) at (0,-2) {};
\node[scale=.8] (t0a) at (0,.3) {$\cC$};
\node[scale=.8] (t0b) at (0,-2.3) {0};
\node[scale=.8] (p1) at (-.5,-1) {$\boldsymbol{[I_1,\varnothing]}$};
\node[scale=.8] (p2) at (.5,-1) {$\boldsymbol{[I_2,A_1]}$};
\node[scale=.8] (t1a) at (-.5,-.4) {$I_1$};
\node[scale=.8] (t1b) at (-.5,-1.6) {$I_0^{(2)}$};
\node[scale=.8] (t2a) at (.5,-.4) {$I_2$};
\node[scale=.8] (t2b) at (.5,-1.6) {$III$};
\draw[red] (p0a) -- (p1);
\draw[red] (p0a) -- (p2);
\draw[red,dashed] (p1) -- (p0b);
\draw[red] (p2) -- (p0b);
\end{scope}
\end{tikzpicture}}
{\caption{\label{CBA1D6}The Hasse diagram for the Coulomb branch of the $(A_1,D_6)$ AD theory.}}
\end{subfloatrow}\hspace{1cm}%
\begin{subfloatrow}
\capbtabbox[7cm]{%
  \renewcommand{\arraystretch}{1.1}
  \begin{tabular}{|c|c|} 
  \hline
  \multicolumn{2}{|c|}{$(A_1,D_6)$}\\
  \hline\hline
  $(\D_u,\D_v)$  &\quad $\left(\frac43,\frac53\right)$\quad{} \\
  $24a$ &  26\\  
  $12c$ & 14 \\
$\ff_k$ & $\suf(2)_{\frac{10}3}$ \\ 
$h$&0\\
$T({\bf2}\bh)$&0\\
\hline\hline
$\cSb$&$L_{(4,5)}(0,1,1)$\\
\hline
  \end{tabular}
}{%
  \caption{\label{CcA1D6}Central charges, Coulomb branch parameters and ECB dimension.}%
}
\end{subfloatrow}}{\caption{\label{TotA1D6}Information about the $(A_1,D_6)$ AD theory.}}
\end{figure}

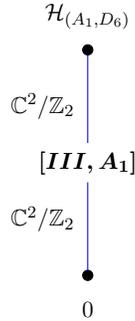
\begin{figure}
\begin{tikzpicture}[decoration={markings,
mark=at position .5 with {\arrow{>}}}]
\begin{scope}[scale=1.5]
\node[bbc,scale=.5] (p0a) at (0,0) {};
\node[bbc,scale=.5] (p0b) at (0,-2) {};
\node[scale=.8] (t0a) at (0,.3) {$\cH_{(A_1,D_6)}$};
\node[scale=.8] (t0b) at (0,-2.3) {0};
\node[scale=.8] (p1) at (-0,-1) {$\boldsymbol{ [III,A_1]}$};
\node[scale=.8] (t1a) at (-0.4,-.5) {$\C^2/\Z_2$};
\node[scale=.8] (t1b) at (-.4,-1.5) {$\C^2/\Z_2$};
\draw[blue] (p0a) -- (p1);
\draw[blue] (p1) -- (p0b);
\end{scope}
\end{tikzpicture}}
{\caption{\label{HBA1D6}The Hasse diagram for the Higgs branch of the $(A_1,D_6)$ AD theory. Following the convention of \cite{Grimminger:2020dmg} we blue for Hasse diagrams of Higgs branches.}
\end{figure}

The Higgs branch of the $(A_1,D_{2n+2})$ can be elegantly written as the intersection of symplectic varieties \cite{Beem:2017ooy}
\beq
\bar{\O_{[n+1,1]}}\cap \cS_{[n,1,1]},
\eeq
where $\O_{[n+1,1]}$ is the subregular nilpotent orbit of $\slf(n+2)$ and $\cS_{[n,1,1]}$ is the Slodowy slice of the nilpotent orbit associated to the $[n,1,1]$ partition. Adapting the notation that we have used to label the stratification on the Coulomb branch to the stratification of nilpotent orbits: $\cS_{[n,1,1]}\cong\mT(\O_{[n,1,1]},\O_{[n+2]})$, that is, $\cS_{[n,1,1]}$ can be identified as the transverse slice of the nilpotent orbit associated to the $[n,1,1]$ into the principal nilpotent orbit of $\slf(n+2)$. As $n$ increases this space can be quite complicated but for $n=2$ is relatively simple. It is two quaternionic dimensional, and it has only three strata with elementary slices $\C^2/\Z_2$ (see figure \ref{HBA1D6}). We then observe that the stratification we find is compatible with the fact the the lower leaf of the Higgs branch extends into the mixed branch of this theory and the low-energy theory living on the second stratum of the Higgs branch is the rank-1 AD theory  $(A_1,D_3)$ which is the same as $(A_1,A_3)$.

\subsection{F-theory construction}

Another interesting set of rank-2 theories are those which arise as the low-energy theory on the worldvolume of a stack of two D3-branes probing an F-theory singularity. In this category we include the by now \emph{classic} $\cN=2$ rank-$n$ instanton SCFTs \cite{Minahan:1996fg,Minahan:1996cj,Banks:1996nj,Dasgupta:1996ij,Sen:1996vd,Beem:2019snk} and the more recently introduced $\cN=2$ S-folds \cite{Apruzzi:2020pmv} which combine F-theory singularity and S-folds \cite{Aharony:2016kai}.

Given the explicit F-theory construction of each theory it is possible to fully characterize the Hasse diagram of the Coulomb branch. In fact the Coulomb branch is geometrically realized as the coordinates of the position of the D3-branes transverse to the F-theory singularity and at rank-2 we expect extra massless states to arise either when the position of the two D3-branes coincides or when one of the two D3-branes probes the F-theory singularity and/or the S-fold. We therefore expect all such theories' Coulomb branch Hasse diagrams to have two intermediate strata. The stratum arising from two coincident D3-branes is topologically a (1,2) knotted stratum ($\cS_1$) with special Kahler structure $\bar{\cS_1}\equiv K_{u}$, with the $\suf(2)$ $\cN=4$ theory supported over it, implying that $\mT_1\equiv I_0^*$. Here by $K_u$ we mean the Kodaira type given by the scaling dimension of the $u$ coordinate of the Coulomb branch which, recall, is the coordinate with lower scaling dimension. The second connected component is, instead, an unknotted one, again corresponding to $\bar{\cS_2}\equiv K_{u}$, and the theory supported on it is the rank-1 version of the F-theory construction under analysis. This in turns implies that also $\mT_2\equiv K_{u}$. Finally all the rank-2 theories constructed this way have an ECB of quaternionic dimension \cite{Apruzzi:2020pmv} $h=1+h_{\rm rank-1}$, where $h_{\rm rank-1}$ is the quaternionic dimension of the ECB of the rank-1 F-theory construction. Let's see now how what we described above plays out in examples. 

\subsubsection{Rank-2 $E_8$ theories}

To start we analyze the rank-2 $E_8$ MN theory but this construction, with appropriate obvious modifications, works as well with any other entry of the rank-2 theories in \cite{Beem:2019snk}. 

\begin{figure}[h!]
\ffigbox{
\begin{subfloatrow}
\ffigbox[7cm][]{
\begin{tikzpicture}[decoration={markings,
mark=at position .5 with {\arrow{>}}}]
\begin{scope}[scale=1.5]
\node[bbc,scale=.5] (p0a) at (0,0) {};
\node[bbc,scale=.5] (p0b) at (0,-2) {};
\node[scale=.8] (t0a) at (0,.3) {$\cC$};
\node[scale=.8] (t0b) at (0,-2.3) {0};
\node[scale=.8] (p1) at (-.5,-1) {$\boldsymbol{[I^*_0,C_1]}$\ \ };
\node[scale=.8] (p2) at (.5,-1) {\ \ $\boldsymbol{[II^*,E_8]}{\times}\H$};
\node[scale=.8] (t1a) at (-.5,-.4) {$I_0^*$};
\node[scale=.8] (t1b) at (-.5,-1.6) {$II^*$};
\node[scale=.8] (t2a) at (.5,-.4) {$II^*$};
\node[scale=.8] (t2b) at (.5,-1.6) {$II^*$};
\draw[red] (p0a) -- (p1);
\draw[red] (p0a) -- (p2);
\draw[red] (p1) -- (p0b);
\draw[red] (p2) -- (p0b);
\end{scope}
\end{tikzpicture}}
{\caption{\label{CBEnr2}The Hasse diagram for the Coulomb branch of the rank-$2$ MN $E_8$ theory.}}
\end{subfloatrow}\hspace{1cm}%
\begin{subfloatrow}
\capbtabbox[7cm]{%
  \renewcommand{\arraystretch}{1.1}
  \begin{tabular}{|c|c|} 
  \hline
  \multicolumn{2}{|c|}{rank-2 $E_8$ MN}\\
  \hline\hline
  $(\D_u,\D_v)$  &\quad (6,12)\quad{} \\
  $24a$ &  263\\  
  $12c$ & 161 \\
$\ff_k$ & $(E_8)_{24}\times \suf(2)_{13}$ \\ 
$h$&1\\
$\boldsymbol{R_{2h}}$&$({\bf 1},{\bf 2})$\\
\hline\hline
$\cSb$&$L_{(1,2)}(0,1,1)$\\
\hline
  \end{tabular}
}{%
  \caption{\label{CcEnr2}Central charges, Coulomb branch parameters and ECB dimension.}%
}
\end{subfloatrow}}{\caption{\label{TotEnr2}Information about the rank-2 MN $E_8$ theory.}}
\end{figure}

Since $\D_u=6$, in this case $K_u\equiv II^*$ and the theory supported on $\cS_2$ is $[II^*,E_8]$. This observation leads to the Hasse diagram of the Coulomb branch of the rank-2 $E_8$ theory, which is shown in figure \ref{CBEnr2}. Since the rank-1 version of this theory, $[II^*,E_8]$, has no ECB, $h=1$. In this example the matching of the flavor symmetry levels is slightly more intricate and deserves a few more words.

This is the first case we encounter in which the flavor symmetry of the rank-2 theory is semi-simple. It is also the first case in which more than one theory living on the various strata carry a non-trivial flavor symmetry. These two facts are connected, in fact each connected component of the stratification ``carries'' a simple factor of the rank-2 flavor symmetry, see \cite{Martone:2020nsy}. Also from table \ref{CcEnr2} it follows that $T_{E_8}({\bf2}\bh)=0$ and $T_{\suf(2)}({\bf2}\bh)=1$. It is a rewarding exercise to check that putting everything together, the correct levels for the $E_8\times\suf(2)$ flavor algebra are perfectly reproduced. 

\subsubsection{$\langle E_6,\Z_2\rangle_{\rm rank-2}$ rank-2 $\cN=2$ S-fold}

The discussion of this theory is slightly more tricky and we will see that for the first time we will be able to use our tools to make predictions about the structure of $\cN=2$ theories. Our results agree with those presented in \cite{Giacomelli:2020jel}.

\begin{figure}[h!]
\ffigbox{
\begin{subfloatrow}
\ffigbox[7cm][]{
\begin{tikzpicture}[decoration={markings,
mark=at position .5 with {\arrow{>}}}]
\begin{scope}[scale=1.5]
\node[bbc,scale=.5] (p0a) at (0,0) {};
\node[bbc,scale=.5] (p0b) at (0,-2) {};
\node[scale=.8] (t0a) at (0,.3) {$\cC$};
\node[scale=.8] (t0b) at (0,-2.3) {0};
\node[scale=.7] (p1) at (-.5,-1) {$\boldsymbol{[I^*_0,C_1]}{\times}\H^5$\quad \ };
\node[scale=.7] (p2) at (.5,-1) {\quad \ $\boldsymbol{[II^*,C_5]}{\times}\H$};
\node[scale=.8] (t1a) at (-.5,-.4) {$I_0^*$};
\node[scale=.8] (t1b) at (-.5,-1.6) {$II^*$};
\node[scale=.8] (t2a) at (.5,-.4) {$II^*$};
\node[scale=.8] (t2b) at (.5,-1.6) {$II^*$};
\draw[red] (p0a) -- (p1);
\draw[red] (p0a) -- (p2);
\draw[red] (p1) -- (p0b);
\draw[red] (p2) -- (p0b);
\end{scope}
\end{tikzpicture}}
{\caption{\label{CBC5r2}The Hasse diagram for the Coulomb branch of the $\langle E_6,\Z_2\rangle_{\rm rank-2}$ $\cN=2$ S-fold.}}
\end{subfloatrow}\hspace{1cm}%
\begin{subfloatrow}
\capbtabbox[7cm]{%
  \renewcommand{\arraystretch}{1.1}
  \begin{tabular}{|c|c|} 
  \hline
  \multicolumn{2}{|c|}{$\langle E_6,\Z_2\rangle_{\rm rank-2}$ $\cN=2$ S-fold}\\
  \hline\hline
  $(\D_u,\D_v)$  &\quad (6,12)\quad{} \\
  $24a$ &  232\\  
  $12c$ & 130 \\
$\ff_k$ & $\spf(4)_{13}\times \suf(2)_{26}$ \\ 
$h$&6\\
$\boldsymbol{R_{2h}}$&$({\bf 8},{\bf 1})\oplus({\bf 1},{\bf 2})\oplus({\bf 1},{\bf 2})$\\
\hline\hline
$\cSb$&$L_{(1,2)}(0,1,1)$\\
\hline
  \end{tabular}
}{%
  \caption{\label{CcC5r2}Central charges, Coulomb branch parameters and ECB dimension.}%
}
\end{subfloatrow}}{\caption{\label{TotC5r2}Information about the $\langle E_6,\Z_2\rangle_{\rm rank-2}$ $\cN=2$ S-fold.}}
\end{figure}
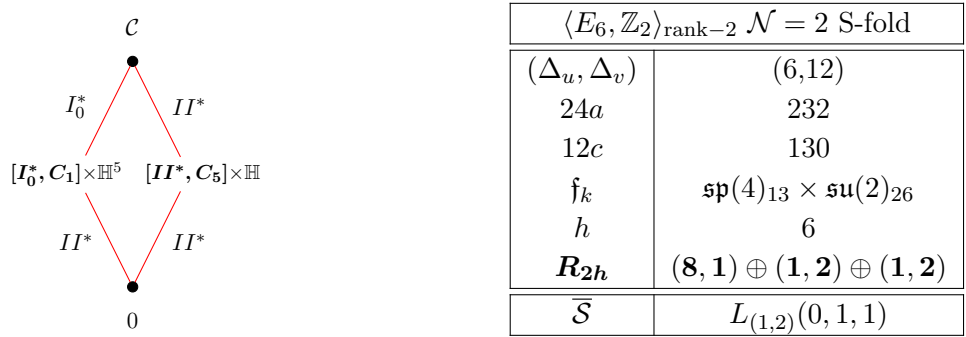

First it is useful to remind the reader that the rank-1 version of this theory is the well known $[II^*,C_5]\equiv \langle E_6,\Z_2\rangle_{\rm rank-1}$. This readily implies the Coulomb branch scaling dimensions of the rank-2 version as well as the complete characterization of the the Hasse diagram of the Coulomb branch of the theory as reported in figure \ref{TotC5r2}. As described in \cite{Apruzzi:2020pmv}, the enhancement of the flavor symmetry $\spf(4)\times \suf(2)\to\spf(5)$ is an accident of the rank-1 version and it is expected that the higher rank version will only carry the flavor symmetry visible from the brane construction. The rank-2 theory can be realized in F-theory as the low-energy limit of the world volume theory on two D3 branes probing an $E_6$ 7-brane singularity combined with a $\Z_2$ S-fold \cite{Aharony:2016kai} which is located at the origin of the space transverse to the D3, This particular theory, $\langle E_6,\Z_2\rangle_{\rm rank-2}$, is also realized in  class-S (see entry 113 at page 13 of \cite{Chacaltana:2015bna}) from which we read off that the flavor symmetry for this theory is $\ff_k=\Sp(4)_{13}\times\suf(2)_{26}$. Let's now quickly review few more details of how the $\cN=2$ S-fold construction works which will be useful below.

The S-fold has a non-trivial action on a $\C^3\times T^2$, where the $T^2$ and one of the $\C$ factors come from, respectively, the torus fiber and the base of the K3 transverse to the D7. The other two $\C$ factors are instead identified with the four dimensions transverse to the D3 but along the worldvolume of the D7. Because of the combination of the D7 and the $\Z_k$ S-fold background the topology of this four-fold is, schematically, \cite{Apruzzi:2020pmv}:
\beq\label{orbSf}
\C^3\times T^2\quad\to\quad\left[\C^2\times (\C\times T^2)\right]/\left[\Z_k\times \Z_{k\D_7}\right]
\eeq
where the $\Z_k$ only acts on the $\C^2$ while the $\Z_{k\D_7}$ on the elliptically fibered K3, or explicitly:
\beq\label{action}
\left[\Z_k\times \Z_{k\D_7}\right]:=\left(
\begin{array}{cccc}
\zeta^{\D_7}&&&\\
&\zeta^{\D_7}&&\\
&&\zeta&\\
&&&\ \zeta\ 
\end{array}
\right)\qquad \zeta^{k\D_7}=1.
\eeq 
Both in \eqref{orbSf} and \eqref{action} the $\D_7$ corresponds to the orbifold action on the K3 induced by the D7 background alone, i.e., $\D_7=3$ for an $E_6$ 7-brane singularity or $2$ for a $D_4$ 7-brane singularity. In \cite{Apruzzi:2020pmv} it was conjectured that the F-theory compactified on an $\cN=2$ S-fold gives a collection of $k(\D_7-1)$ hypermultiplets. Our analysis is consistent with this conjecture.

Let's now proceed with a careful analysis of the moduli space of the specific $\langle E_6,\Z_2\rangle_{\rm rank-2}$ example. As shown already in figure \ref{CBC5r2}, the stratification of the Coulomb branch perfectly reproduces the D-brane picture. The reproduction of the central charges of the theory fixes the quaternionic dimension of the ECB of this theory to 6 which is compatible with the counting in \cite{Apruzzi:2020pmv}. In fact, on a generic point of the Coulomb branch, the D3 branes provide two hypermuliplets to which we need to add the contribution of the F-theory compactified on the S-fold which indeed provides the remaining four free hypers. 

The flavor symmetry also works out nicely. We notice that the the stratification of the Coulomb branch carries a larger flavor symmetry: $\ff_{\rm CB}=\spf(5)\times\suf(2)$. Clearly $\spf(4)\times \suf(2)'\subset \spf(5)$ with index of embedding 1. Then the level of the $\spf(4)$ can be perfectly reproduced by applying \eqref{kcc} to the $\spf(4)\subset\spf(5)$. The geometric $\suf(2)_{UV}$, which acts on the $\C^2/\Z_2$ transverse to each D3 along the D7, is instead the diagonal combination of the $\suf(2)'$ commutant of $\spf(4)$ in $\spf(5)$, and the $\suf(2)$ from the $\cN=4$ theory.  Again summing the contributions of both, we reproduce the $k_{\suf(2)_{UV}}=26$ on the nose.

Let's conclude with a quick discussion of the Higgs branch of this theory. This analysis draws extensively from what was done in \cite{Apruzzi:2020pmv, CCLMW2020}. Thus far we considered the motion of the two D3s along the direction transverse to the D7.  To grasp the structure of the Higgs branch we need to instead analyze the motion of the D3 in the direction along the D7. This space is topologically $\C^2/\Z_2$ with the S-fold ``sitting'' at the origin of it. Moving the D3 brane away from the origin we then obtain the following Higgsings:
\beq\label{HBE6r2}
\langle E_6,\Z_2\rangle \xrightarrow{\C^2/\Z_2} 
[IV^*,E_6]\times[II^*,C_5] \xrightarrow{\C^2/\Z_2}[IV^*,E_6]\times[IV^*,E_6]\times\H^4 \xrightarrow{\ef_6\times\ef_6} 
\H^{24} ,
\eeq
where $\ef_6$ represents the minimal nilpotent orbit of the $E_6$ Lie algebra. To check the consistency of this pattern of Higgsings, we can use the technique developed in \cite{Beem:2019tfp, Beem:2019snk, CCLMW2020}, construct the free-field realization of the corresponding chiral algebras and perform a quick central charge check. The only subtlety of this calculation is to identify the correct level of the $\suf(2)$ that needs to be Higgsed at, say, the first step in \eqref{HBE6r2} which is not the ``full'' 26 of the $\suf(2)$ of the rank-2 theory. One way to do that is by noticing that after step-1 the low-energy theory on the Higgs branch has an $\spf(5)_7\supset\suf(2)_7$ and therefore conjecture that the $\suf(2)_{26}=[\suf(2)_7\times \suf(2)_{19}]_{\rm diag}$ is the diagonal combination of two $\suf(2)$'s and the one that drives the first Higgsing has level $19$. With this observation, we can perfectly reproduce the central charges also from a Higgs branch analysis.

\subsection{Appearance of discrete gauging}\label{subsec:Discrete}

A somewhat unexpected phenomenon that our detailed analysis of the Coulomb branch of rank-2 theories reveals, is the appearance of the discretely gauged version of rank-1 geometries \cite{Argyres:2016yzz} on the Coulomb branch of rank-2 theories which are not discretely gauged. As it is by now well-known, the discretely gauged theory violates the Shapere-Tachikawa relation between the values of the $(a,c)$ central charges and the scaling dimensions of the Coulomb branch coordinates \cite{Shapere:2008zf}. We therefore need to take this into account when using our formulae \eqref{acc}-\eqref{kcc} to reproduce the central charges of the rank-2 theory from the Coulomb branch stratification. 

\subsubsection{$\D=(4,6)$ and $\ff_k=\spf(6)_8$}\label{sec:sp6}

\begin{figure}
\begin{tikzpicture}[decoration={markings,
mark=at position .5 with {\arrow{>}}}]
\begin{scope}[scale=1.5]
\node[bbc,scale=.5] (p0a) at (0,0) {};
\node[bbc,scale=.5] (p0b) at (0,-3) {};
\node[scale=.8] (t0a) at (0,.3) {$\cH_{\spf(6)_8}$};
\node[scale=.8] (t0b) at (0,-3.3) {0};
\node[scale=.8] (t0c) at (.2,-.5) {$\ef_6$};
\node[scale=.8] (p1) at (-0,-1) {$\boldsymbol{ [IV^*,E_6]}$};
\node[scale=.8] (t1b) at (.2,-1.5) {$\cf_5$};
\node[scale=.8] (p2) at (-0,-2) {$\boldsymbol{[II^*,C_5]}$};
\node[scale=.8] (t2b) at (.2,-2.5) {$\cf_6$};
\draw[blue] (p0a) -- (p1);
\draw[blue] (p1) -- (p2);
\draw[blue] (p2) -- (p0b);
\end{scope}
\end{tikzpicture}}
{\caption{\label{HBSp6}The Hasse diagram for the Higgs branch of the $\spf(7)_9$ theory.}
\end{figure}
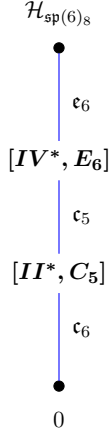

This theory has a class-S realization, see e.g. \cite{Chacaltana:2013oka}, from which the data reported in table \ref{CcSp6} is taken. This theory is conjectured to be the second non-trivial entry of an infinite series of four dimensional $\cN=2$ SCFTs with flavor symmetry $\spf(n+3)$ (for $n=1$ the theory is simply a bunch of free hypers) \cite{Zafrir:2016wkk}. The Higgs branches of this infinite series are worked out in details using the techniques of magnetic quivers in \cite{Bourget:2020asf} and in the context of 3d $\cN=4$ in \cite{Grimminger:2020dmg} where the corresponding theory is named $X_6$. Our analysis perfectly matches with their predictions suggesting that this 4d theory flows to  the $X_6$ upon circle compactification. We report the Hasse diagram of the Higgs branch of the $\spf(6)_8$ theory in figure \ref{HBSp6} (which is a decorated version of the one in \cite{Grimminger:2020dmg}). The theory living on each stratum of the Higgs branch can be easily guessed by following the rules in \cite{Bourget:2019aer, Grimminger:2020dmg} and matching the transverse slices to each stratum with the Higgs branch supported on it. 

\begin{figure}[h!]
\ffigbox{
\begin{subfloatrow}
\ffigbox[7cm][]{
\begin{tikzpicture}[decoration={markings,
mark=at position .5 with {\arrow{>}}}]
\begin{scope}[scale=1.5]
\node[bbc,scale=.5] (p0a) at (0,0) {};
\node[bbc,scale=.5] (p0b) at (0,-2) {};
\node[scale=.8] (t0a) at (0,.3) {$\cC$};
\node[scale=.8] (t0b) at (0,-2.3) {0};
\node[scale=.8] (p1) at (-.8,-1) {$\boldsymbol{[I_1,\varnothing]}$};
\node[scale=.8] (p2) at (.8,-1) {$\boldsymbol{[I^*_6,C_6]_{\Z_2}}$};
\node[scale=.8] (p3) at (0,-1) {$\boldsymbol{[I_1,\varnothing]}$};
\node[scale=.8] (t1a) at (-.6,-.4) {$I_1$};
\node[scale=.8] (t1b) at (-.6,-1.6) {$I^*_0$};
\node[scale=.8] (t2a) at (-.2,-.6) {$I_1$};
\node[scale=.8] (t2b) at (-.2,-1.5) {$I^*_0$};
\node[scale=.8] (t3a) at (.6,-.4) {$I_6^*$};
\node[scale=.8] (t3b) at (.6,-1.6) {$II^*$};
\draw[red] (p0a) -- (p1);
\draw[red] (p0a) -- (p2);
\draw[red] (p0a) -- (p3);
\draw[red,dashed] (p1) -- (p0b);
\draw[red] (p2) -- (p0b);
\draw[red,dashed] (p3) -- (p0b);
\node[bbc,scale=.5] (p0a) at (2.5,0) {};
\node[bbc,scale=.5] (p0b) at (2.5,-2) {};
\node[scale=.8] (t0a) at (2.5,.3) {$\cC$};
\node[scale=.8] (t0b) at (2.5,-2.3) {0};
\node[scale=.8] (p1) at (2,-1) {$\boldsymbol{[II,\varnothing]}$};
\node[scale=.8] (p2) at (3,-1) {$\boldsymbol{[I_6^*,C_6]_{\Z_2}}$};
\node[scale=.8] (t1a) at (2,-.4) {$II$};
\node[scale=.8] (t1b) at (2,-1.6) {$I_0^*$};
\node[scale=.8] (t2a) at (3,-.4) {$I_6^*$};
\node[scale=.8] (t2b) at (3,-1.6) {$II^*$};
\draw[red] (p0a) -- (p1);
\draw[red] (p0a) -- (p2);
\draw[red] (p1) -- (p0b);
\draw[red] (p2) -- (p0b);
\end{scope}
\end{tikzpicture}}
{\caption{\label{CBSp6}The Hasse diagram for the Coulomb branch of the $\spf(6)_8$ $\cN=2$ SCFT.}}
\end{subfloatrow}\hspace{1cm}%
\begin{subfloatrow}
\capbtabbox[7cm]{%
  \renewcommand{\arraystretch}{1.1}
  \begin{tabular}{|c|c|} 
  \hline
  \multicolumn{2}{|c|}{$\D=(4,6)$ {\rm w/} $\spf(6)_8$}\\
  \hline\hline
  $(\D_u,\D_v)$  &\quad (4,6)\quad{} \\
  $24a$ & 130\\  
  $12c$ & 76 \\
$\ff_k$ & $\spf(6)_8$ \\ 
$h$&0\\
$T({\bf2}\bh)$&0\\
\hline\hline
$\cSb$&$L_{(2,3)}(1,0,2)$\\
\hline
  \end{tabular}
}{%
  \caption{\label{CcSp6}Central charges, Coulomb branch parameters and ECB dimension.}%
}
\end{subfloatrow}}{\caption{\label{TotSp6} Information about the $\spf(6)_8$ $\cN=2$ theory.}}
\end{figure}

In \cite{Grimminger:2020dmg} the Coulomb branch of the theory was also analyzed using a generalization of the notion of inversion of the Hasse diagram. This analysis led to the conjecture that indeed a discretely gauged theory appears on the 3d Coulomb branch of the $X_6$ theory by partial Higgsing. In the notation of \cite{Grimminger:2020dmg} the discretely gauged theory appearing on the Coulomb branch of the $X_6$ is the $O(2) - C_6$ which corresponds in our notation to the $\Z_2$ gauged version of a $\cN=2$ $U(1)$ gauge theory with flavor symmetry $\spf(6)$. This arises by gauging a $U(1)$ with twelve hypermultiplets ($[I_{12},A_{11}]$) with charge one and is precisely the $[I_6^*,C_6]_{\Z_2}$ theory appearing in figure \ref{CBSp6} \cite{Argyres:2016yzz}.

Let us now look more carefully at the stratification of the rank-2 four dimensional theory. We have already identified that there is a stratum supporting the $[I_6^*,C_6]_{\Z_2}$ but have not yet clarified the stratum and in particular whether it is a knotted or unknotted one. Because the theory living on the first stratum of the Higgs branch is a rank-1 theory, a criterion that is useful in this context, discussed in the next section, is that the Coulomb branch scaling dimension of the theory living on the Higgs branch stratum, has to divide the Coulomb branch scaling dimension of the uniformizing parameter of the stratum on the Coulomb branch. Since the Coulomb branch of the $[II^*,C_5]$ is 6, it picks as the only possibility the fact that the discretely gauged theory has to be supported on the $u=0$ stratum. Since the theory is not a product, by conjecture \ref{knot} the Coulomb branch has to have at least another, knotted, stratum. Using \eqref{acc} and \eqref{ccc} and matching with the central charges of the theory, the only two compatible interpretations are two $(2,3)$ knotted strata supporting an $[I_1,\varnothing]$ or a single one supporting a $[II,\varnothing]$. Since we have no argument favoring one or the other, we report both stratifications in figure \ref{CBSp6}. The matching of the level of the $\spf(6)$ works straightforwardly. As we mentioned above, there are extra subtleties involved when we attempt to reproduce the central charge of discretely gauged theory using Coulomb branch geometry considerations, to get around this problem in computing the central charges, we use the data of the $[I_{12},A_{11}]$ rather than the $[I_6^*,C_6]_{\Z_2}$.

Finally, notice that the structure of the moduli space we found is perfectly consistent with what was proposed in \cite{Grimminger:2020dmg}. Indeed after dimensional reduction, the special Kahler structure of the Coulomb branch is ``lifted'' to a hyperkahler space which schematically coincides with the total space of the SW geometry of the four dimensional Coulomb branch. Since the total space of the $I_1/II$ singularity is in fact not singular, it ``disappears'' under compactification. We therefore claim that the 3d reduction of this $\spf(6)_8$ theory is precisely the $X_6$ theory discussed in \cite{Grimminger:2020dmg}. The analysis of \cite{Grimminger:2020dmg} extends to $X_N$ for any $N$ and brings forward a precise conjecture for its moduli space and its structure of the partial Higgsing. We claim that indeed this series of 3d $\cN=4$ $X_N$ theories exist. In the next section we characterize the properties for the four dimensional lift of the $X_7$ theory, which is also a rank-2 theory. Our analysis can be easily extended to generic $N$.

\subsubsection{$\D=(3,4)$ and $\ff_k=\suf(2)_8\times \spf(4)_6$}\label{sec:su2sp4}

\begin{figure}[t!]
\ffigbox{
\begin{subfloatrow}
\ffigbox[7cm][]{
\begin{tikzpicture}[decoration={markings,
mark=at position .5 with {\arrow{>}}}]
\begin{scope}[scale=1.5]
\node[bbc,scale=.5] (p0a) at (0,0) {};
\node[bbc,scale=.5] (p0b) at (0,-2) {};
\node[scale=.8] (t0a) at (0,.3) {$\cC$};
\node[scale=.8] (t0b) at (0,-2.3) {0};
\node[scale=.8] (p1) at (-.8,-1) {$\boldsymbol{[I_1,\varnothing]}$};
\node[scale=.8] (p2) at (.8,-1) {$\boldsymbol{[I^*_4,C_4]_{\Z_2}}$};
\node[scale=.8] (p3) at (0,-1) {$\boldsymbol{[I_2,A_1]}$};
\node[scale=.8] (t1a) at (-.6,-.4) {$I_1$};
\node[scale=.8] (t1b) at (-.6,-1.6) {$I_0$};
\node[scale=.8] (t2a) at (.2,-.6) {$I_2$};
\node[scale=.8] (t2b) at (.2,-1.5) {$IV^*$};
\node[scale=.8] (t3a) at (.6,-.4) {$I^*_4$};
\node[scale=.8] (t3b) at (.6,-1.6) {$III^*$};
\draw[red] (p0a) -- (p1);
\draw[red] (p0a) -- (p2);
\draw[red] (p0a) -- (p3);
\draw[red,dashed] (p1) -- (p0b);
\draw[red] (p2) -- (p0b);
\draw[red] (p3) -- (p0b);
\end{scope}
\end{tikzpicture}}
{\caption{\label{CBSU2Sp4}The Hasse diagram for the Coulomb branch of the $\suf(2)_8\times \spf(4)_6$ $\cN=2$ SCFT.}}
\end{subfloatrow}\hspace{1cm}%
\begin{subfloatrow}
\capbtabbox[7cm]{%
  \renewcommand{\arraystretch}{1.1}
  \begin{tabular}{|c|c|} 
  \hline
  \multicolumn{2}{|c|}{$\suf(2)_8\times \spf(4)_6$}\\
  \hline\hline
  $(\D_u,\D_v)$  &\quad (3,4)\quad{} \\
  $24a$ &  84\\  
  $12c$ & 48 \\
$\ff_k$ & $\suf(2)_8\times \spf(4)_6$ \\ 
$h$&0\\
$T({\bf2}\bh)$&0\\
\hline\hline
$\cSb$&$L_{(3,4)}(1,1,1)$\\
\hline
  \end{tabular}
}{%
  \caption{\label{CcSU2Sp4}Central charges, Coulomb branch parameters and ECB dimension.}%
}
\end{subfloatrow}}{\caption{\label{TotSU2Sp4} Information about the $\suf(2)_8\times\spf(4)_6$ $\cN=2$ SCFT.}}
\end{figure}

\begin{figure}[b!]
\begin{tikzpicture}[decoration={markings,
mark=at position .5 with {\arrow{>}}}]
\begin{scope}[scale=1.5]
\node[bbc,scale=.5] (p0b) at (0,-3) {};
\node[bbc,scale=.5] (p0a) at (0,-1) {};
\node[scale=.8] (t0a) at (0,-.7) {$\cH_{O(2)-C_4}$};
\node[scale=.8] (t0b) at (0,-3.3) {0};
\node[scale=.8] (t1b) at (.2,-1.5) {$\cf_3$};
\node[bbc,scale=.5] (p2) at (0,-2) {};
\node[scale=.8] (t2b) at (.2,-2.5) {$\cf_4$};
\draw[blue] (p0a) -- (p2);
\draw[blue] (p2) -- (p0b);
\end{scope}
\end{tikzpicture}}
{\caption{\label{I4C4}The Hasse diagram for the Higgs branch of the $[I_4^*,C_4]_{\mathbb{Z}_2}$ or $O(2) - C_4$ theory.}
\end{figure}
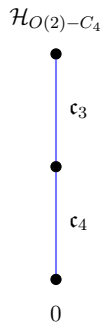

This theory has been realized in class-S, see, e.g., \cite{Chacaltana:2012ch}, and again it presents the interesting feature that the low energy theory supported on one of the strata is a discretely gauged rank-1 theory which carries the $\spf(4)$ non-abelian factor of the flavor symmetry of the rank-2 theory at the origin. The Higgs branch of the $[I_4^*,C_4]_{\mathbb{Z}_2}$ theory has been worked out in \cite{Bourget:2019aer} to be a symplectic singularity with three strata; see fig. \ref{I4C4} for its Hasse diagram. From this analysis, it follows that the theory supported on the first symplectic leaf of the Higgs branch sticking out from the singular strata on the Coulomb branch where the $[I_4^*,C_4]_{\mathbb{Z}_2}$ is supported has to have a $\cf_3$ as first leaf of its Higgs branch. From the total dimension of the Higgs branch of the rank-2 theory, which is 12 quaternionic dimensional, we can also deduce that the second leaf of the Higgs branch of such theory has to be five quaternionic dimensional. This strongly suggests that the theory which ``lives'' on the first symplectic leaf of $\cH_{O(2)-C_4}$ is indeed $[III^*,C_3A_1]$ which has itself Coulomb branch scaling dimension 4 which has to divide the scaling dimension of the uniformizing parameter of the stratum where the discretely gauged theory is supported. We therefore conclude that the $[I_4^*,C_4]_{\mathbb{Z}_2}$ is supported on the $u=0$ unknotted strata. 

The second theory, realizing the $\suf(2)$ flavor factor, is an $[I_2,A_1]$ which is instead supported on the $v=0$ unknotted stratum. This follows from the fact that the theory which is supported on the $\af_1$ Higgs stratum sticking out of the component supporting the $[I_2,A_1]$ is a $[IV^*,E_6]$. To summarize the discussion we report the Hasse diagram of the Higgs branch of the rank-2 theory in figure \ref{I4C4}. Finally the reader can check that our considerations perfectly, and somewhat remarkably, reproduce both central charges and flavor levels.

The theory that we just analyzed can be obtained as a circle reduction of a 5d SCFT and it is predicted to be the second non-trivial entry of an infinite series of four dimensional $\cN=2$ SCFT with flavor symmetry $\spf(n+1)\times \suf(2)$ \cite{Zafrir:2016wkk} (for $n=1$ the theory is simply a bunch of free hypermultiplets). The detailed analysis of the HB Hasse diagram of this series of theories is performed in \cite{Bourget:2020asf}. For $n=3$ we find perfect agreement with the the Hasse diagram in figure \ref{HBSu2Sp4}. 

\begin{figure}
\begin{tikzpicture}[decoration={markings,
mark=at position .5 with {\arrow{>}}}]
\begin{scope}[scale=1.5]
\node[bbc,scale=.5] (p0a) at (0,0) {};
\node[bbc,scale=.5] (p0b) at (0,-3) {};
\node[scale=.8] (t0a) at (0,.3) {$\cH_{\suf(2)_8\times \spf(4)_6}$};
\node[scale=.8] (t0b) at (0,-3.3) {0};
\node[scale=.8] (t0c) at (.4,-.35) {$\df_4$};
\node[scale=.8] (p1) at (.6,-1) {$\boldsymbol{ [I_0^*,D_4]}$};
\node[scale=.8] (t1b) at (.8,-1.5) {$\cf_3$};
\node[scale=.8] (p2) at (.6,-2) {$\boldsymbol{[III^*,C_3A_1]}$};
\node[scale=.8] (t2b) at (.4,-2.65) {$\cf_4$};
\node[scale=.8] (2t0c) at (-.4,-2.4) {$\af_1$};
\node[scale=.8] (2p1) at (-.6,-1.5) {$\boldsymbol{ [IV^*,E_6]}$};
\node[scale=.8] (2t1b) at (-.4,-.6) {$\ef_6$};
\draw[blue] (p0a) -- (2p1);
\draw[blue] (2p1) -- (p0b);
\draw[blue] (p0a) -- (p1);
\draw[blue] (p1) -- (p2);
\draw[blue] (p2) -- (p0b);
\end{scope}
\end{tikzpicture}}
{\caption{\label{HBSu2Sp4}The Hasse diagram for the Higgs branch of the $\suf(2)_8\times \spf(4)_6$ theory.}
\end{figure}

\subsection{New predictions}

Now that we have understood in detail how to reconstruct the UV information from the Coulomb branch, as well as carrying out some basics Higgs branch/chiral algebra consistency checks, we can use our techniques to predict new properties of rank-2 theories. 

\subsubsection{$\langle D_4,\Z_2\rangle_{\rm rank-2}$ with $\ff_k=\spf(2)_9\times \suf(2)_{16}\times\suf(2)'_{18}$}

Let us first start with the systematic analysis of the rank-2 version of the $\cN=2$ $\Z_2$ S-folds. Here we go beyond the mere check of the consistency of the predictions in \cite{Apruzzi:2020pmv}, and we compute the levels of the flavor symmetry groups of these theories. This theory, along with a systematic study of higher rank $\cN=2$ $S$-folds, including the one discussed in the next subsection, is analyzed in \cite{Giacomelli:2020jel}. Our results perfectly agree with theirs.

\begin{figure}[h!]
\ffigbox{
\begin{subfloatrow}
\ffigbox[7cm][]{
\begin{tikzpicture}[decoration={markings,
mark=at position .5 with {\arrow{>}}}]
\begin{scope}[scale=1.5]
\node[bbc,scale=.5] (p0a) at (0,0) {};
\node[bbc,scale=.5] (p0b) at (0,-2) {};
\node[scale=.8] (t0a) at (0,.3) {$\cC$};
\node[scale=.8] (t0b) at (0,-2.3) {0};
\node[scale=.7] (p1) at (-.5,-1) {$\boldsymbol{[I^*_0,C_1]}{\times}\H^3$\qquad\ };
\node[scale=.7] (p2) at (.5,-1) {\qquad\ $\boldsymbol{[III^*,C_3A_1]}{\times}\H$};
\node[scale=.8] (t1a) at (-.5,-.4) {$I_0^*$};
\node[scale=.8] (t1b) at (-.5,-1.6) {$III^*$};
\node[scale=.8] (t2a) at (.5,-.4) {$III^*$};
\node[scale=.8] (t2b) at (.5,-1.6) {$III^*$};
\draw[red] (p0a) -- (p1);
\draw[red] (p0a) -- (p2);
\draw[red] (p1) -- (p0b);
\draw[red] (p2) -- (p0b);
\end{scope}
\end{tikzpicture}}
{\caption{\label{CBD4r2}The Hasse diagram for the Coulomb branch of the $\langle D_4,\Z_2\rangle_{\rm rank-2}$ $\cN=2$ S-fold.}}
\end{subfloatrow}\hspace{1cm}%
\begin{subfloatrow}
\capbtabbox[7cm]{%
  \renewcommand{\arraystretch}{1.1}
  \begin{tabular}{|c|c|} 
  \hline
  \multicolumn{2}{|c|}{$\langle D_4,\Z_2\rangle_{\rm rank-2}$ $\cN=2$ S-fold}\\
  \hline\hline
  $(\D_u,\D_v)$  &\quad (4,8)\quad{} \\
  $24a$ &  146\\  
  $12c$ & 80 \\
$\ff_k$ & $\spf(2)_{9}\times \suf(2)_{10}\times\suf(2)_{18}$ \\ 
$h$&4\\
$\boldsymbol{R_{2h}}$&$({\bf 4},{\bf 1},{\bf 1})\oplus({\bf 1},{\bf 1},{\bf 2})\oplus({\bf 1},{\bf 1},{\bf 2})$\\
\hline\hline
$\cSb$&$L_{(1,2)}(0,1,1)$\\
\hline
  \end{tabular}
}{%
  \caption{\label{CcD4r2}Central charges, Coulomb branch parameters and ECB dimension.}%
}
\end{subfloatrow}}{\caption{\label{TotD4r2}Information about the $\langle D_4,\Z_2\rangle_{\rm rank-2}$ $\cN=2$ S-fold.}}
\end{figure}

The analysis is very much analogous to the previous one. The rank-1 version of this theory is the $[III^*,C_3A_1]$, see corresponding entry in table \ref{tab:theories}, and which readily implies the stratification in figure \ref{CBD4r2}. From the F-theory analysis, we expect the flavor symmetry of this theory to be $\spf(2)\times\suf(2)\times\suf(2)'$, where we label with the prime the geometric $\suf(2)$ which arises from the action on the two complex dimensional space transverse to the D3 but along the D7. As before, on the Coulomb branch we find an enlarged $\spf(3)\times\suf(2)_1\times\suf(2)_2$ where the second $\suf(2)$ is the one arising from the rank-1 $\cN=4$ theory. We identify the $\spf(2)\subset\spf(3)$ as the $\spf(2)$ of the rank-2 theory. This $\spf(2)$ on the Coulomb branch has index of embedding one in the $\spf(3)$ and therefore, from table \ref{tab:theories}, it has level $k^{\rm IR}_{\spf(2)}=5$. Plugging things into \eqref{kcc} we predict that the level of the $\spf(2)$ is:
\beq
k^{\rm UV}_{\spf(2)}=\frac{8(5-1)}4+1=9.
\eeq  
The identification of the two $\suf(2)$s is a bit trickier. In fact, the commutant of the $\spf(2)$ into $\spf(3)$ introduces a third $\suf(2)_3$ on the Coulomb branch. A careful analysis of the decomposition of the ECB of the rank-1 $[III^*,C_3A_1]$ theory from table \ref{tab:theories} shows that $\suf(2)_3$ does indeed act on $\C^2/\Z_2$ transverse to the D3 and therefore should contribute, along with $\suf(2)_2$ that also act on the ECB, to the geometric $\suf(2)'$. On the other side, the $\suf(2)_1$ is the only factor that doesn't act on the ECB and can be readily identified with the UV non-geometric $\suf(2)$:
\beq
\suf(2)\equiv\suf(2)_3\quad\Rightarrow\quad k^{\rm UV}_{\suf(2)}=\frac{8(8)}4=16.
\eeq
The identification of the geometric $\suf(2)'$ proceeds in a completely analogous way to the $E_6$:
\beq
\suf(2)=\left[\suf(2)_3\times\suf(2)_2\right]_{\rm diag}\quad \Rightarrow\quad k_{\suf(2)}=\frac{8(5-1)}4+1+\frac{8(3-1)}2+1=18.
\eeq

We can also repeat here the Higgsing analysis above which leads to:
\beq\label{HBd4r2}
\langle D_4,\Z_2\rangle\xrightarrow{\C^2/\Z_2} [I_0^*,D_4]\times[III^*,C_3A_1]\xrightarrow{\C^2/\Z_2}[I_0^*,D_4]\times[I_0^*,D_4]\times\H^2\xrightarrow{\df_4\times\df_4} \H^{10}
\eeq
which again can be checked by a chiral algebra computation after realizing that the $\suf(2)$ level corresponding to the first Higgsing is $k_{\suf(2)}=18-5=13$.

\subsubsection{$\langle A_2,\Z_2\rangle_{\rm rank-2}$ with $\ff_k=\suf(2)_7\times \suf(2)'_{14} \times U(1)$}

\begin{figure}[h!]
\ffigbox{
\begin{subfloatrow}
\ffigbox[7cm][]{
\begin{tikzpicture}[decoration={markings,
mark=at position .5 with {\arrow{>}}}]
\begin{scope}[scale=1.5]
\node[bbc,scale=.5] (p0a) at (0,0) {};
\node[bbc,scale=.5] (p0b) at (0,-2) {};
\node[scale=.8] (t0a) at (0,.3) {$\cC$};
\node[scale=.8] (t0b) at (0,-2.3) {0};
\node[scale=.7] (p1) at (-.5,-1) {$\boldsymbol{[I^*_0,C_1]}{\times}\H^2$\quad\ };
\node[scale=.7] (p2) at (.5,-1) {\quad\ $\boldsymbol{[IV^*,C_2]}{\times}\H$};
\node[scale=.8] (t1a) at (-.5,-.4) {$I_0^*$};
\node[scale=.8] (t1b) at (-.5,-1.6) {$IV^*$};
\node[scale=.8] (t2a) at (.5,-.4) {$IV^*$};
\node[scale=.8] (t2b) at (.5,-1.6) {$IV^*$};
\draw[red] (p0a) -- (p1);
\draw[red] (p0a) -- (p2);
\draw[red] (p1) -- (p0b);
\draw[red] (p2) -- (p0b);
\end{scope}
\end{tikzpicture}}
{\caption{\label{CBA2r2}The Hasse diagram for the Coulomb branch of the $\langle A_2,\Z_2\rangle_{\rm rank-2}$ $\cN=2$ S-fold.}}
\end{subfloatrow}\hspace{1cm}%
\begin{subfloatrow}
\capbtabbox[7cm]{%
  \renewcommand{\arraystretch}{1.1}
  \begin{tabular}{|c|c|} 
  \hline
  \multicolumn{2}{|c|}{$\langle A_2,\Z_2\rangle_{\rm rank-2}$ $\cN=2$ S-fold}\\
  \hline\hline
  $(\D_u,\D_v)$  &\quad (3,6)\quad{} \\
  $24a$ &  103\\  
  $12c$ & 55 \\
$\ff_k$ & $\suf(2)_{7}\times \suf(2)_{14}\times U_1$ \\ 
$h$&3\\
$\boldsymbol{R_{2h}}$&$({\bf 2},{\bf 1})_0\oplus({\bf 1},{\bf 2})_0\oplus({\bf 1},{\bf 2})_0$\\
\hline\hline
$\cSb$&$L_{(1,2)}(0,1,1)$\\
\hline
  \end{tabular}
}{%
  \caption{\label{CcA2r2}Central charges, Coulomb branch parameters and ECB dimension.}%
}
\end{subfloatrow}}{\caption{\label{TotA2r2}Information about the $\langle A_2,\Z_2\rangle_{\rm rank-2}$ $\cN=2$ S-fold.}}
\end{figure}
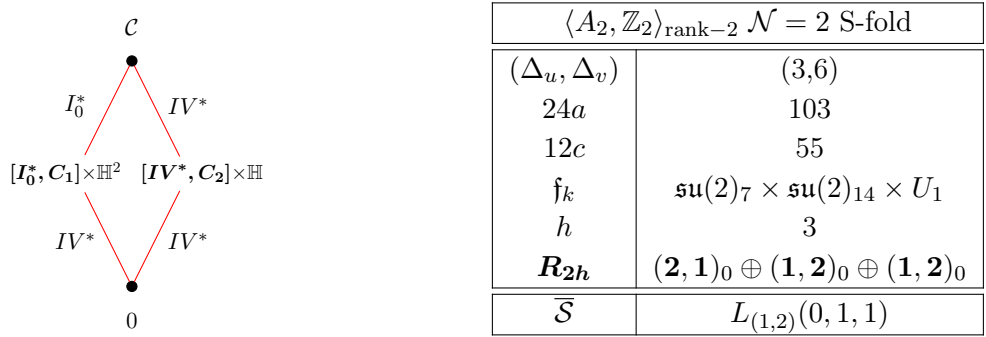

The analysis here is very much analogous to the previous one. The stratification in figure \ref{CBA2r2} can be readily obtained by noticing that $\langle A_2,\Z_2\rangle_{\rm rank-1}=[IV^*,C_2]$, see table \ref{tab:theories}. Plugging in the corresponding information, the $c$ and $a$ predicted in \cite{Apruzzi:2020pmv} can be perfectly reproduced by applying \eqref{acc} and \eqref{ccc} to the stratification in figure \ref{CBA2r2}. We will now go over quickly to the careful identification of the flavor symmetries.

The flavor symmetry on the Coulomb branch is $\spf(2)\times\suf(2)$ and contains three $\suf(2)$ factor $\suf(2)_1\times\suf(2)_2\times\suf(2)$, where the first two are contained in the $\spf(2)$. We notice that the $\suf(2)$ from the $[I_0^*,C_1]$ has to contribute to the geometric $\suf(2)'$ while the first two factors have the same level and therefore is completely up to us which contributes to which UV flavor symmetry:
\begin{align}
\suf(2)\equiv\suf(2)_1\quad&\Rightarrow\quad k_{\suf(2)}=\frac{6(4-1)}3+1=7\\
\suf(2)'\equiv\left[\suf(2)_2\times\suf(2)\right]\quad&\Rightarrow\quad k_{\suf(2)'}=\frac{6(4-1)}3+1\frac{6(3-1)}2+1=14.
\end{align}

Doing an analysis as above leads to the following Higgsing:
\beq\label{HBA2r2}
\langle A_2,\Z_2\rangle\xrightarrow{\C^2/\Z_2} [IV,A_2]\times[IV^*,C_2]\xrightarrow{\C^2/\Z_2}[IV,A_2]\times[IV,A_2]\times \H\xrightarrow{\af_2\times\af_2} \H^{3}
\eeq
which are consistent with chiral algebra calculation where the first Higgsing is activated by an $\suf(2)$ at level $k_{\suf(2)}=14-4=10$.

\subsubsection{$\D=(6,8)$ and $\ff_k=\spf(7)_9$}\label{sec:sp7}

Finally, let's present the discussion of the moduli space of the four dimensional lift of the 3d $\cN=4$ $X_7$ theory pointed out at the end of \cite{Grimminger:2020dmg} in the context of 3d $\cN=4$ theories. This theory was also predicted as compactification of a six dimensional $D_6$ (2,0)  with twisted punctures in \cite{Ohmori:2018ona} and is the third non-trivial entry of the infinite series mentioned at the start of subsection \ref{sec:sp6}. Our results perfectly match their predictions. 

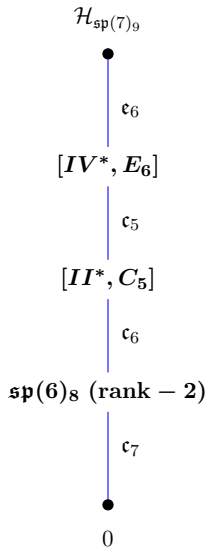
\begin{figure}
\begin{tikzpicture}[decoration={markings,
mark=at position .5 with {\arrow{>}}}]
\begin{scope}[scale=1.5]
\node[bbc,scale=.5] (p0a) at (0,0) {};
\node[bbc,scale=.5] (p0b) at (0,-4) {};
\node[scale=.8] (t0a) at (0,.3) {$\cH_{\spf(7)_9}$};
\node[scale=.8] (t0b) at (0,-4.3) {0};
\node[scale=.8] (t0c) at (.2,-.5) {$\ef_6$};
\node[scale=.8] (p1) at (-0,-1) {$\boldsymbol{ [IV^*,E_6]}$};
\node[scale=.8] (t1b) at (.2,-1.5) {$\cf_5$};
\node[scale=.8] (p2) at (-0,-2) {$\boldsymbol{ [II^*,C_5]}$};
\node[scale=.8] (t2b) at (.2,-2.5) {$\cf_6$};
\node[scale=.8] (p3) at (-0,-3) {$\boldsymbol{ \spf(6)_8\ ({\rm rank-2})}$};
\node[scale=.8] (t3b) at (.2,-3.5) {$\cf_7$};
\draw[blue] (p0a) -- (p1);
\draw[blue] (p1) -- (p2);
\draw[blue] (p2) -- (p3);
\draw[blue] (p3) -- (p0b);
\end{scope}
\end{tikzpicture}}
{\caption{\label{HBSp7}The Hasse diagram for the Higgs branch of the $\spf(7)_9$ theory.}
\end{figure}

Let's start from reproducing the Hasse diagram of the Higgs branch where we decorate it with the extra information about the theory living on each stratum, which again can easily be achieved by matching the transverse slice to each stratum with the Higgs branch of the theory living on it; see figure \ref{HBSp7}. As it is described in \cite{Grimminger:2020dmg}, this theory should have a seven quaternionic dimensional ECB which extends over the entire second stratum of the Higgs branch. This readily implies two things. First that the theory living on the second stratum of the Higgs branch has to be a rank-2 theory (compatible with our assignment of theory in figure \ref{HBSp7}) and secondly that the there has to be a locus on the Coulomb branch supporting a theory with a seven quaternionic dimensional ECB. A natural candidate is an $\cN=2$ $\suf(2)$ gauge theory with seven hypermultiplets in the adjoint, which we will indicate as $[I_{24}^*,C_7]$ as it carries an $\spf(7)$ flavor symmetry. The information about this theory can be found in the $[I_n^*,C_{\frac{n+4}4}]$ entry in table \ref{tab:theories} with $n=24$.


A back-of-the-envelop chiral algebra computation predicts that a $\cf_7$ fibration of $\spf(6)_8$ theory with $12c=76$ should give rise to an $\spf(7)_9$ flavor symmetry and with central charge 12$c$=107. These are indeed the results which are reported in table \ref{CcSp7}. We want to now show that the same values can be correctly reproduced from the Coulomb branch. We have already identified that the theory supported on the $v=0$ stratum is a $[I_{24}^*,C_7]$. Since the theory is not a product theory, we expect at least a second stratum ($\cS_2$) corresponding to a knotted singularity and the structure of the Higgs branch suggests that the theory supported on such stratum should have itself a trivial Higgs branch. A natural and minimal guess is that the on $\cS_2$ there is an $[I_1,\varnothing]$ therefore leading to the Coulomb branch Hasse diagram in figure \ref{CBSp7}. It is remarkable that by plugging $h=7$ and the correct data for the theories supported on the two connected components, we can perfectly reproduce the central charges predicted by the chiral algebra analysis.

\begin{figure}[h!]
\ffigbox{
\begin{subfloatrow}
\ffigbox[7cm][]{
\begin{tikzpicture}[decoration={markings,
mark=at position .5 with {\arrow{>}}}]
\begin{scope}[scale=1.5]
\node[bbc,scale=.5] (p0a) at (0,0) {};
\node[bbc,scale=.5] (p0b) at (0,-2) {};
\node[scale=.8] (t0a) at (0,.3) {$\cC$};
\node[scale=.8] (t0b) at (0,-2.3) {0};
\node[scale=.8] (p1) at (-.5,-1) {$\boldsymbol{[I_1,\varnothing]}{\times}\H^7$\ \ };
\node[scale=.8] (p2) at (.5,-1) {\ \ $\boldsymbol{[I^*_{24},C_7]}$};
\node[scale=.8] (t1a) at (-.5,-.4) {$I_1$};
\node[scale=.8] (t1b) at (-.5,-1.6) {$I_0^*$};
\node[scale=.8] (t2a) at (.5,-.4) {$I_{24}^*$};
\node[scale=.8] (t2b) at (.5,-1.6) {$II^*$};
\draw[red] (p0a) -- (p1);
\draw[red] (p0a) -- (p2);
\draw[red,dashed] (p1) -- (p0b);
\draw[red] (p2) -- (p0b);
\end{scope}
\end{tikzpicture}}
{\caption{\label{CBSp7}The Hasse diagram for the Coulomb branch of the $\spf(7)_9$ $\cN=2$ SCFT.}}
\end{subfloatrow}\hspace{1cm}%
\begin{subfloatrow}
\capbtabbox[7cm]{%
  \renewcommand{\arraystretch}{1.1}
  \begin{tabular}{|c|c|} 
  \hline
  \multicolumn{2}{|c|}{$\spf(7)_9$}\\
  \hline\hline
  $(\D_u,\D_v)$  &\quad (2,3)\quad{} \\
  $24a$ &  185\\  
  $12c$ & 107 \\
$\ff_k$ & $\spf(7)_9$ \\ 
$h$&7\\
$\boldsymbol{R_{2h}}$&${\bf14}$\\
\hline\hline
$\cSb$&$L_{(3,4)}(0,1,1)$\\
\hline
  \end{tabular}
}{%
  \caption{\label{CcSp7}Central charges, flavor level, Coulomb branch parameters and ECB dimension.}%
}
\end{subfloatrow}}{\caption{\label{TotSp7}Information about the $\spf(7)_9$ theory.}}
\end{figure}

After this extremely encouraging consistency check, let's compute the level of the flavor symmetry. In this case there is a perfect matching between the flavor symmetry of the UV and the one visible on the Coulomb branch. We therefore need to only plug into \eqref{kcc} the correct information:
\beq
k_{\spf(7)}=\frac{8(3-1)}2+1=9
\eeq
finding a perfectly consistent story. 

Finally notice that, as in the $X_6$ case discussed above, the structure of the moduli space that we found is perfectly consistent with that discussed in \cite{Grimminger:2020dmg} and we therefore claim that the 3d reduction of this $\spf(7)_9$ theory is precisely the $X_7$ theory. The discussion here generalizes to all $N$ following the observations in \cite{Grimminger:2020dmg} with the appropriate modifications which are relevant to lift the discussion to four dimensions.

\section{Bringing it all together: stratification of the full moduli space}\label{sec:FullModuli}

Thus far we have discussed the stratification of the Coulomb branch and Higgs branch separately, but these two branches only represent a subvariety of the full moduli space and a systematic analysis of the full moduli space is the natural extension to pursue. Even in simple cases it is apparent that in bringing together the stratification of the Higgs and Coulomb branches we find non-trivial consistency conditions which, if understood in detail, can further constrain the picture presented so far. We will present this analysis here but rather than being systematic in analyzing the type of constraints which can arise, we will give a brief, non-technical and somewhat heuristic outline of some of the patterns we have noticed. After that, we will describe how these come about in concrete examples. An analysis of this sort was performed in \cite{Grimminger:2020dmg} for three dimensional $\cN=4$ theory and a Hasse diagram of the full moduli for such theories was presented. As is the case in 3d, the Hasse diagram of the full moduli space of four dimensional $\cN=2$ SCFTs is non-planar and therefore the Hasse diagram will necessarily be represented in three dimensions. 

Call a generic point of a mixed branch stratum $\cM_i$, $\bm\in\cM_i$. We will assume that it can be locally trivialized as a product of Coulomb and Higgs components $\bm=(\bu,\bh)$. Then the main leverage that an analysis of the mixed branch provides is the fact that $\bm$ can be ``reached'' both by turning on the $\bu$ Coulomb branch moduli of a low-energy theory supported along the Higgs branch at $(0,\bh)$ or by turning on a Higgs branch moduli $\bh$ of a low-energy theory supported along the Coulomb branch at $(\bu,0)$. Therefore the stratum, at least locally, is isomorphic to the cartesian product of the Coulomb branch of the theory supported on the Higgs branch stratum of $(0,\bh)$, which we will label as $\cC_\bh$, and the Higgs branch of the theory supported on the Coulomb branch stratum of $(\bu,0)$ which we will instead label $\cH_\bu$: 
\beq\label{M=CxH}
\cM_i\sim \cC_\bh\times \cH_\bu
\eeq 
We call $\cC_\bh$ the \emph{Coulomb branch component} and the $\cH_\bu$ the \emph{Higgs branch component} of the mixed branch stratum.\footnote{This point is more subtle than one might think. Indeed, the Coulomb branch component of a mixed branch stratum actually gives information about the low-energy description on the Higgs branch of the superconformal theory, while the Higgs component constraints the low-energy description on the Coulomb branch.}  As we will see in examples, the global structure is obtained via quotienting the product \eqref{M=CxH} by the action of a finite, often cyclic, group. This structure ties the theories supported on Higgs branch and Coulomb branch strata in an interesting way though we don't understand these constraints in general.

An immediate result of this line of argument can be obtained if an ECB is present. Recall that an ECB implies that the low energy theory has a decoupled sector of free hypermultiplets everywhere on the Coulomb branch.  By turning on a vev for these hypermultiplets, we can move to a mixed branch stratum which is fibered over every point of the Coulomb branch. Therefore we conclude that the Coulomb branch component of the ECB is of the same complex dimension as the Coulomb branch of the superconformal theory. Or in other words, the existence of an ECB implies that it is possible, along a Higgs branch direction, to Higgs the superconformal field theory  to a theory of the same rank.  


From the analysis of many theories of rank 2, we observe that a $h$ quaternionic dimensional ECB has the following structure
\beq
\text{ECB}\cong \Big(\tilde \cC\times \H^h\Big)/\G
\eeq
where $\tilde \cC$ is a covering of $\cC$, the Coulomb branch of the superconformal field theory, $\H^h\cong\H^{\otimes^h}$ is a tensor product of $h$ factors of the quaternions and $\G$ is a finite group which does not act irreducibly on $\tilde \cC$.\footnote{Shortly before completing this work, we were told by S. Giacomelli, C. Meneghelli and W. Peelaers that in their analysis \cite{Giacomelli:2020jel} they find examples of a richer structure. We thank them for their comments.} This is a generalization of the structure of the ECB of rank-1 theories \cite{CCLMW2020}. We will illustrate this structure in two examples below.

\subsection{Without ECB}
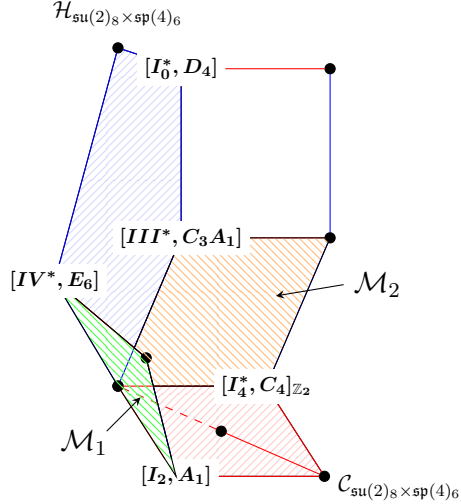
\begin{figure}
\begin{tikzpicture}[decoration={markings,
mark=at position .5 with {\arrow{>}}}]
\begin{scope}[scale=1.5]
\draw[pattern=north east lines, pattern color=blue!15] (.14447,-2.66075) -- (.56,-1.685) -- (.56,-.185) -- (0,0) -- (-.56,-2.065) -- cycle;
\draw[pattern=north east lines, pattern color=red!20] (.3131,-3) -- (1.315,-3) -- (1.83,-3.8) -- (.515,-3.8) -- cycle;
\node[bbc,scale=.5] (p0a) at (0,0) {};
\node[bbc,scale=.5] (p0b) at (0,-3) {};
\node[scale=.8] (t0a) at (0,.3) {$\cH_{\suf(2)_8\times \spf(4)_6}$};
\node[scale=.7,fill=white] (p1) at (.56,-.185) {$\boldsymbol{ [I_0^*,D_4]}$}; 
\draw[blue] (p0a) -- (-.56,-2.065);
\draw[blue] (-.56,-2.065) -- (p0b);
\draw[blue] (p0a) -- (p1);
\draw[blue] (p1) -- (.56,-1.685);
\draw[blue] (.56,-1.685) -- (p0b);
\node[bbc,scale=.5] (Cp0a) at (1.83,-3.8) {};
\node[bbc,scale=.5] (Cp0b) at (0,-3) {};
\node[scale=.8] (Ct0a) at (2.5,-3.9) {$\cC_{\suf(2)_8\times \spf(4)_6}$};
\node[scale=.7,fill=white] (Cp2) at (.515,-3.8) {$\boldsymbol{ [I_2,A_1]}$};
\node[bbc,scale=.5] (Cp3) at (.915,-3.4) {};
\node[bbc,scale=.5] (2Cp1) at (1.88,-1.685) {};
\node[bbc,scale=.5] (3Cp1) at (1.88,-.185) {};
\node[bbc,scale=.5] (2Cp2) at (.25,-2.75) {};
\draw[red] (.56,-1.685) -- (2Cp1);
\draw[red] (p1) -- (3Cp1);
\draw[red] (Cp0a) -- (1.315,-3);
\draw[red] (Cp0a) -- (Cp2);
\draw[red] (1.315,-3) -- (Cp0b);
\draw[red] (Cp2) -- (Cp0b);
\draw[blue] (2Cp1) -- (1.315,-3);
\draw[blue] (3Cp1) -- (2Cp1);
\draw[red] (2Cp2) -- (-.56,-2.065);
\draw[blue] (2Cp2) -- (Cp2);
\draw[red] (Cp0a) -- (Cp3);
\draw[red,dashed] (Cp3) -- (Cp0b);
\draw[pattern=north west lines, pattern color=orange!50] (0.14447,-2.66075) -- (.25,-2.75) -- (.3131,-3) -- (1.315,-3) -- (1.88,-1.685) -- (.56,-1.685) -- cycle;
\draw[pattern=north west lines, pattern color=green!70] (0,-3) -- (.515,-3.8) -- (.25,-2.75) -- (-.56,-2.065) -- cycle;
\node[scale=.7,fill=white] (p2) at (.56,-1.685) {$\boldsymbol{[III^*,C_3A_1]}$}; 
\node[scale=.7,fill=white] (Cp1) at (1.315,-3) {$\boldsymbol{ [I_4^*,C_4]_{\Z_2}}$};
\node[scale=.7,fill=white] (2p1) at (-.56,-2.065) {$\boldsymbol{ [IV^*,E_6]}$}; 
\node (M1) at (-0.3,-3.5) {$\cM_1$};
\node (M2) at (2.3,-2.1) {$\cM_2$};
\draw[->,>=stealth] (2,-2.1) -- (1.4,-2.2);
\draw[->,>=stealth,bend right] (-.1,-3.4) -- (0.2,-3.1);
\end{scope}
\end{tikzpicture}}
{\caption{\label{MbSu2Sp4}The Hasse diagram for the full moduli space of the $\suf(2)_8\times \spf(4)_6$ theory. We have highlighted in green $(\cM_1$) and orange ($\cM_2$) the two mixed branch strata discussed in the text.}
\end{figure}

To start, let us analyze the full moduli space of the $\suf(2)_8\times \spf(4)_6$ theory which has been discussed above in section \ref{sec:su2sp4}. The conjectural Hasse diagram for the full moduli space is reported in figure \ref{MbSu2Sp4}. 

As described above, the theories supported on the two unknotted strata of the Coulomb branch are a $\U(1)$ gauge theory with two massless hypermultiplets ($[I_2,A_1]$) and a $O(2)$ gauge theory with flavor symmetry $C_4$ ($[I_4^*,C_4]_{\Z_2}$). To understand the nice constraining picture provided by the analysis of the entire moduli space, it is important to also track the special Kahler structure of the strata. From the Hasse diagram of the Coulomb branch of the theory, see figure \ref{CBSU2Sp4}, we see that the $[I_2,A_1]$ is supported over a $IV^*$ stratum ($v=0$) while the discretely gauged theory is supported over a $III^*$ ($u=0$). This implies that the scaling dimension of the uniformizing parameter are, respectively, $3$ and $4$. Both strata support theories with a non-trivial Higgs branch and therefore there is a mixed branch sticking out of both. We will adopt the following notation:
\begin{subequations}
\begin{align}\label{MB1T1}
\cM_1&\cong\big(\widetilde{IV^*}\times \af_1\big)/\G_1\\\label{MB2T1}
\cM_2&\cong\big(\widetilde{III^*}\times \cf_4\big)/\G_2
\end{align}
\end{subequations}
where $\af_1\cong \H/\Z_2$ and $\cf_4\cong \H^4/\Z_2$ are respectively the minimal nilpotent orbit of $\suf(2)$ and $\spf(4)$. $\cM_1$ and $\cM_2$ are highlighted in green and orange respectively in figure \ref{MbSu2Sp4}. We are going to use the overall constraints of the full moduli space to determine $\G_1$ and $\G_2$.

First we need to determine whether the theory has a non-trivial ECB, that is, decoupled hypermultiplets on the generic point of its Coulomb branch. We can reach such a point by moving along one of the two strata discussed above and then turn on the Coulomb branch parameter of the theory supported on either. This makes it clear that an ECB for the superconformal theory should also appear as ECB of the rank-1 theory supported on the Coulomb branch strata. But since both the $[I_2,A_1]$ and $[I_4^*,C_4]_{\Z_2}$ theory don't have an ECB, we infer that this theory does not have an ECB and therefore nowhere on its Higgs branch the $\suf(2)_8\times \spf(4)_6$  is higgsed to a rank-2 theory. This is compatible with the Higgs branch depicted in figure \ref{HBSu2Sp4} and therefore the two mixed branch strata \eqref{MB1T1} and \eqref{MB2T1} are not part of a larger ECB.

Let's now see how the entire moduli space comes together. From the figure \ref{HBSu2Sp4} the two theories which are supported on the second two, disconnected, Higgs branch strata are $[IV^*,E_6]$ and $[III^*,C_3A_1]$. But from staring at figure \ref{MbSu2Sp4}, these two Higgs branch strata arise, respectively, from the intersection of $\cM_1$ and $\cM_2$ with the whole Higgs branch. This is useful information, since the covering of $\widetilde{IV}^*$ and $\widetilde{III}^*$ are given respectively by the Coulomb branch of $[IV^*,E_6]$ and $[III^*,C_3A_1]$. We therefore immediately conclude that $\widetilde{IV}^*\cong IV^*$ as well as $\widetilde{III}^*\cong III^*$, therefore both $\G_1$ and $\G_2$ are trivial. Notice that, aside from basic Higgs branch reasoning which led to the Hasse diagram in figure \ref{HBSu2Sp4}, we can leverage the fact that the scaling dimension of the covering space is related to the original Coulomb branch by quotienting by an integer to conclude that it would have been impossible for the $[III^*,C_3A_1]$ to be supported on $\cM_1$ and similarly for $[IV^*,E_6]$ on $\cM_2$.  By inspection of figure \ref{MbSu2Sp4}, the mixed branch stratum $\cM_2$ has a neighboring stratum which has instead a non-trivial global twisting which will be analyzed in more detail in \cite{CCLMW2020}.

\subsection{With ECB}

Here we will discuss two theories with a non-trivial ECB. We will start with a somewhat standard lagrangian example and then discuss the full moduli space of the new theory $\spf(7)_9$.

\subsubsection{A lagrangian example}

\begin{figure}
\begin{tikzpicture}[decoration={markings,
mark=at position .5 with {\arrow{>}}}]
\begin{scope}[scale=1.5]
\node[bbc,scale=.5] (p0a) at (0,-1) {};
\node[bbc,scale=.5] (p0b) at (0,-4) {};
\node[scale=.8] (t0a) at (0,-.7) {$\cH_{\gf_2\ w/\ \text{4}\, 7}$};
\node[scale=.7] (t0b) at (-.6,-4.1) {$\boldsymbol{g_2\ w/\ \text{4}\, 7}$};
\node[scale=.8] (t1b) at (-.2,-1.5) {$\df_4$};
\node[scale=.7] (ph2) at (-0,-2) {$\boldsymbol{ [I_0^*,D_4]}$};
\node[scale=.8] (t2b) at (-.2,-2.5) {$\af_5$};
\node[scale=.7] (ph3) at (-0,-3) {$\boldsymbol{ \suf(3)\ w/\ \text{6}\, 3}$};
\node[scale=.8] (t3b) at (-.2,-3.5) {$\cf_4$};
\draw[blue] (p0a) -- (ph2);
\draw[blue] (ph2) -- (ph3);
\draw[blue] (ph3) -- (p0b);
\node[bbc,scale=.5] (p0a) at (2,-4.8) {};
\node[scale=.8] (1t2) at (1.5,-5) {$I_{12}^*$};
\node[bbc,scale=.5] (p0b) at (0,-4) {};
\node[scale=.8] (t0a) at (2.55,-4.9) {$\cC_{\boldsymbol{ \gf_2\, w/\, \text{4}\, 7}}$};
\node[bbc,scale=.5] (p1) at (1.2,-4) {};
\node[scale=.7] (p2) at (.8,-4.8) {$\boldsymbol{ [I_0^*,C_4]}$};
\node[scale=.8] (1t2) at (.2,-4.5) {$I_0^*$};
\node[bbc,scale=.5] (pp3) at (1,-4.4) {};
\draw[red] (p0a) -- (p1);
\draw[red] (p0a) -- (p2);
\draw[red] (p0a) -- (pp3);
\draw[red,dashed] (pp3) -- (p0b);
\draw[red,dashed] (p1) -- (p0b);
\draw[red] (p2) -- (p0b);
\node[bbc,scale=.5] (2p0b) at (2,-3.8) {};
\node[bbc,scale=.5] (2p1) at (1.2,-3) {};
\node[scale=.8] (t0a) at (2.6,-3.85) {$\cC_{\boldsymbol{ \suf(3)\, w/\, \text{6}\, 3}}$};
\node[bbc,scale=.5] (2p2) at (.8,-3.8) {};
\node[bbc,scale=.5] (2pp3) at (1,-3.4) {};
\draw[red,dashed] (ph3) -- (2p1);
\draw[red] (ph3) -- (2p2);
\draw[red,dashed] (ph3) -- (2pp3);
\draw[red] (2pp3) -- (2p0b);
\draw[red] (2p1) -- (2p0b);
\draw[red] (2p2) -- (2p0b);
\draw[blue] (pp3) -- (2pp3);
\draw[blue] (p1) -- (2p1);
\draw[blue] (p2) -- (2p2);
\draw[blue] (2p0b) -- (p0a);
\node[scale=.8] (3t0a) at (1.25,-2.8) {$\cC_{\boldsymbol{[I_0^*,D_4]}}$};
\node[bbc,scale=.5] (3p2) at (.8,-2.8) {};
\draw[red] (ph2) -- (3p2);
\draw[blue] (2p2) -- (3p2);
\end{scope}
\end{tikzpicture}}
{\caption{\label{Totg2}The Hasse diagram for the full moduli space of the $\gf_2$ theory with 4 hypermultiplets in the ${\bf 7}$.}
\end{figure}
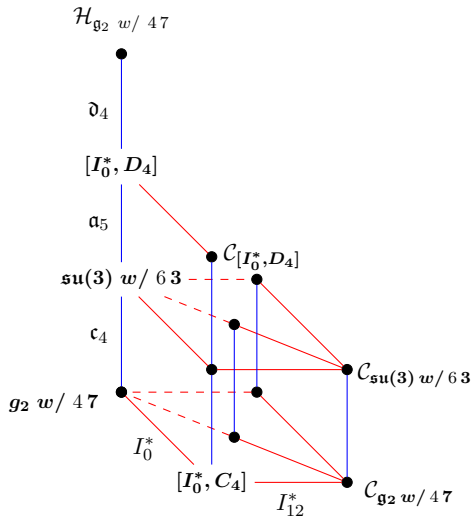

Let's start from the study of the lagrangian case, a SCFT with $\gf_2$ gauge algebra and four hypermultiplets in the ${\bf 7}$. The stratification of the Coulomb branch has been discussed in subsection \ref{sec:g2} above while the stratification of the Higgs branch is discussed, for example, in \cite{Bourget:2019aer}. This theory has a four quaternionic dimensional ECB. Turning on a vev for the free hypers fibered over the Coulomb branch, the theory is Higgsed to an $\suf(3)$ gauge theory with six fundamentals which was also discussed in the previous section. The three dimensional Hasse diagram is presented in figure \ref{Totg2}.

From the Coulomb branch Hasse diagram of the $\gf_2$ (see figure \ref{CBG247}) the unknotted, $v=0$, stratum supports an $\cN=2$ $\suf(2)$ gauge theory with four massless hypermultiplets in the ${\bf 3}$ of $\suf(2)$, also indicated as $[I_0^*,C_4]$. This theory has a Higgs branch, $\cH_{[I_{12}^*,C_4]}\cong \cf_4$ and therefore we expect a mixed branch stratum to stick out of the Coulomb branch strata. Following the same reasoning as the previous example, we can readily conclude that this theory has also an ECB. In fact each adjoint hypermultiplet in the $[I_{12}^*,C_4]$ contains a weight zero element which contributes a quaternionic dimension to the ECB, perfectly consistent with our expectations. This also imply that the low-energy theory on each of the two other strata, indicated with a dashed line, is a  $\U(1)$ gauge theory with a single massless hyper of charge 1 as well four decoupled hypermultiplets, indicated as $[I_1,\varnothing]{\times}\H^4$.

In this example we can study the subtle structure of the ECB which will lead to an understanding of why by turning on the vev of the free hypers on $\cC_{\gf_2}$ the scaling dimension of one of the two scaling dimension halves ($\D_{\gf_2}=(2,6)$ and $\D_{\suf(3)}=(2,3)$). Call ($\tilde u,\tilde v$) the coordinates of the covering $\tilde \cC_{\gf_2}$, therefore $p\D(\tilde u)=2$ and $q\D(\tilde v)=6$ with $p,q\in\Z$.  Consider the following action:
\beq\label{Z2act}
\Z_2:\quad \tilde \cC_{\gf_2}\times\H^4\ \ni\ (\tilde u, \tilde v, h)\quad \to \quad(\tilde u,-\tilde v,-h) 
\eeq
Let's now consider the space  $(\tilde{\cC}_{\gf_2}\times\H^4)/\Z_2$. The $\Z_2$ action just defined is clearly not free, and its fixed loci are:
\begin{subequations}
\begin{align}\label{fixloc1}
\tilde v=0&\quad  \sim\quad  \tilde u \times \H^4/\Z_2\equiv \tilde u\times \cf_4\\\label{fixloc2}
h=0&\quad \sim\quad  \tilde u \times( \tilde v/\Z_2)\sim \tilde u \times \tilde v^2
\end{align}
\end{subequations}
Observe that in \eqref{fixloc1} $\cf_4$ precisely reproduces the Higgs branch of the theory supported over the unknotted stratum $v=0$ and that in \eqref{fixloc2}, ($\tilde u, \tilde v^2$) can represent the Coulomb branch coordinates of $\cC_{\gf_2}$ if we identify $\tilde \cC_{\gf_2}\cong \cC_{\suf(3)}$, that is $p=1$ and $q=2$ above. This remarkably reproduces the structure of the full moduli space in figure \ref{Totg2} therefore we make the claim that:
\beq
\text{ECB}_{\gf_2}\quad\equiv\quad (\cC_{\suf(3)}\times\H^4)/\Z_2
\eeq
where the $\Z_2$ action is defined in \eqref{Z2act}. We leave it up to the reader to work out the remaining details of the full moduli space Hasse diagram and check that all the self-consistency conditions that we have outlined work out nicely in figure \ref{Totg2}.

\subsubsection{A non-lagrangian example}

\begin{figure}[t!]
\begin{tikzpicture}[decoration={markings,
mark=at position .5 with {\arrow{>}}}]
\begin{scope}[scale=1.5]
\node[bbc,scale=.5] (p0a) at (0,0) {};
\node[bbc,scale=.5] (p0b) at (0,-4) {};
\node[scale=.8] (t0a) at (0,.3) {$\cH_{\spf(7)_9}$};
\node[scale=.8] (t0b) at (-.2,-4.2) {$\boldsymbol{\spf(7)_9}$};
\node[scale=.8] (t0c) at (-.2,-.5) {$\ef_6$};
\node[scale=.8] (ph1) at (-0,-1) {$\boldsymbol{ [IV^*,E_6]}$};
\node[scale=.8] (t1b) at (-.2,-1.5) {$\cf_5$};
\node[scale=.8] (ph2) at (-0,-2) {$\boldsymbol{ [II^*,C_5]}$};
\node[scale=.8] (t2b) at (-.2,-2.5) {$\cf_6$};
\node[scale=.8] (ph3) at (-0,-3) {$\boldsymbol{ \spf(6)_8}$};
\node[scale=.8] (t3b) at (-.2,-3.5) {$\cf_7$};
\draw[blue] (p0a) -- (ph1);
\draw[blue] (ph1) -- (ph2);
\draw[blue] (ph2) -- (ph3);
\draw[blue] (ph3) -- (p0b);
\node[bbc,scale=.5] (p0a) at (2,-4.8) {};
\node[scale=.8] (1t2) at (1.5,-5) {$I_{24}^*$};
\node[bbc,scale=.5] (p0b) at (0,-4) {};
\node[scale=.8] (t0a) at (2.45,-4.9) {$\cC_{\boldsymbol{ \spf(7)_9}}$};
\node[bbc,scale=.5] (p1) at (1.2,-4) {};
\node[bbc,scale=.5] (p2) at (.8,-4.8) {};
\node[scale=.8] (1t2) at (.2,-4.5) {$I_0^*$};
\node[bbc,scale=.5] (pp3) at (1,-4.4) {};
\draw[red] (p0a) -- (p1);
\draw[red] (p0a) -- (p2);
\draw[red] (p0a) -- (pp3);
\draw[red,dashed] (pp3) -- (p0b);
\draw[red,dashed] (p1) -- (p0b);
\draw[red] (p2) -- (p0b);
\node[bbc,scale=.5] (2p0b) at (2,-3.8) {};
\node[bbc,scale=.5] (2p1) at (1.2,-3) {};
\node[scale=.8] (t0a) at (2.45,-3.85) {$\cC_{\boldsymbol{\spf(6)_8}}$};
\node[bbc,scale=.5] (2p2) at (.8,-3.8) {};
\node[bbc,scale=.5] (2pp3) at (1,-3.4) {};
\draw[red,dashed] (ph3) -- (2p1);
\draw[red] (ph3) -- (2p2);
\draw[red,dashed] (ph3) -- (2pp3);
\draw[red] (2pp3) -- (2p0b);
\draw[red] (2p1) -- (2p0b);
\draw[red] (2p2) -- (2p0b);
\draw[blue] (pp3) -- (2pp3);
\draw[blue] (p1) -- (2p1);
\draw[blue] (p2) -- (2p2);
\draw[blue] (2p0b) -- (p0a);
\node[scale=.8] (3t0a) at (1.3,-2.8) {$\cC_{\boldsymbol{[II^*,C_5]}}$};
\node[bbc,scale=.5] (3p2) at (.8,-2.8) {};
\draw[red] (ph2) -- (3p2);
\draw[blue] (2p2) -- (3p2);
\node[scale=.8] (4t0a) at (1.3,-1.8) {$\cC_{\boldsymbol{[IV^*,E_6]}}$};
\node[bbc,scale=.5] (4p2) at (.8,-1.8) {};
\draw[red] (ph1) -- (4p2);
\draw[blue] (3p2) -- (4p2);
\end{scope}
\end{tikzpicture}}
{\caption{\label{TotSp7}The Hasse diagram for the full moduli space of the $\spf(7)_9$ theory.}
\end{figure}
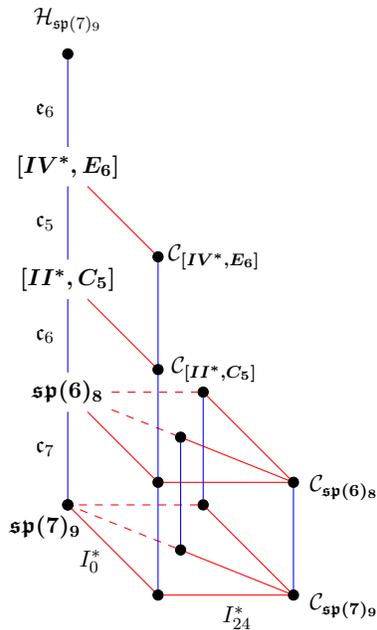

Let's now analyze a non-Lagrangian case. We pick the theory that we have discussed in subsection \ref{sec:sp7} the $\spf(7)_9$. This discussion will be brief, many of the salient features have already been discussed in the previous two subsections.

The full moduli space of the 3d limit of this theory has been already discussed in \cite{Grimminger:2020dmg}, where its full moduli space as a three dimensional $\cN=4$ theory is also presented. Our and their Hasse diagrams are perfectly consistent. Here we can again describe the structure of the ECB. We find a structure very similar to the previous case. Rather than reproducing the same steps we simply quote the results:
\beq
\text{ECB}_{\spf(7)_9}\quad:\quad(\cC_{\spf(6)_8}\times \H^7)/\Z_2
\eeq
where the action of the $\Z_2$ is basically analogous to \eqref{Z2act}. The moduli space Hasse diagram in this case is even more involved than in the previous case and contains an intricate set of strata which all fit beautifully together.   The reader will appropriately appreciate this beauty after having worked out the details of figure \ref{TotSp7}.

These examples show that the full moduli space structure can quickly become complex but, as we have discussed, it remains very constrained. 

\section{Conclusion and open questions}\label{sec:Conclusions}

In this paper we have formalized the special Kahler stratification induced on the Coulomb branches of generic $\cN=2$ SCFTs from the structure of their singular locus. This stratification is very reminiscent of the stratification of symplectic singularities which apply to Higgs branches of $\cN=2$ SCFTs in four dimensions. The special Kahler stratification is both more constrained and richer.  It is more constrained because the complex dimension of the strata jumps precisely by one at each step and a full list of allowed elementary slices is known, while an analogous list remains an open question for symplectic singularities.  And it is richer because strata supporting $\U(1)$ gauge theories with massless hypers and trivial Higgs branch they only inherit a \emph{loose special Kahler structure}.   

After explaining this structure in full generality for theories of arbitrary rank, we have worked out explicitly how this works in a large number of examples of rank-2 SCFTs. We were able to apply the newly derived central charge formulae \cite{Martone:2020nsy} to beautifully reproduce the properties of the rank-2 theories from their rank-1 building blocks in all cases, in doing so deriving many new properties of rank-2 SCFTs and their moduli spaces and low-energy description, and checked that the $\cN=2$ UV-IR simple flavor condition \cite{Martone:2020nsy} applies to all examples we have studied.

This work represents a considerable step forward in implementing our classification program of four dimensional $\cN=2$ SCFTs beyond rank 1. At its core our program of probing the space of $\cN=2$ SCFTs at a given rank $r$, is divided into two steps \cite{Argyres:2020nrr}:
\begin{itemize}
\item[1.] Classify all scale invariant Coulomb branch geometries of complex dimension $r$.
\item[2.] Classify all relevant deformations of the entries in item $1$.
\end{itemize} 
The work presented here lays the foundation for progress in regards to item $1$. To systematize this picture there are, at least, two ways of proceeding.

\paragraph{Revisiting the genus 2 scale invariant analysis.} 

Our understanding of properties of $\cN=2$ SCFTs, and their Coulomb branch structure, has improved dramatically since the first attempt at classifying rank-2 geometries by one of the authors \cite{Argyres:2005pp,Argyres:2005wx}. One the main limitations of that approach was its mathematical complexity which resulted in large systems of high-order polynomial equations which could not be solved with current algorithms in a reasonable amount of computing time. It is likely possible to circumvent this obstacle, perhaps completely, by leveraging our current knowledge. In particular, in the analysis of \cite{Argyres:2005pp,Argyres:2005wx} the possible scaling dimensions of the rank-2 Coulomb branch parameters were left as unknowns to be solved for. By now we know that the set of allowed pairs of scaling dimensions is indeed finite \cite{Argyres:2018zay,Caorsi:2018zsq,Argyres:2018urp} and can be inputed from the start. We believe that this will, along with a clever use of scale invariance, considerably simplify the analysis and could perhaps lead to a first principles classification of scale invariant Coulomb branch geometries\footnote{With the assumptions of a freely generated Coulomb branch chiral ring and a principal Dirac pairing.} at rank 2 presented as two parameter families of genus 2 Seiberg Witten (SW) curves. If our study is successful, we will be able to straightforwardly apply the insights provided by the quantum discriminant \cite{Martone:2020nsy} of these geometries to make remarkable progress on their mass deformations as well. We will report a partial update on this search in an upcoming publication \cite{Argyres:2020}.

\paragraph{Generalization to rank-$r$.} 

The method just outlined, becomes most likely unfeasible at rank higher than 2. This is due to the fact that curves for genus equal to or larger than 3 are no longer all hyperelliptic. There is then no practical way to algebraically parametrize all SW curves as families of curves of appropriate genus.  A more practical, and perhaps still very constraining, method might arise by the application of the stratification analysis in the following way. The maximal transverse slice stemming out of any given stratum is associated to the Coulomb branch of the theory supported on the stratum.  As we discussed here, the singular locus for theories of rank higher than one has in general multiple connected components.  This is equivalent, in the Hasse diagram representation, to multiple edges connected to a single node. It is then plausible that for theories of rank strictly higher than 2, the consistency of the nesting arising from the theories supported on the various singular strata provides strong constraints on what theories are allowed and in turn on the set of scale invariant geometries at higher ranks.

We unfortunately have not implemented this method to any significant extent, and it will only be feasible once our understanding of the space of rank-2 theories is systematized.

\acknowledgments We would like to thank Fabio Apruzzi, Antoine Bourget, Simone Giacomelli, Julius Grimminger, Amihay Hanany, Carlo Meneghelli, Sakura Schafer-Nameki, Marcus Sperling, Wolfger Peelaers, Yuji Tachikawa, Gabi Zafrir, and Zhenghao Zhong for many helpful discussions and insightful comments. PCA is supported by DOE grant DE-SC0011784, and MM is supported by NSF grants PHY-1151392 and PHY-1620610.

\bibliographystyle{JHEP}

\end{document}